\documentclass[%
reprint,
nofootinbib,
 amsmath,amssymb,
 aps,
prd,
]{revtex4-2}
\usepackage{xcolor}
\usepackage{graphicx}
\usepackage{dcolumn}
\usepackage{bm}

\usepackage{mathtools}
\usepackage[normalem]{ulem} 
\allowdisplaybreaks

\def\beq{\begin{equation}}
\def\eeq{\end{equation}}
\def\bea{\begin{eqnarray}}
\def\eea{\end{eqnarray}}

\usepackage[colorlinks,
    linkcolor=blue,
    citecolor=teal,
    urlcolor=blue]{hyperref}

\begin{document}


\title{Revisiting the Matter Creation Process: \\Observational Constraints on Gravitationally Induced Dark Energy\\ and the Hubble Tension}

\author{Tiziano Schiavone}
\email{tiziano.schiavone@sissa.it}
\affiliation{SISSA-International School for Advanced Studies, Via Bonomea 265, 34136 Trieste, Italy}
\affiliation{INFN, Sezione di Trieste, Via Valerio 2, I-34127 Trieste, Italy}
\affiliation{Institute of Space Sciences (ICE,CSIC), C. Can Magrans s/n, 08193 Barcelona, Spain}
 
\author{Mariaveronica De Angelis}%
\email{mdeangel@ucm.es}
\affiliation{Departamento de Físíca Teórica, Facultad de Ciencias Físicas, Universidad Complutense de Madrid, 28040 Madrid, Spain}

\author{Luis A. Escamilla}
\email{torresl@itu.edu.tr}
\affiliation{Department of Physics, Istanbul Technical University, Maslak 34469 Istanbul, T\"{u}rkiye}

\author{Giovanni Montani}
\email{giovanni.montani@enea.it}
\affiliation{ENEA, Nuclear Department, C.R. Frascati, Via E. Fermi 45, 00044 Frascati, Italy}
\affiliation{Physics Department, Sapienza University of Rome, P.le A. Moro 5, 00185 Roma, Italy}

\author{Eleonora Di Valentino}
\email{e.divalentino@sheffield.ac.uk}
\affiliation{School of Mathematical and Physical Sciences, University of Sheffield, Hounsfield Road, Sheffield, S3 7RH, South Yorkshire, United Kingdom}

\date{\today}

\begin{abstract}
The Hubble tension and the unknown origin of dark energy motivate the exploration of alternative mechanisms for late-time cosmic acceleration. We investigate gravitationally induced particle creation (PC) as a non-equilibrium process that can effectively mimic dynamical dark energy. Within the thermodynamic framework of open systems, we adopt an agnostic approach to the extra created component, leaving its equation-of-state parameter $w_E$ free. We consider four phenomenological parametrisations of the PC rate, allowing deviations from the standard cosmological model ($\Lambda$CDM) only at late times ($0<z<3$). The PC models are constrained using a joint analysis of cosmic chronometers, Type Ia supernovae, local $H_0$ measurements, baryon acoustic oscillations, and cosmic microwave background data. The constraints on $w_E$ are consistent with dark energy, while particle creation of pressureless matter is disfavoured. All PC scenarios provide fits comparable to $\Lambda$CDM, with one showing effective dynamical dark-energy behaviour. When early- and late-time datasets are analysed separately, the PC models reduce the Hubble tension to $\simeq 2.4\,\sigma$--$3\,\sigma$, compared to $4.3\,\sigma$ in $\Lambda$CDM. Gravitationally induced dark energy thus offers a consistent late-time extension of $\Lambda$CDM and a viable theoretical framework for dynamical dark energy.
\end{abstract}

\maketitle




\section{Introduction}

The standard cosmological model, the $\Lambda$CDM paradigm, provides an excellent description of a wide range of observational data, including measurements of the cosmic microwave background~\cite{WMAP:2012fli,Planck:2018vyg,Planck:2018nkj,ACT:2020gnv,AtacamaCosmologyTelescope:2025blo,SPT-3G:2025bzu}, large-scale structure~\cite{Beutler:2011hx,Ross:2014qpa,BOSS:2016wmc,eBOSS:2020yzd,DES:2021wwk,DESI:2025zgx,DESI:2025wyn,Wright:2025xka}, and the expansion history of the Universe~\cite{Brout:2022vxf,Scolnic:2021amr,DES:2024jxu,Moresco:2016mzx,Moresco:2023zys}. Despite its remarkable empirical success, the model relies on the presence of a cosmological constant (or dark energy) to explain the observed late-time acceleration of the Universe, whose fundamental physical origin remains theoretically unexplained~\cite{Weinberg:1988cp}. This limitation has motivated the investigation of alternative mechanisms, among which gravitationally induced particle creation has emerged as a particularly appealing scenario.


The idea that particle creation may occur in an expanding Universe has a well-established theoretical foundation in quantum field theory in curved spacetime. In this framework, the time dependence of the gravitational background leads to a mismatch between vacuum states defined at different epochs, resulting in the spontaneous production of particles (gravitational particle production). This mechanism was first systematically investigated in the context of scalar fields in an expanding and isotropic Universe~\cite{Parker:1968mv,Parker:1969au}, and later extended to anisotropic cosmologies~\cite{Zeldovich:1971mw}. These studies demonstrated that a time-dependent gravitational field can itself act as a source of particle production through vacuum polarisation effects, providing a fundamental physical mechanism for matter creation. Although these models were initially developed to describe the early Universe, particle creation may persist throughout cosmic history and play a role at late times.

In a cosmological context, however, one is typically not concerned with the detailed microphysical description of individual particles, but rather with the macroscopic properties of the cosmic fluid, based on an averaged description of the underlying microphysical processes. The standard approach relies on a coarse-grained, effective description in which the relevant quantities are the averaged energy density, pressure, and particle number density of the fluid components. Thus, gravitational particle production can be formulated within the framework of non-equilibrium thermodynamics and open-system cosmology~\cite{Prigogine:1989zz,Calvao:1991wg}. Within this framework, particle creation is incorporated phenomenologically as a non-equilibrium process affecting the continuity equation of the cosmic fluid, leading to an effective negative pressure term. Therefore, this mechanism can naturally drive cosmic acceleration without the need for an explicit dark-energy component.

This approach has been widely employed in the literature, particularly in scenarios involving the particle creation of cold dark matter~\cite{Freaza:2002ic,Pan:2016jli,Cardenas:2020grl,Elizalde:2024rvg}.
This choice is motivated by several arguments: particle creation could offer an explanation for the origin of dark matter itself, naturally generate the negative pressure required for accelerated expansion, and potentially alleviate the coincidence problem~\cite{Zlatev:1998tr} by treating dark matter and dark energy as different manifestations of a single unified component.

More recently, the particle creation framework has been extended beyond cold dark matter to include exotic fluids and generalized relativistic species. For instance,~\citet{CostaNetto:2025vew} performed a thermodynamic analysis of dark energy modeled as a relativistic fluid undergoing particle creation and annihilation. A more general multi-fluid extension was proposed by~\citet{Trevisani:2023wpw}, who studied gravitationally induced particle production in a decoupled cosmology involving baryons, photons, neutrinos, and dark matter. Additional developments include the gravitational production of a reduced relativistic gas~\cite{Lima:2025yza} and connections with entropic cosmology~\cite{Gohar:2020bod}, suggesting intriguing links between information-theoretic concepts and matter creation.

The versatility of the particle creation framework is further illustrated by its successful implementation within modified gravity theories. Matter creation naturally arises in scalar--tensor and $f(R,T)$ models~\cite{Cipriano:2023yhv}, energy--momentum squared gravity~\cite{Cipriano:2024jng,Akarsu:2023nyl}, and $f(R)$ theories~\cite{Singh:2024nsh}, often leading to distinctive phenomenological signatures. Related scenarios have been explored in $f(R)$ models with entropic production~\cite{Pinto:2022tlu}, scalar--tensor theories with evolving vacuum energy~\cite{Montani:2024xys}, and effective fluid descriptions in non-Riemannian geometries~\cite{Marciu:2024gqv}. These models are increasingly being tested against cosmological observations, particularly those probing the background expansion and the growth of large-scale structure~\cite{Ganjizadeh:2022mxe,Bouali:2023fid}. From a thermodynamic perspective, their internal consistency is supported by the validity of the generalized second law~\cite{Cardenas:2023zmn}, while phenomenologically they share similarities with bulk viscous models~\cite{Cardenas:2020exv}, exotic fluids such as Chaplygin gases~\cite{Cardenas:2025sqf,Bolotin:2020qbx}, and unified dark energy scenarios that may exhibit phantom-like behavior.

Motivated by these developments, we extend the investigation of particle creation mechanisms in the context of late-time cosmology. In this work, we contribute to this growing line of research by generalising cosmological models based on 
the effective fluid description of an underlying gravitationally induced particle creation. Rather than focusing on cold dark matter, we propose the gravitational production of an unknown particle species, characterised by an arbitrary equation-of-state parameter treated as a free variable. 
This approach is intentionally phenomenological and agnostic, aiming to constrain particle-creation models directly with observational data without assuming the nature of the created particles.
We 
consider negligible creation rates for radiation, baryons, and cold dark matter (\textit{i.e.}, their particle numbers remain conserved as the Universe expands), while the created species effectively behaves as a dark energy component. We then explore the implications of this mechanism for late-time cosmic acceleration and examine its potential role in addressing the Hubble tension, providing an alternative interpretation of the observed expansion history of the Universe.

The Hubble tension, referring to the discrepancy between local and early-Universe measurements of the Hubble constant $H_0$, has been extensively documented in the literature~\cite{Verde:2019ivm,Vagnozzi:2019ezj,DiValentino:2020zio,DiValentino:2021izs,Perivolaropoulos:2021jda,Schoneberg:2021qvd,Shah:2021onj,Abdalla:2022yfr,DiValentino:2022fjm,Kamionkowski:2022pkx,Giare:2023xoc,Hu:2023jqc,Verde:2023lmm,DiValentino:2024yew,Perivolaropoulos:2024yxv,CosmoVerseNetwork:2025alb}. More specifically, the Planck collaboration reports $H_0^{\mathrm{P}} = (67.36 \pm 0.54)\,\mathrm{km\,s^{-1}\,Mpc^{-1}}$ from cosmic microwave background radiation (CMB) data~\cite{Planck:2018vyg} (TT, TE, EE + lowE + lensing), while the SH0ES collaboration finds a significantly higher local value, $H_0^{\mathrm{loc}} = (73.04 \pm 1.04)\,\mathrm{km\,s^{-1}\,Mpc^{-1}}$, based on Cepheid-calibrated Type~Ia supernovae (SNe Ia)~\cite{Riess:2021jrx}, corresponding to a $\sim5\sigma$ discrepancy. More recently, this tension has increased to the $7.1\sigma$ level when combining the latest CMB measurements from Planck+SPT+ACT~\cite{SPT-3G:2025bzu} with the updated local measurement from the H0DN network~\cite{H0DN:2025lyy}, which provides a new consensus value for the distance-ladder determinations.

In response to the Hubble tension, a wide variety of models have been proposed, 
including those invoking gravitational particle production~\cite{Elizalde:2024rvg,Erdem:2024vsr,Erdem:2025xtr}, modified gravity~\cite{Odintsov:2020qzd,Schiavone:2022shz,Schiavone:2022wvq,Nojiri:2022ski,Montani:2023xpd,Escamilla:2024xmz,Schiavone:2024heb,Odintsov:2024woi,Montani:2025jkk,Efstratiou:2025iqi,DOnofrio:2025cuk,Valletta:2025bgu,Montani:2025nmz}, 
self-interacting scalar fields~\cite{Montani:2023ywn}, bulk viscosity~\cite{Montani:2024ntj,Navone:2025gxr}, 
and interactions within the dark sector~\cite{Kumar:2016zpg,Murgia:2016ccp,Kumar:2017dnp,DiValentino:2017iww,Kumar:2021eev,Gao:2021xnk,Pan:2023mie,Benisty:2024lmj,Yang:2020uga,Forconi:2023hsj,Pourtsidou:2016ico,DiValentino:2020vnx,DiValentino:2020leo,Nunes:2021zzi,Yang:2018uae,vonMarttens:2019ixw,Lucca:2020zjb,Gao:2022ahg,Zhai:2023yny,Bernui:2023byc,Hoerning:2023hks,Giare:2024ytc,Escamilla:2023shf,vanderWesthuizen:2023hcl,Silva:2024ift,DiValentino:2019ffd,Li:2024qso,Pooya:2024wsq,Halder:2024uao,Castello:2023zjr,Yao:2023jau,Mishra:2023ueo,Nunes:2016dlj,Silva:2025hxw,Yang:2025uyv,vanderWesthuizen:2025rip,Montani:2024pou,Zhang:2025dwu,Li:2025owk,Li:2026xaz}. 
Other studies have revisited the statistical assumptions and binning procedures used in cosmological data analyses, 
suggesting that methodological biases might contribute to the observed discrepancy~\cite{Dainotti:2023bwq,Dainotti:2023ebr,Dainotti:2024gca,Bargiacchi:2023jse,Dainotti:2024aha}. 
Recent investigations have also pointed to a possible redshift evolution in the inferred values of $H_0$, $\Omega_{m0}$, 
or the absolute magnitude of Type~Ia supernovae within standard $\Lambda$CDM fits across different redshift bins~\cite{H0LiCOW:2019pvv,Kazantzidis:2020tko,Krishnan:2020obg,dainottiApJ-H0(z),DainottiGalaxies-H0(z),Schiavone:2022shz,Colgain:2022tql,Dainotti:2023yrk,Jia:2022ycc,Malekjani:2023ple,Colgain:2022rxy,Liu:2024vlt,DeSimone:2024lvy,Montani:2023ywn,Schiavone:2024heb,Dainotti:2025qxz,Wang:2025xvi,Jia:2024wix,Jia:2025poj,Kalita:2025jqz,Valletta:2025bgu,Dai:2026pvx}. 
These trends have been interpreted as potential evidence for a local underdensity, 
departures from General Relativity, or the need for a dynamical dark energy component, \textit{e.g.}, the $w_0w_a$CDM model according to the Chevallier-Polarski-Linder (CPL) parameterization~\cite{Chevallier:2000qy,Linder:2002et}.

In this context, recent studies~\cite{Fazzari:2025mww,Montani:2025rcy,Navone:2025gxr} have investigated matter creation as a viable mechanism to address the Hubble tension in connection with evolving dark energy models. These results further support the idea that gravitationally induced particle production may provide a unified, thermodynamically consistent framework capable of explaining both late-time acceleration and the observed tensions in cosmological data.

The paper is organized as follows. In Sect.~\ref{sec:particle-creation-in-cosmology}, we introduce the thermodynamic framework of open systems in cosmology. In Sect.~\ref{sec:particle-creation-models}, we present four phenomenological parametrizations of the particle-creation rate describing the late Universe.
Sect.~\ref{sec:constraints} discusses the theoretical constraints on the model parameters, while Sect.~\ref{sec:datasets-methodology} describes the observational datasets and statistical methodology employed. The main cosmological results are presented in Sect.~\ref{sec:cosmological-results}, with particular emphasis on late-time acceleration, the effective dark energy behavior, and the Hubble tension. Finally, Sect.~\ref{sec:conclusions} summarizes our conclusions and outlines possible directions for future work.


\section{Particle creation in cosmology} \label{sec:particle-creation-in-cosmology}

In this section, we develop the thermodynamic framework describing gravitationally induced particle creation, which provides the theoretical foundation for the cosmological models investigated in the following sections. The idea that a time-dependent gravitational field can lead to matter production can be phenomenologically formulated within the non-equilibrium thermodynamic approach developed in Refs.~\cite{Calvao:1991wg,Pan:2016jli,Cardenas:2020grl,Elizalde:2024rvg}. 

A key assumption of this framework is the existence of a particle species whose total number $N$ within a physical volume $V$ is not conserved. 
Consequently, the standard particle number conservation equation is modified to
\begin{equation}
\nabla_\mu N^\mu = n \Gamma \,, \label{eq:particlecreat}
\end{equation}
where $\nabla_\mu$ denotes the covariant derivative compatible with the underlying spacetime metric (to be specified later), 
the coordinate indices are $\mu = \{0, \dots, 3\}$, $N^\mu = n u^\mu$ is the particle flux vector, 
$n = N/V$ is the number density, $u^\mu$ is the particle four-velocity, 
and $\Gamma$ represents the particle creation (or annihilation) rate. 
The source term $n \Gamma$ therefore quantifies the rate of change of the particle number per unit volume due to non-conservation. 
Using the definition of $N^\mu$, Eq.~\eqref{eq:particlecreat} can equivalently be written as
\begin{equation}
\partial_\mu n\, u^\mu + \Theta n = n \Gamma \,, \label{eq:evolution-n}
\end{equation}
where $\Theta \equiv \nabla_\mu u^\mu$ is the expansion scalar. 
Eq.~\eqref{eq:evolution-n} follows directly from taking the covariant divergence of the particle flux and is commonly employed in the study of relativistic fluids (see, e.g.~\cite{Pan:2016jli}).

Since the functional form of $\Gamma$ is not known \textit{a priori}, different particle creation models can be explored 
and subsequently constrained by observational data. 
When particle number is not conserved, the first and second laws of thermodynamics for open systems lead to the generalized relation
\begin{equation}
    dU = -p\,dV + T\,dS + \mu\,dN \,,
    \label{mcm1}
\end{equation}
where $U$ is the internal energy, $p$ is the pressure, $T$ is the temperature, $S$ is the entropy, and $\mu$ is the chemical potential. Writing $U=\rho V$ and $S=\sigma N$, with $\rho$ the energy density and $\sigma$ the specific entropy per particle, the chemical potential follows from the Gibbs relation~\cite{Lima:2007kk}, yielding $\mu=(\rho+p)V/N-T\sigma$. Substituting this expression into Eq.~\eqref{mcm1} gives
\begin{equation}
    d\rho = -(\rho + p)\left(1 - \frac{d\ln N}{d\ln V}\right)\frac{dV}{V} + T\,\frac{N}{V}\,d\sigma \,.
    \label{mcm2}
\end{equation}
In contrast to a isentropic system, we only require the specific entropy to be conserved, \textit{i.e.}, $d\sigma=0$. Since $\sigma=S/N$, this implies $S\propto N$, indicating that particle creation increases the total entropy. This assumption corresponds to adiabatic particle production, where entropy increases due to the growth in particle number rather than changes in the entropy per particle, which is a standard condition in cosmological particle-creation models. 

We now consider a fiducial volume in a spatially flat, homogeneous, and isotropic Universe, consistent with \textit{Planck} satellite observations~\cite{Planck:2018vyg}. 
The spacetime geometry on large cosmological scales is described by the Friedmann-Leme\^itre-Robertson-Walker (FLRW) line element~\cite{Weinberg:2008zzc}:
\begin{equation}
    ds^2 = dt^2 - a^2(t)\left[dr^2 + r^2\left(d\theta^2 + \sin^2{\theta}\,d\phi^2\right)\right] \,,
    \label{mcm4}
\end{equation}
where $t$ is the comoving time, $\{r,\theta,\phi\}$ are spherical coordinates, and $a(t)$ is the scale factor. 
We adopt natural units ($c = 1$). 
Throughout this work we assume General Relativity as the underlying theory of gravity within a FLRW spacetime. Then, the dynamics of cosmic expansion in the FLRW metric are governed by the well-known Friedmann equation~\cite{Weinberg:2008zzc},\footnote{We stress that Eq.~\eqref{eq:Friedmann} retains its standard form since we do not modify the gravitational sector, and particle-creation effects are fully encoded in a modified conservation equation for the cosmic fluid, or equivalently in an effective redefinition of the energy-momentum tensor, as detailed later in this section. The Friedmann equation has been widely adopted in the literature on cosmological particle creation within General Relativity~\cite{Prigogine:1989zz,Calvao:1991wg,Pan:2016jli,Cardenas:2020grl,Elizalde:2024rvg}.}
\begin{equation}
    H^2 \equiv \left(\frac{\dot{a}}{a}\right)^2 = \frac{\chi}{3}\rho \,,
    \label{eq:Friedmann}
\end{equation}
where $\chi = 8\pi G$ is the Einstein constant, $H = H(t)$ is the Hubble parameter, and an overdot denotes differentiation with respect to time. 
The total energy density $\rho = \rho_m + \rho_r + \rho_E$ 
includes matter (\textit{e.g.}, baryons and dark matter), radiation, 
and an extra component $E$ associated with particle creation. 
We assume that only this additional component exhibits a non-vanishing particle creation rate (\textit{i.e.}, $\Gamma_E \neq 0$), 
while $\Gamma_m = \Gamma_r = 0$. 
For simplicity, we drop the subscript $E$ and denote its creation rate by $\Gamma$. 
We exclude a cosmological constant, as our goal is to attribute the observed cosmic acceleration to the effects of particle production.

In the FLRW background, Eq.~\eqref{eq:evolution-n} for the extra species becomes
\begin{equation}
    \dot{n}_E + 3H n_E = \Gamma n_E  \,,
\end{equation}
where, in the comoving gauge, $u_E^\mu = \delta^\mu_t = (1,0,0,0)$ and the scalar expansion is $\Theta = 3H$. Note that the particle number within a physical volume $V$ is not conserved due to both cosmic expansion and particle creation. In particular, while the expansion alone would preserve the particle number within a comoving volume, the presence of a nonzero creation rate $\Gamma$ implies that particle number is not conserved even in a comoving volume.
If $\Gamma = 0$ or $\Gamma \ll 3H$, particle creation is negligible and the standard conservation law in an expanding Universe is recovered. 
For $\Gamma > 0$ ($\Gamma < 0$), particles are produced (annihilated). 

Assuming that the specific entropy per particle remains constant within a fiducial volume in a homogeneous and isotropic Universe, 
Eq.~\eqref{mcm2} reduces to
\begin{equation}
    \dot{\rho}_E = 
    -\left(\rho_E + p_E\right) \left(1 - \frac{d\ln N_E}{d\ln V}\right) \frac{d\ln V}{dt} \,,
    \label{mcm3}
\end{equation}
where $V = a^3 V_0$ in an expanding Universe, with $V_0$ a constant comoving volume. 
This leads to the generalised continuity equation for the extra component,
\begin{equation}
    \dot{\rho}_E + 3H\left(\rho_E + p_E\right) = \Gamma\left(\rho_E + p_E\right) \,,
    \label{eq:continuity-with-Gamma}
\end{equation}
where the particle creation rate is defined as
\begin{equation}
    \Gamma \equiv \frac{\dot{N}_E}{N_E} \,.
    \label{eq:def-Gamma}
\end{equation}
The case $\Gamma = 0$ recovers the standard conservation equation for the extra component.
Meanwhile, the matter and radiation components are assumed to be conserved separately, following the usual continuity equations:
\begin{equation}
    \dot{\rho}_m + 3H\left(\rho_m + p_m\right) = 0 \,, \qquad
    \dot{\rho}_r + 3H\left(\rho_r + p_r\right) = 0 \,.
    \label{eq:continuity-matter-and-radiation}
\end{equation}

To study late-time cosmic acceleration, we begin by differentiating Eq.~\eqref{eq:Friedmann} with respect to time, and combining it with Eqs.~\eqref{eq:continuity-with-Gamma} and~\eqref{eq:continuity-matter-and-radiation}. 
This yields the modified Raychaudhuri equation,
\begin{equation}
    \dot{H} = -\frac{\chi}{2}\left(\rho + p\right) + \frac{\chi\,\Gamma}{6H}\left(\rho_E + p_E\right) \,.
\end{equation}
Using $\dot{H} = \ddot{a}/a - H^2$, we derive the cosmic acceleration equation including the effects of particle creation,
\begin{equation}
    \frac{\ddot{a}}{a} = -\frac{\chi}{6}\left(\rho_m + 2\rho_r\right)
    - \frac{\chi \rho_E}{6}\left[ 1 + 3w_E - \frac{\Gamma}{H}\left(1 + w_E\right)\right] \,,
    \label{eq:cosmic-accel}
\end{equation}
where we assume a barotropic equation of state $p_i = w_i \rho_i$ for each component, 
with $w_m = 0$ for pressureless matter, $w_r = 1/3$ for radiation, and $w_E$ treated as a constant free parameter for the extra species. The assumption of constant $w_E$ represents a minimal phenomenological choice that reduces the number of free parameters while capturing deviations from matter and radiation. 

The second term on the right-hand side of Eq.~\eqref{eq:cosmic-accel} represents the contribution of the extra component and particle production. 
This term can drive acceleration (\textit{i.e.}, dark-energy-like behavior) even in the absence of exotic fluids with intrinsic negative pressure. 
Assuming $\Gamma, H, \rho_i > 0$, the condition for present-day acceleration is
\begin{equation}
    w_E\left(3 - \frac{\Gamma_0}{H_0}\right) < \frac{\Gamma_0}{H_0} - 1 \,,
    \label{eq:condition-wE}
\end{equation}
where the subscript “$0$” denotes present-day values. 
The direction of the inequality depends on the sign of the factor $\left(3 - \frac{\Gamma_0}{H_0}\right)$, which is not known \textit{a priori}. 
If $\Gamma = 0$, the standard condition for dark energy, $w_E < -1/3$, is recovered.

An alternative way to describe the non-equilibrium thermodynamic effects of particle creation 
is to introduce an effective pressure term, denoted by $\Pi$, into the energy-momentum tensor of the extra fluid. 
The modified tensor takes the form
\begin{equation}
    T_{\mu\nu}^{(E)} = \left(\rho_E + p_E + \Pi\right) u^E_\mu u^E_\nu + \left(p_E + \Pi\right) g_{\mu\nu} \,.
    \label{eq:modified-energy-momentum-tensor}
\end{equation}
The non-equilibrium pressure is related to the particle creation rate by
\begin{equation}
    \Pi = -\frac{\Gamma}{3H}\left(\rho_E + p_E\right) \,.
    \label{eq:non-equilibrium-pressure}
\end{equation}
This effective negative pressure provides a macroscopic description of irreversible particle-production processes within an expanding Universe. The effects of particle creation are entirely incorporated at the level of the energy--momentum tensor $T_{\mu\nu}^{(E)}$. This approach is consistent with the Bianchi identities, and hence with the covariant conservation of $T_{\mu\nu}^{(E)}$, without any violation of the prescriptions of General Relativity. Indeed, it can be shown that
the continuity equation~\eqref{eq:continuity-with-Gamma}
can be equivalently derived from the condition $\nabla_\nu T^{\mu\nu}_{(E)} = 0$, using the modified tensor~\eqref{eq:modified-energy-momentum-tensor} and the non-equilibrium pressure~\eqref{eq:non-equilibrium-pressure}. 
Thus, particle creation can effectively generate a negative pressure, 
thereby contributing to the accelerated expansion of the Universe.

To conclude this section, if the extra component is interpreted as cold dark matter 
and only one species is considered while neglecting relativistic particles at late times, 
our results reduce to those obtained in~\cite{Freaza:2002ic,Pan:2016jli,Cardenas:2020grl,Elizalde:2024rvg} 
(with “E” corresponding to “m” and $w_m = 0$). 
A similar treatment applies to the case of light relativistic particles, 
by identifying “E” with “r” and setting $w_r = 1/3$. 
More generally, we consider the gravitational production of particles characterized by an unspecified 
equation-of-state parameter $w_E$, which may effectively mimic dark energy in the late Universe. 
The strength of this approach lies in its agnosticism regarding the nature of the created particles while remaining rooted in a well-defined thermodynamic and relativistic framework, 
leaving $w_E$ as a free parameter to be constrained by observations. 
In Sect.~\ref{sec:cosmological-results}, we derive observational constraints on $w_E$ 
and on the associated cosmological parameters for different particle creation models.


\renewcommand{\arraystretch}{1.3} 
\begin{table} \begin{centering} \begin{tabular}{|c|l|l|c|c|} \hline \multicolumn{5}{|c|}{PC Models}\tabularnewline 
\hline Model & $\Gamma$: PC rate & $\#$ extra params & $g(z)$ & $\Lambda$CDM limit \tabularnewline 
\hline PC1 & \, $3 \beta H_0 \left(\frac{H}{H_0}\right)^{\alpha}$ & \quad 3: $\alpha$, $\beta$, $w_E$ & N & $\beta=0$ \tabularnewline 
\hline PC2 & \, $3\mu H$ & \quad 1: $\xi$ & A & $\mu = 0\;(\xi = 0)$\tabularnewline
\hline PC3 & \, $3\gamma H_0$ & \quad 2: $\gamma$, $w_E$ & N & $\gamma = 0$ \tabularnewline 
\hline PC4 & \, $3\gamma H_0+3\mu H$ & \quad 3: $\gamma$, $\mu$, $w_E$ & N  & $\gamma = \mu = 0$\tabularnewline
\hline \end{tabular}
\caption{The functional form of the particle–creation rates and the number of additional free parameters entering the dimensionless Hubble function $E(z)$ for each PC model are listed in the second and third columns, respectively. Note that, for the PC2 model, the only extra parameter affecting the cosmological dynamics is $\xi = 3(1+w_E)(1-\mu)$. The
fourth column indicates whether the continuity equation for the dimensionless energy density $g(z)$ admits an analytical solution (A) or requires a numerical one (N). The fifth column indicates the conditions for recovering $\Lambda$CDM for each PC model together with the condition $w_{ E}=-1$.} 
\label{TableGamma} \par\end{centering} \end{table}

\section{Phenomenology of the particle creation process with different particle production rates}\label{sec:particle-creation-models}


In this section, we investigate different scenarios of gravitationally induced particle production, each characterised by a specific parametrisation of the particle-creation rate $\Gamma$. Motivated by theoretical studies in quantum field theory in curved spacetime~\cite{Parker:1968mv,Parker:1969au,Zeldovich:1971mw}, a natural assumption is to relate $\Gamma$ to the Hubble function~\cite{Cardenas:2020grl,Elizalde:2024rvg}, reflecting the role of the expanding background as the source of particle production.

Since $\Lambda$CDM provides an excellent description of current observations, especially at high redshift, any viable extension should recover its behaviour in this regime. Therefore, we follow a restrictive approach by considering minimal and slight deviations from $\Lambda$CDM at late times, driven by particle creation: we assume that the particle-creation process operates exclusively in the late Universe ($0 < z < 3$), while at higher redshifts we impose that the standard $\Lambda$CDM behaviour is recovered. Our aim is to investigate the impact of these deviations from $\Lambda$CDM while preserving the successful $\Lambda$CDM description of early cosmology, still providing the observed late-time acceleration, and potentially alleviating the Hubble tension.

The cosmological dynamics, governed by the Friedmann equation~\eqref{eq:Friedmann}, can be equivalently expressed in terms of the dimensionless Hubble function $E(z)$, defined through $H(z)=H_0 E(z)$, such that
\begin{equation}
    E(z) = \sqrt{\Omega_{m0}(1+z)^3 + \Omega_{r0}(1+z)^4 + \Omega_{E0}\,g(z)} \,.
    \label{eq:new-Hubble-parameter}
\end{equation}
The redshift variable is defined as $z(t) = a_0/a(t) - 1$, with $a_0 = 1$ denoting the present-day value of the scale factor. 
The cosmological density parameters today are defined as $\Omega_{i0} = \rho_{i0}/\rho_{c0}$, 
where $i$ corresponds to matter, radiation, or the extra component, 
and $\rho_{c0} = 3H_0^2/\chi$ is the present critical energy density. 
We express the energy density of the extra component as $\rho_E(z) = \rho_{E0}\,g(z)$, 
where $g(0) = 1$ by construction. 
The function $g(z)$ is determined by the continuity equation, which depends on the specific choice of $\Gamma$, 
and can be obtained either analytically or numerically. 
Evaluating Eq.~\eqref{eq:new-Hubble-parameter} at $z = 0$ recovers the standard normalization condition 
$\Omega_{m0} + \Omega_{r0} + \Omega_{E0} = 1$. 
Within this framework, the extra component due to the particle creation process behaves as a dynamical dark energy candidate through the function $g(z)$.

The continuity equation (see Eq.~\eqref{eq:continuity-with-Gamma}) can be rewritten in terms of the redshift as
\begin{equation}
    \rho_E^\prime(z) = \frac{3(1+w_E)}{1+z}\left(1 - \frac{\Gamma}{3H}\right)\rho_E(z)\,,
    \label{eq:continuity-with-Gamma-in-z}
\end{equation}
or equivalently, using the relations $H(z) = H_0 E(z)$ and $\rho_E(z) = \rho_{E0}\,g(z)$,
\begin{equation}
    g^\prime(z) = \frac{3(1+w_E)}{1+z}\left(1 - \frac{\Gamma}{3H_0 E(z)}\right)g(z)\,,
    \label{eq:continuity-g-with-Gamma-in-z}
\end{equation}
where the prime denotes differentiation with respect to redshift. 
The chain rule $\dot{(...)} = -(1+z)H(... )^\prime$, obtained from the definitions of redshift and the Hubble parameter, 
is used to express time derivatives in terms of redshift derivatives.

Eqs~\eqref{eq:new-Hubble-parameter} and~\eqref{eq:continuity-g-with-Gamma-in-z} together form a coupled system 
for the two unknown functions $E(z)$ and $g(z)$. 
Once a form for the particle creation rate $\Gamma$ is chosen
and the initial condition $g(0) = 1$ is imposed, 
this system can be solved to determine the cosmological evolution of the model. In practice, this system is solved in the redshift range $0< z < 3$, where deviations from $\Lambda$CDM are allowed. At higher redshifts ($z\geq 3$), we impose by construction that the model exactly matches the $\Lambda$CDM expansion history by fixing the dimensionless Hubble function $E(z)$ to its standard form in $\Lambda$CDM. This effectively switches off the evolution of $g(z)$ in this regime, ensuring consistency with early-Universe observables.

The four PC models considered in this work, together with their particle–creation rates and main features, are summarised in Table~\ref{TableGamma} and described in detail below. The adopted parametrisations are intended to capture generic deviations from standard cosmology while remaining sufficiently simple to allow for direct observational constraints. Furthermore, considering different PC models and treating them independently allows us to test whether current data favour simpler parametrisations (with a reduced number of free parameters) and to isolate the role of specific terms in $\Gamma$ (\textit{e.g.}, constant, linear, or power-law dependence on the Hubble function $H$) in driving deviations from $\Lambda$CDM.

\subsection{PC1 model}
We begin by considering the following ansatz for the PC rate, as proposed in~\cite{Freaza:2002ic,Cardenas:2020grl,Elizalde:2024rvg}
\begin{equation}
    \Gamma = 3 \beta H_0 \left(\frac{H}{H_0}\right)^{\alpha} \,,
    \label{eq:condition-Gamma}
\end{equation}
where $\alpha$ and $\beta$ are dimensionless constants. 
This choice describes a dynamical and continuous creation process, 
in which the particle production rate evolves with the expansion rate of the Universe. 
Equivalently, the number of created particles increases with cosmic time as the Universe expands.

Substituting Eq.~\eqref{eq:condition-Gamma} into Eq.~\eqref{eq:continuity-g-with-Gamma-in-z} yields
\begin{equation}
    g^{\prime}(z) = 3\,\frac{1+w_E}{1+z}\left[1 - \beta\,E(z)^{\alpha-1}\right] g(z)\,,
    \label{eq:continuity2}
\end{equation}
where $E(z)$ is defined in Eq.~\eqref{eq:new-Hubble-parameter}. 
This model is characterised by three free parameters, $\alpha$, $\beta$, and $w_E$, 
and the function $g(z)$ is obtained numerically for $0 < z < 3$ from the initial condition $g(0) = 1$. The $\Lambda$CDM limit is recovered when $\beta\rightarrow 0$ and $w_E\rightarrow -1$ for finite values of $\alpha$.

\subsection{PC2 model}
We now consider an alternative parametrisation of the particle creation rate
\begin{equation}
    \Gamma = 3\mu H \,,
\end{equation}
in which it scales linearly with the expansion rate of the Universe. 
For this choice, the continuity equation admits an exact analytical solution:
\begin{equation}
    g(z) = (1+z)^{\xi} \,,
\end{equation}
where $\xi = 3(1+w_E)(1-\mu)$. 
The corresponding Hubble function is
\begin{equation}
    E(z) = \sqrt{\Omega_{m0}(1+z)^3 + \Omega_{r0}(1+z)^4 + \Omega_{E0}(1+z)^{\xi}} \,.
\end{equation}
This formulation is equivalent to a $w$CDM scenario~\cite{Turner:1997npq,Linder:2002et,DESI:2025zgx} with a constant effective equation-of-state parameter for dark energy,
\begin{equation}
    w^{\mathrm{eff}}_{\mathrm{DE}} = -1 + \frac{\xi}{3} \,.
\end{equation}
Moreover, the PC1 model reduces to this PC2 case when $\alpha = 1$ and $\beta = \mu$. The $\Lambda$CDM limit is recovered when $\mu\rightarrow 0$ and $w_E\rightarrow -1$, or equivalently $\xi \rightarrow 0$.

\subsection{PC3 model}
We now consider the case of a constant particle creation rate,
\begin{equation}
    \Gamma = 3\gamma H_0 \,.
\end{equation}
In this scenario, the continuity equation becomes
\begin{equation}
    g^\prime(z) = 3\,\frac{1+w_E}{1+z}\left(1 - \frac{\gamma}{E(z)}\right) g(z) \,.
\end{equation}
This model is characterised by two free parameters, $\gamma$ and $w_E$. 
As in the PC1 setup, the function $g(z)$ is obtained numerically for $0 < z < 3$ from the initial condition $g(0) = 1$, 
in combination with Eq.~\eqref{eq:new-Hubble-parameter}. 
Note that the PC1 model reduces to PC3 when $\alpha = 0$ and $\beta = \gamma$. Trivially, the $\Lambda$CDM limit is achieved when $\gamma\rightarrow 0$ and $w_E\rightarrow -1$.

\subsection{PC4 model}
Finally, we examine a mixed ansatz that combines both constant and time-dependent contributions
\begin{equation}
    \Gamma = 3\gamma H_0 + 3\mu H \,.
\end{equation}
With this choice, the continuity equation becomes
\begin{equation}
    g^\prime(z) = 3\,\frac{1+w_E}{1+z}\left(1 - \mu - \frac{\gamma}{E(z)}\right) g(z) \,.
\end{equation}
This PC4 model involves three free parameters, $\gamma$, $\mu$, and $w_E$. 
The function $g(z)$ is obtained numerically for $0 < z < 3$ from the initial condition $g(0) = 1$, 
together with Eq.~\eqref{eq:new-Hubble-parameter}. The $\Lambda$CDM limit is recovered when $\gamma\rightarrow 0$, $\mu\rightarrow0$, and $w_E\rightarrow -1$.


\section{Theoretical constraints for cosmic acceleration} \label{sec:constraints}

In this section, we examine the cosmological implications of particle-creation mechanisms and compare them with the standard cosmological-constant paradigm. Specifically, we analyse the deceleration parameter and derive the effective dark-energy equation of state.

\subsection{Cosmic acceleration}
The dynamics induced by the four PC models discussed above must be able to explain the current accelerated expansion by satisfying the condition in Eq.~\eqref{eq:condition-wE}. For each model, this requirement translates into the following constraints:
\begin{align}
    & \textrm{PC1:} \qquad  w_E\, (1-\beta) < \beta - \frac{1}{3}\,,\label{eq:condition-wE-model1}\\
    & \textrm{PC2:} \qquad  w_E\, (1-\mu) < \mu - \frac{1}{3} \,,\label{eq:condition-wE-model2}\\
    & \textrm{PC3:} \qquad  w_E\, (1-\gamma) < \gamma - \frac{1}{3} \,,\label{eq:condition-wE-model3} \\
    & \textrm{PC4:} \qquad  w_E\, (1-\gamma-\mu) < \gamma + \mu - \frac{1}{3} \label{eq:condition-wE-model4} \,.
\end{align}
To study the transition from a decelerated to an accelerated expansion of the Universe, we use the general definition of the deceleration parameter
\begin{align}
    q(z) &= -\frac{\ddot{a}}{a H^2}=-\Big(1+\frac{\dot{H}}{H^2}\Big)\nonumber\\
    &= -1+\frac{1}{2}\left[\frac{1+z}{E^2(z)} \frac{d}{dz}\Big(E^2(z)\Big)\right]\,.\label{eq:deceleration-parameter-general-definition}
\end{align}
Assuming an extra component induced by the gravitational field, with the Hubble function given in Eq.~\eqref{eq:new-Hubble-parameter}, the deceleration parameter becomes
\begin{widetext}
\begin{equation}
    q(z)= -1+\frac{3}{2}\left[\frac{\Omega_{m0}\left(1+z\right)^3 + \frac{4}{3}\Omega_{r0}\left(1+z\right)^4 + \Omega_{E0} g(z) \left(1+w_E\right)}{\Omega_{m0}\left(1+z\right)^3 + \Omega_{r0}\left(1+z\right)^4 + \Omega_{E0} g(z)}\right]
    - \frac{\Gamma}{2H}\,\frac{\Omega_{E0} g(z)\left(1+w_E\right)}{\Omega_{m0}\left(1+z\right)^3 + \Omega_{r0}\left(1+z\right)^4 + \Omega_{E0} g(z)}\,.
    \label{eq:deceleration-parameter-with-Gamma}
\end{equation}
\end{widetext}

By considering the $\Gamma$-rates of the PC models, we can rewrite the last term of Eq.~\eqref{eq:deceleration-parameter-with-Gamma} as:
\begin{align}
    & \textrm{PC1:} \quad -\frac{3\beta}{2} \frac{\Omega_{E0} g(z)\left(1+w_E\right)}{\left[E(z)\right]^{3-\alpha}}\,, \\
    & \textrm{PC2:} \quad -\frac{3\mu}{2} \frac{\Omega_{E0} g(z)\left(1+w_E\right)}{E^2(z)}\,,\\
    & \textrm{PC3:} \quad -\frac{3\gamma}{2} \frac{\Omega_{E0} g(z)\left(1+w_E\right)}{E^3(z)}\,, \\
    & \textrm{PC4:} \quad -\frac{3}{2} \left(\frac{\gamma}{E(z)}+\mu\right)\frac{\Omega_{E0} g(z)\left(1+w_E\right)}{E^2(z)} \,.
\end{align}
The present-day value ($z=0$) of the deceleration parameter is given by:
\begin{align}
    q_0 =& -\frac{\ddot{a}}{a H^2}\bigg|_{t=t_0}
    = -\left(1+\frac{\dot{H}}{H^2}\right)_{t=t_0}
    = -1+\frac{1}{2}\frac{dE^2}{dz}\bigg|_{z=0}\nonumber\\
    =&\, q_0^{\Lambda\textrm{CDM}}
    + \frac{1}{2}\,\Omega_{E0}\left(1+w_E\right)
    \left(3-\frac{\Gamma_0}{H_0}\right)\,.
    \label{eq:deceleration-parameter-today-with-Gamma}
\end{align}
In the $\Lambda$CDM model the deceleration parameter is given by
$q_0^{\Lambda\textrm{CDM}} = -1 + \frac{3}{2}\Omega_{m0} + 2\Omega_{r0}$. 
In the PC framework, deviations from this baseline arise due to the presence of the additional component and the specific form of the particle-creation rate. Specifically, for each PC model, the expression for $q_0$ becomes:
\begin{align}
    & \textrm{PC1:} \quad  q_0 = q_0^{\Lambda\textrm{CDM}} + \frac{3}{2}\,\Omega_{E0}\left(1+w_E\right)\left(1-\beta\right)\,, \label{eq:q0-PC1} \\
    & \textrm{PC2:} \quad  q_0 = q_0^{\Lambda\textrm{CDM}} + \frac{1}{2}\,\Omega_{E0}\,\xi\,, \label{eq:q0-PC2} \\
    & \textrm{PC3:} \quad  q_0 = q_0^{\Lambda\textrm{CDM}} + \frac{3}{2}\,\Omega_{E0}\left(1+w_E\right)\left(1-\gamma\right)\,, \label{eq:q0-PC3} \\
    & \textrm{PC4:} \quad  q_0 = q_0^{\Lambda\textrm{CDM}} + \frac{3}{2}\,\Omega_{E0}\left(1+w_E\right)\left(1-\gamma-\mu\right)\,. \label{eq:q0-PC4}
\end{align}

\subsection{Effective equation-of-state parameter for dark energy}

The particle-creation process can be reformulated in terms of an effective dark-energy equation of state $w^{\textrm{eff}}_{\textrm{DE}}(z)$. Indeed, the modified Friedmann equations take the form
\begin{align}
    H^2 &= \frac{\chi}{3}\left(\rho_m+\rho_r+\rho^{\textrm{eff}}_{\textrm{DE}}\right)\,, \label{eq:Friedmann-effective-DE}\\
    \frac{\ddot{a}}{a}&=-\frac{\chi}{6}\left(\rho_m+2\rho_r\right)
    -\frac{\chi\,\rho^{\textrm{eff}}_{\textrm{DE}}}{6}\left[1+3\,w^{\textrm{eff}}_{\textrm{DE}}(z)\right]\,, \label{eq:acceleration-effective-DE}
\end{align}
where the effective dark-energy density evolves as
\begin{equation}
    \rho^{\textrm{eff}}_{\textrm{DE}}(z)
    = \rho^{\textrm{eff}}_{\textrm{DE0}}
    \exp{\left[3\int_0^z \frac{ds}{1+s}\left(1+w^{\textrm{eff}}_{\textrm{DE}}(s)\right)\right]}\,,
    \label{eq:rho_DE_effective}
\end{equation}
where $\rho^{\textrm{eff}}_{\textrm{DE0}}$ denotes its present-day value. 

Therefore, by comparing Eq.~\eqref{eq:Friedmann} with Eq.~\eqref{eq:Friedmann-effective-DE}, and Eq.~\eqref{eq:cosmic-accel} with Eq.~\eqref{eq:acceleration-effective-DE}, one can readily express the effective equation-of-state parameter for dark energy as
\begin{equation}
    w^{\textrm{eff}}_{\textrm{DE}}(z)
    = w_E - \frac{\Gamma}{3H}\left(1+w_E\right)\,.
    \label{eq:w-eff-DE}
\end{equation}
Taking the logarithm of Eq.~\eqref{eq:w-eff-DE}, differentiating with respect to redshift, and using the relation $\rho^{\textrm{eff}}_{\textrm{DE}}(z)=\rho_E(z)=\rho_{E0} g(z)$, we obtain an alternative expression for the effective equation of state:
\begin{equation}
    w^{\textrm{eff}}_{\textrm{DE}}(z) = -1 + \frac{1+z}{3}\frac{g^\prime(z)}{g(z)}\,,
\end{equation}
which is particularly convenient when the functional form of $g(z)$ is known.  
For clarity, note that $w_E$ is constant and characterizes the intrinsic equation of state of the created component, whereas $w^{\textrm{eff}}_{\textrm{DE}}(z)$ is redshift-dependent and effectively incorporates the impact of gravitationally induced particle production. 
Importantly, the particle-creation formalism naturally provides a dynamical dark-energy behaviour. In our work, this effect does not originate from a time dependence of $w_E$, but from the redshift-dependent particle-creation rate $\Gamma(z)$ and the Hubble rate, as evidenced by Eq.~\eqref{eq:w-eff-DE}. Therefore, even assuming a constant $w_E$ does not preclude an effective dynamical dark-energy behaviour.

Using Eq.~\eqref{eq:w-eff-DE}, we can write the effective equation-of-state parameter for each PC model introduced in Sect.~\ref{sec:particle-creation-models}:
\begin{align}
    & \textrm{PC1:} \quad  
    w^{\textrm{eff}}_{\textrm{DE}}(z) = 
    w_E - \beta \left[E(z)\right]^{\alpha-1} \left(1+w_E\right)\,, 
    \label{eq:w-eff-DE-model1}\\[4pt]
    & \textrm{PC2:} \quad  
    w^{\textrm{eff}}_{\textrm{DE}}(z) = 
    -1 + \frac{\xi}{3}\,, 
    \label{eq:w-eff-DE-model2}\\[4pt]
    & \textrm{PC3:} \quad  
    w^{\textrm{eff}}_{\textrm{DE}}(z) = 
    w_E - \frac{\gamma}{E(z)} \left(1+w_E\right)\,, 
    \label{eq:w-eff-DE-model3}\\[4pt]
    & \textrm{PC4:} \quad  
    w^{\textrm{eff}}_{\textrm{DE}}(z) = 
    w_E - \left[\frac{\gamma}{E(z)}+\mu\right]\left(1+w_E\right)\,, 
    \label{eq:w-eff-DE-model4}
\end{align}
It is important to note that $w^{\textrm{eff}}_{\textrm{DE}}(0) < -1/3$ is required to consistently account for the current cosmic acceleration.


\section{Datasets and Methodology} \label{sec:datasets-methodology}

In this work we make use of \texttt{SimpleMC}~\cite{simplemc} as the sampling code. We employ Nested Sampling~\cite{2004AIPC..735..395S} as implemented in the \texttt{dynesty} Python library~\cite{2020MNRAS.493.3132S}, which also provides the Bayesian Evidence needed to compare the PC models against the $\Lambda$CDM and $w_0w_a$CDM baselines. For most analyses we used 1024 live points and an accuracy of $0.01$ (which serves as the convergence criterion for nested sampling), ensuring good precision in the estimation of the evidence and posterior distributions. For the joint analysis \texttt{CC+SN+SH0ES+BAO+CMB}, we instead adopted 500 live points for each model, which was sufficient to achieve numerical convergence while significantly reducing the computational cost. 

Our particle-creation (PC) models primarily affect late-time cosmology, in particular the background expansion history. In our setup, the particle-creation mechanism is assumed to operate only in the redshift range $0<z<3$, as specified in Sect.~\ref{sec:particle-creation-models}, and standard $\Lambda$CDM behaviour is recovered at higher redshift. For this reason, we make use of background-only datasets that predominantly probe the low-redshift Universe ($z \lesssim 2$).

The datasets used are as follows: a catalogue of 15 Cosmic Chronometers, including their covariance matrix~\cite{Moresco:2020fbm} (hereafter \texttt{CC}); the PantheonPlus sample of Type Ia Supernovae (SNe Ia)~\cite{Scolnic:2021amr,Brout:2022vxf}, consisting of 1701 light curves from 1550 unique SNe Ia used to measure the distance modulus (referred to as \texttt{SN}); the latest DESI DR2 Baryon Acoustic Oscillation measurements~\cite{DESI:2025zgx,DESI:2025fii,DESI:2025qqy}, hereafter \texttt{BAO}, which contains information on the transverse distance $D_M(z)/r_d$, the radial distance $D_H(z)/r_d$ and the volume-averaged distance $D_V(z)/r_d$ (where $r_{d}$ is the comoving size of the sound horizon at the drag epoch); the SH0ES calibration for the PantheonPlus catalogue~\cite{Riess:2021jrx}, denoted \texttt{SH0ES}; 
and a compressed (also referred to as \textit{geometrical} or \textit{background}) CMB likelihood, where the CMB is treated as a BAO measurement at $z \approx 1100$ using the Planck 2018 data release, which will be referred to as \texttt{CMB}. The CMB background information can be condensed into three quantities: $\omega_{\rm b}$ (physical baryon density parameter), $\omega_{\rm m}$ (physical matter density parameter) and $D_{\rm M}(z\sim 1100)/r_{d}$ (for a discussion on how to adopt this approach, please refer to~\cite{Planck:2016tof}). 
For the PantheonPlus sample, we consider 1590 data points, corresponding to the effective number of independent distance-modulus measurements entering the likelihood, after accounting for the combination of multiple light curves and the impact of peculiar velocities.

As the datasets are independent from each other, when combining them for the parameter inference procedure we can obtain the joint $\chi^2$ by simply adding the individual ones. As an example, for the case where the combination \texttt{BAO+CMB} is used, we would have:
\begin{equation}
    \chi^2_{\tt total} = \chi^2_{\text{BAO}} + \chi^2_{\text{CMB}}.
\end{equation}
We then compute the $\chi^2$ contribution from each dataset separately (\texttt{CC}, \texttt{SN}, \texttt{SH0ES}, \texttt{BAO}, and \texttt{CMB}) in order to monitor the behaviour of the fits and to identify possible tensions within specific probes. However, for clarity and consistency in the presentation of results, we report only the total $\chi^2$ for each model and dataset combination.

To investigate the Hubble tension within the PC framework, we analyse two separate combinations of these datasets: the late-time set \texttt{CC+SN+SH0ES} and the high-redshift set \texttt{BAO+CMB}. We then combine all probes into \texttt{CC+SN+SH0ES+BAO+CMB} to maximise constraining power. Results for all three cases will be presented in the following section.

\begin{figure*}
    \includegraphics[scale=0.35]{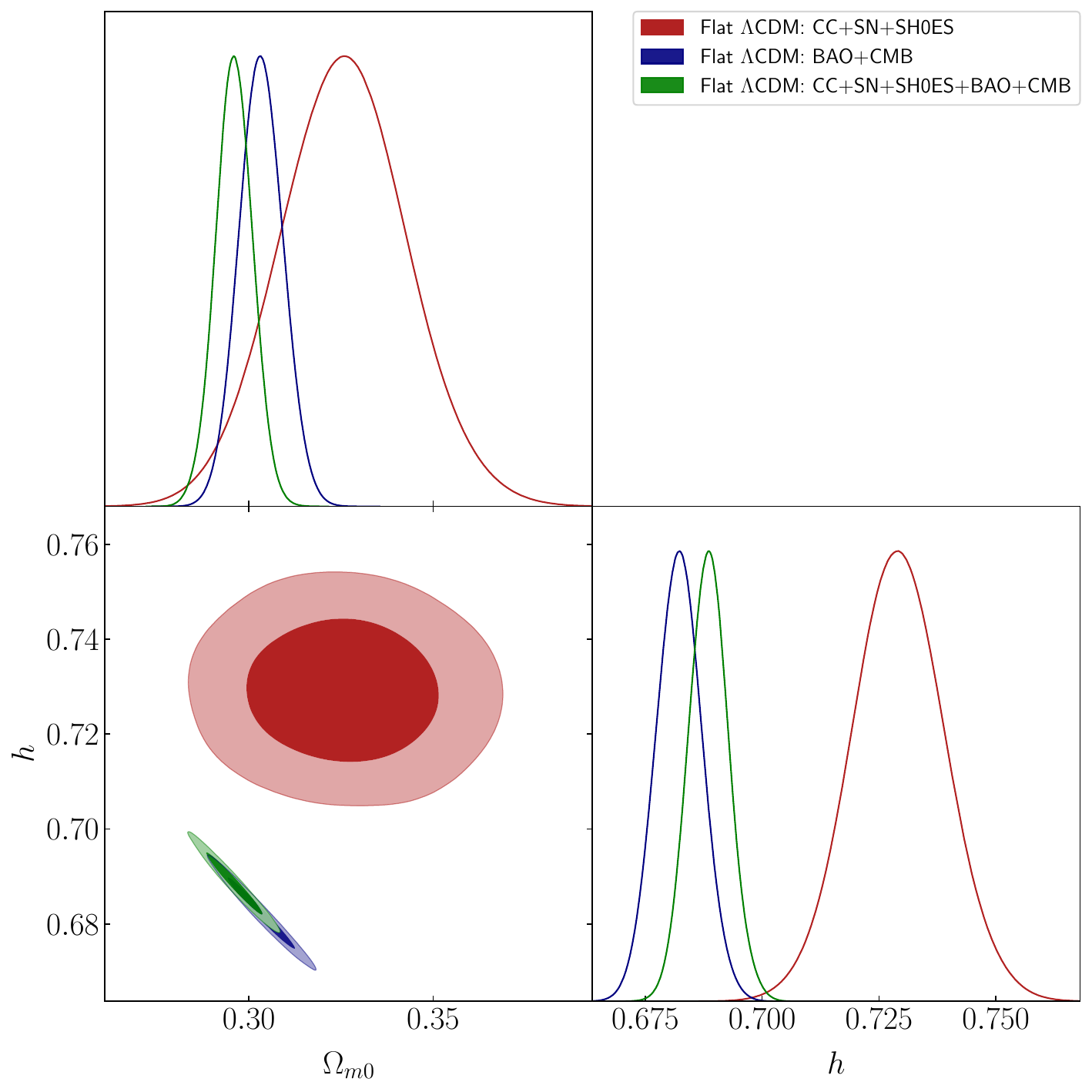} \includegraphics[scale=0.35]{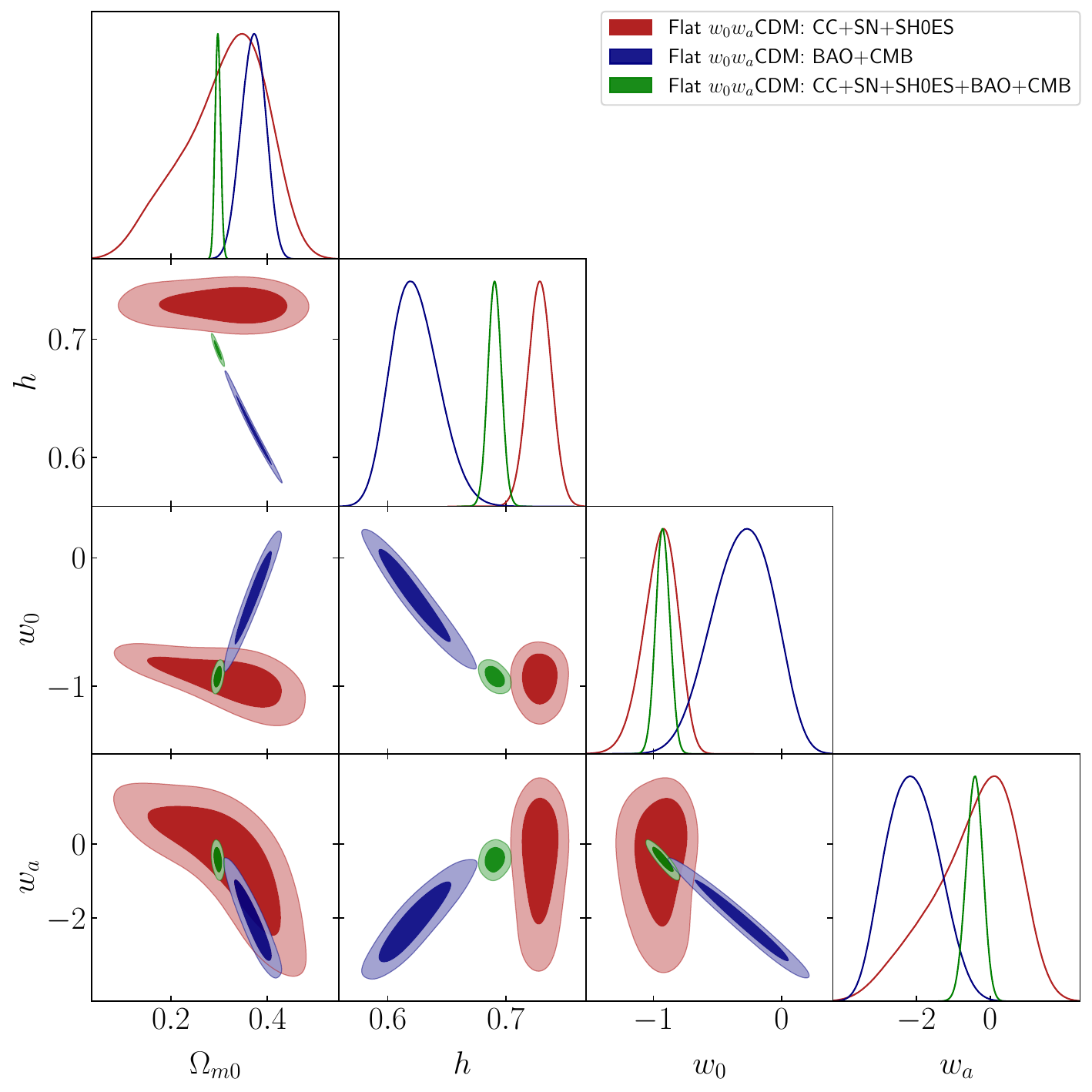} \caption{Triangular plots showing the 1D and 2D posterior distributions for the $\Lambda$CDM (left panels) and $w_0w_a$CDM (right panels) models, obtained using the \texttt{CC+SN+SH0ES}, \texttt{BAO+CMB}, and full \texttt{CC+SN+SH0ES+BAO+CMB} dataset combinations.} \label{fig:contour-LCDM-and-w0waCDM}
\end{figure*}

The priors for all parameters used in the analysis are chosen to be flat (agnostic) and can be found in Table~\ref{TablePriors}. These were chosen following a preliminary exploration of the PC models, aimed at evaluating the impact of different parameter values on the Hubble function, and the effective dark-energy equation of state (see Appendix~\ref{sec:appendix}). 
For reference, we also use the flat (zero-curvature) $\Lambda$CDM and $w_0w_a$CDM models; Fig.~\ref{fig:contour-LCDM-and-w0waCDM} displays the marginalised posteriors for the background parameters 
$h \equiv \frac{H_0}{100\,\textrm{km s}^{-1}\,\textrm{Mpc}^{-1}}$ 
and $\Omega_{m0}$ obtained using the combined \texttt{CC+BAO+SN+SH0ES+CMB} dataset.

Although the contribution of radiation ($\Omega_{r0}$) 
to the normalised Friedmann equation is negligible at low redshift, it is nevertheless included in our analysis for consistency. Since we are interested in quantifying slight deviations from $\Lambda$CDM induced by the particle-creation mechanism, it is important to retain all standard cosmological components (including radiation) and isolate the impact of the non-standard contribution associated with particle creation. This ensures that any departure from the standard expansion history can be robustly attributed to the particle-creation sector rather than to an incomplete background modeling. 
To this end, we fix the CMB temperature and neutrino contributions and write the present-day radiation density parameter as
\begin{equation}
    \Omega_{r0} = \Omega_{\gamma0} \left[1+\frac{7}{8}\left(\frac{4}{11}\right)^{4/3} N_{\textrm{eff}} \right],
\end{equation}
with the photon density parameter given by
\begin{equation}
    \Omega_{\gamma0} = \frac{8\pi^5}{15} 
    \frac{(K_B T_\gamma)^4}{h_P^3 c^5} 
    \frac{8 \pi G}{3 H_0^2},
\end{equation}
where $N_{\textrm{eff}} = 3.046$ is the effective number of relativistic species,  
$T_\gamma = 2.7255\,\textrm{K}$ is the CMB temperature,  
$K_B$ is the Boltzmann constant,  
$h_P$ the Planck constant,  
$c$ the speed of light,  
and $H_0 = 100\,h\,\textrm{km s}^{-1}\,\textrm{Mpc}^{-1}$.  
Using numerical values, we obtain
\begin{equation}
    \Omega_{r0} = 4.183 \times 10^{-5} \, h^{-2}.
    \label{eq:Omega_rad}
\end{equation}
Thus, once the posterior distribution of $h$ is known, $\Omega_{r0}$ may be treated as a derived parameter.

Regarding the baryonic contribution to the total energy density of the Universe, it cannot be neglected at low redshifts in the same manner as radiation, given that $\Omega_{b0} \sim 0.05$ while $\Omega_{r0} \sim 10^{-5}$. We require the baryon density to adhere to the constraints imposed by Big Bang Nucleosynthesis (BBN), following \cite{Schoneberg:2024ifp}. Consequently, although we adopt an agnostic prior of $\Omega_{b0} h^2 \in [0.02, 0.025]$, the posterior is predominantly driven by the BBN likelihood towards $0.02218 \pm 0.00055$. We therefore deem it unnecessary to report this parameter alongside the other free parameters in this study, as its behaviour remains largely unchanged due to the restrictive nature of BBN.

Finally, we perform a model comparison making use of the Bayesian Evidence. Following the definition of Bayes’ theorem,
\begin{equation}
    P(u|D,M)= \frac{L(D|u,M)P(u|M)}{P(D|M)},
\end{equation}
where $u$ is the vector of free parameters to estimate, $M$ is the model (also referred to as the \textit{hypothesis}), $D$ is the data, $P(u|D,M)$ is the posterior probability distribution, $L(D|u,M)$ is the \textit{likelihood}, $P(u|M)$ is the prior distribution, and
$P(D|M)$ is commonly referred to as the \textit{Bayesian evidence}.
During a parameter inference procedure, one uses a sampler code to find the posterior distribution of a model’s parameters using data. This posterior distribution is obtained by sampling values of the parameters that maximize the likelihood. By relating the $\chi^2$, whose value decreases as the fit to the data improves, to the likelihood as
\begin{equation}
    L \propto e^{-\frac{\chi^2}{2}},
\end{equation}
we can begin searching for the parameter values that minimize (maximize) the $\chi^2$ ($L$) to sample the posterior distribution, since $P(u|D,M)\propto L$. The $\chi^2$ for each dataset is defined as
\begin{equation}
    \chi^2_{\tt data} = (d_{i,m}-d_{i,\tt data})\, C^{-1}_{ij,\tt data}\, (d_{j,m}-d_{j,\tt data}),
\end{equation}
where $d_{\tt data}$ are the observable data at a certain redshift/scale factor, $d_m$ are the model predictions at that same redshift/scale factor, and $C_{\tt data}$ is the covariance matrix associated with the data.
If the parameter inference procedure is performed for more than one model, the models can be quantitatively compared through their Bayesian evidence. For two competing models, $M_1$ and $M_2$, the Bayes factor is defined as
\begin{equation}
    B_{12} \equiv \frac{P(D|M_1)}{P(D|M_2)}.
\end{equation}
In practice, it is convenient to consider the logarithm of the Bayes factor, $\ln B_{12} = \ln P(D|M_1) - \ln P(D|M_2)$, which quantifies the statistical preference of model $M_1$ relative to $M_2$. A value $B_{12}>1$ ($B_{12}<1$), or equivalently $\ln B_{12} > 0$ ($\ln B_{12} < 0$), indicates that the data favour model $M_1$ over $M_2$ (or $M_2$ over $M_1$). The strength of the evidence can be interpreted using the revised empirical Jeffreys scale~\cite{Trotta:2008qt}: it is considered inconclusive if $0<\left|\ln B_{12}\right|<1$, weak if $1.0<\left|\ln B_{12}\right|<2.5$, moderate if $2.5<\left|\ln B_{12}\right|<5.0$, and strong if $\left|\ln B_{12}\right|>5.0$.

In this work, adopting the $\Lambda$CDM model as the reference baseline, we compute the quantity $\ln B_{M_i-\Lambda\mathrm{CDM}} = \ln P(D|M_i) - \ln P(D|\Lambda\mathrm{CDM})$ for each alternative model $M_i$. Within this convention, $\ln B_{M_i-\Lambda\mathrm{CDM}} > 0$ indicates a statistical preference for the model $M_i$ over $\Lambda$CDM.

For completeness, we also report the reduced chi-square, defined as
\begin{equation}
    \chi^2_{\text{red}} \equiv \frac{\chi^2}{\nu}\,, 
\end{equation}
where $\nu=N_{\textrm{data}}-N_{\textrm{par}}$ denotes the number of degrees of freedom, given by the difference between the number of data points and the number of free model parameters. $\chi^2_{\text{red}}\approx 1$ indicates a statistically acceptable fit.
 
The total $\chi^2_{\textrm{min}}$, the reduced $\chi^2_{\text{red}}$, 
the Bayesian evidence, and the Bayes factor are provided in Tables~\ref{TableConstraintsHigh},~\ref{TableConstraintsLow}, and~\ref{TableConstraintsFull}. The number of data points for each datasets considered in this work are summarized here for convenience: $N_{\textrm{data}}=$ 15 (CC), 1590 (SNe), 1 (SH0ES), 13 (BAOs), and 3 (CMB), respectively.

\begin{table*} \begin{centering} \begin{tabular}{|c|c|c|c|c|c|c|c|c|c|c|} \hline \multicolumn{11}{|c|}{Priors}\tabularnewline 
\hline Model & $h$ & $\Omega_{m0}$ & $w_0$ & $w_a$ & $w_E$ & $\alpha$ & $\beta$ & $\xi$ & $\gamma$ & $\mu$ \tabularnewline 
\hline $\Lambda$CDM & $\left(0.4, 0.9\right)$ & $\left(0.1, 0.5\right)$ & $-$ & $-$ & $-$ & $-$ & $-$ & $-$ & $-$ & $-$ \tabularnewline 
\hline $w_0w_a$CDM & $\left(0.4, 0.9\right)$ & $\left(0.1, 0.5\right)$ & $\left(-3.0, 1.0\right)$ & $\left(-3.5, 2.0\right)$ & $-$ & $-$ & $-$ & $-$ & $-$ & $-$ \tabularnewline
\hline PC1 & $\left(0.4, 0.9\right)$ & $\left(0.1, 0.5\right)$ & $-$ & $-$ & $\left(-3.0, 1.0\right)$ & $\left(0.0, 10.0\right)$ & $\left(-2.0, 2.0\right)$ & $-$ & $-$ & $-$ \tabularnewline 
\hline PC2 & $\left(0.4, 0.9\right)$ & $\left(0.1, 0.5\right)$ & $-$ & $-$ & $-$ & $-$ & $-$ & $\left(-1.0, 0.5\right)$ & $-$ & $-$ \tabularnewline 
\hline PC3 & $\left(0.4, 0.9\right)$ & $\left(0.1, 0.5\right)$ & $-$ & $-$ & $\left(-3.0, 1.0\right)$ & $-$ & $-$ & $-$ & $\left(-3.0, 3.0\right)$ & $-$ \tabularnewline 
\hline PC4 & $\left(0.4, 0.9\right)$ & $\left(0.1, 0.5\right)$ & $-$ & $-$ & $\left(-3.0, 1.0\right)$ & $-$ & $-$ & $-$ & $\left(-3.0, 3.0\right)$ & $\left(-3.0, 3.0\right)$ \tabularnewline
\hline \end{tabular} 
\caption{Flat uniform priors used for each model.} \label{TablePriors} \par\end{centering} \end{table*}

\begin{table*} \begin{centering} \begin{tabular}{|c|l|l|l|c|c|c|c|} \hline \multicolumn{8}{|c|}{Constraints on cosmological parameters from CC+SN+SH0ES}\tabularnewline 
\hline Model & $H_0$ & $\Omega_{m0}$ & \quad Extra parameters & $\chi^2_\text{min}$ & $\chi^2_{\text{red}}$ & $\ln P(D|M_i)$ & $\ln B$ \\ $M_i$ & $\,\left(\textrm{km s}^{-1}\,\textrm{Mpc}^{-1}\right)$ & & & & & & \tabularnewline 
\hline $\Lambda$CDM & $72.92\pm 0.99$ \quad [$72.92$] & $0.326\pm 0.017$ \quad [$0.325$] & \quad $-$ & $1411.403$ & $0.880$ & $-712.45$ & $-$\tabularnewline 
\hline $w_0w_a$CDM & $72.88\pm0.99$ \quad [$73.00$] & $0.312^{+0.100}_{-0.066}$ \quad [$0.197$] & \quad $w_0=-0.95^{+0.15}_{-0.12}$ \quad [$-0.84$] & $1411.047$ & $0.881$ & $-715.29$ & $-2.84$ \\ & & & \quad $w_a=-0.42^{+1.40}_{-0.83}$ \quad [$0.63$] & & & & \tabularnewline
\hline PC1 & $72.94\pm 0.94$ \quad [$73.04$] & $0.331^{+0.056}_{-0.047}$ \quad [$0.352$] & \quad $w_E=-0.89^{+0.29}_{-0.37}$ \quad [$-1.02$] & $1411.024$ & $0.881$ & $-714.37$ & $-1.92$ \\ & & & \quad $\alpha=3.3^{+1.6}_{-3.1}$ \quad [$9.7$] & & & & \\ & & & \quad $\beta=0.56^{+1.00}_{-0.52}$ \quad [$-0.92$] & & & & \tabularnewline 
\hline PC2 & $72.92\pm0.99$ \quad [$72.89$] & $0.319^{+0.039}_{-0.045}$ \quad [$0.303$] & \quad $\xi=0.03^{+0.38}_{-0.25}$ \quad [$-0.18$] & $1411.168$ & $0.880$ & $-712.66$ & $-0.21$ \tabularnewline 
\hline PC3 & $72.90\pm0.99$ \quad [$72.85$] & $0.309^{+0.057}_{-0.038}$ \quad [$0.240$] & \quad $w_E=-1.04^{+0.50}_{-0.40}$ \quad [$-0.35$] & $1411.111$ & $0.881$ & $-714.12$ & $-1.67$ \\ & & & \quad $\gamma=0.69^{+1.10}_{-0.62}$ \quad [$0.79$] & & & & \tabularnewline 
\hline PC4 & $72.91\pm 0.94$ \quad [$73.09$] & $0.309^{+0.057}_{-0.041}$ \quad [$0.240$] & \quad $w_E=-1.00\pm 0.47$ \quad [$-1.39$] & $1411.108$ & $0.881$ & $-714.29$ & $-1.84$ \\ & & & \quad $\gamma=0.3\pm 1.4$ \quad [$-1.4$] & & & & \\ & & & \quad $\mu=0.4^{+1.7}_{-1.4}$ \quad [$2.7$] & & & & \tabularnewline
\hline \end{tabular}
\caption{Mean values and $68\,\%$ confidence-level uncertainties for the cosmological parameters in the $\Lambda$CDM, $w_0w_a$CDM, and PC models constrained with the \texttt{CC+SN+SH0ES} datasets. The corresponding best-fit values are reported in square brackets.
The last four columns list the $\chi^2_\text{min}$, the reduced $\chi^2_\textrm{red}$, the logarithm of the Bayesian evidence (marginal likelihood) $P(D|M_i)$ obtained with nested sampling, and the logarithm of the Bayes factor $B\equiv B_{M_i-\Lambda \textrm{CDM}}$ for each model $M_i$.}
\label{TableConstraintsHigh} \par\end{centering} \end{table*}

\begin{table*} \begin{centering} \begin{tabular}{|c|l|l|l|c|c|c|c|} \hline \multicolumn{8}{|c|}{Constraints on cosmological parameters from BAO+CMB}\tabularnewline 
\hline Model & $H_0$ & $\Omega_{m0}$ & \quad Extra parameters & $\chi^2_\text{min}$ & $\chi^2_{\text{red}}$ & $\ln P(D|M_i)$ & $\ln B$ \\ $M_i$ & $\,\left(\textrm{km s}^{-1}\,\textrm{Mpc}^{-1}\right)$ & & & & & & \tabularnewline 
\hline $\Lambda$CDM & $68.22\pm 0.49$ \quad [$68.22$] & $0.303\pm 0.006$ \quad [$0.303$] & \quad $-$ & $14.827$ & $1.059$ & $-19.98$ & $-$ \tabularnewline 
\hline $w_0w_a$CDM & $62.3^{+1.8}_{-2.2}$ \quad [$62.5$] & $0.371\pm 0.025$ \quad [$0.368$] & \quad $w_0=-0.30^{+0.26}_{-0.22}$ \quad [$-0.33$] & $5.996$ & $0.500$ & $-19.23$ & $0.75$ \\ & & & \quad $w_a=-2.10^{+0.68}_{-0.76}$ \quad [$-2.00$] & & & & \tabularnewline
\hline PC1 & $67.2^{+2.2}_{-1.4}$ \quad [$64.3$] & $0.316^{+0.014}_{-0.023}$ \quad [$0.345$] & \quad $w_E=-1.05^{+0.76}_{-0.56}$ \quad [$-1.99$] & $6.856$ & $0.623$ & $-20.13$ & $-0.15$ \\ & & & \quad $\alpha=1.69^{+0.99}_{-1.60}$ \quad [$0.01$] & & & & \\ & & & \quad $\beta=0.77^{+0.62}_{-0.41}$ \quad [$1.46$] & & & & \tabularnewline 
\hline PC2 & $68.72 \pm 1.01$ \quad [$68.69$] & $0.300\pm 0.009$ \quad [$0.300$] & \quad $\xi=0.01\pm 0.12$ \quad [$0.01$] & $11.129$ & $0.856$ & $-19.32$ & $0.66$ \tabularnewline 
\hline PC3 & $65.5^{+2.7}_{-2.4}$ \quad [$64.6$] & $0.334^{+0.024}_{-0.032}$ \quad [$0.342$] & \quad $w_E=-1.73^{+0.64}_{-0.50}$ \quad [$-1.92$] & $6.831$ & $0.569$ & $-19.01$ & $0.97$ \\ & & & \quad $\gamma=1.39^{+0.43}_{-0.27}$ \quad [$1.46$] & & & & \tabularnewline 
\hline PC4 & $65.8^{+2.8}_{-2.3}$ \quad [$64.5$] & $0.331^{+0.023}_{-0.032}$ \quad [$0.344$] & \quad $w_E=-1.03\pm 0.78$ \quad [$-1.99$] & $6.806$ & $0.619$ & $-19.22$ & $0.76$ \\ & & & \quad $\gamma=0.12\pm 1.84$ \quad [$1.40$] & & & & \\ & & & \quad $\mu=0.85 \pm 1.34$ \quad [$0.04$] & & & & \tabularnewline
\hline \end{tabular}
\caption{Same as Table~\ref{TableConstraintsHigh}, but with parameter constraints obtained using the \texttt{BAO+CMB} datasets.}
\label{TableConstraintsLow} \par\end{centering} \end{table*}

\begin{table*} \begin{centering} \begin{tabular}{|c|l|l|l|c|c|c|c|} \hline \multicolumn{8}{|c|}{Constraints on cosmological parameters from CC+SN+SH0ES+BAO+CMB}\tabularnewline 
\hline Model & $H_0$ & $\Omega_{m0}$ & \quad Extra parameters & $\chi^2_\text{min}$ & $\chi^2_\textrm{red}$ & $\ln P(D|M_i)$ & $\ln B$ \\ $M_i$ & $\,\left(\textrm{km s}^{-1}\,\textrm{Mpc}^{-1}\right)$ & & & & & & \tabularnewline 
\hline $\Lambda$CDM & $68.85\pm 0.43$ \quad [$68.86$] & $0.296\pm0.005$ \quad [$0.296$] & \quad $-$ & $1448.026$ & $0.894$ & $-736.62$ & $-$ \tabularnewline 
\hline $w_0w_a$CDM & $69.06\pm 0.56$ \quad [$69.04$] & $0.297\pm 0.006$ \quad [$0.297$] & \quad $w_0=-0.93\pm 0.05$ \quad [$-0.93$] & $1443.728$ & $0.892$ & $-740.06$ & $-3.44$ \\ & & & \quad $w_a=-0.43\pm 0.22$ \quad [$-0.36$] & & & & \tabularnewline
\hline PC1 & $69.32\pm 0.49$ \quad [$69.28$] & $0.294\pm 0.005$  \quad [$0.295$] & \quad $w_E=-0.98^{+0.21}_{-0.17}$ \quad [$-0.86$] & $1439.288$ & $0.890$ & $-736.15$ & $0.47$ \\ & & & \quad $\alpha=2.1^{+1.2}_{-1.8}$ \quad [$2.1$] & & & & \\ & & & \quad $\beta=0.62^{+0.75}_{-0.43}$ \quad [$0.85$] & & & & \tabularnewline 
\hline PC2 & $69.34\pm 0.55$ \quad [$69.35$] & $0.293\pm 0.005$ \quad [$0.293$] & \quad $\xi=-0.011\pm 0.073$ \quad [$-0.010$] & $1440.975$ & $0.890$ & $-735.26$ & $1.36$ \tabularnewline 
\hline PC3 & $69.28\pm 0.50$ \quad [$69.23$] & $0.294\pm 0.005$ \quad [$0.295$] & \quad $w_E=-1.12^{+0.20}_{-0.11}$ \quad [$-1.24$] & $1439.463$ & $0.890$ & $-735.63$ & $0.99$ \\ & & & \quad $\gamma=0.98^{+0.78}_{-0.44}$ \quad [$1.17$] & & & & \tabularnewline 
\hline PC4 & $69.28\pm0.49$ \quad [$69.27$] & $0.294\pm 0.005$ \quad [$0.295$] & \quad $w_E=-1.02^{+0.19}_{-0.17}$ \quad [$-1.10$] & $1439.420$ & $0.890$ & $-735.57$ & $1.05$ \\ & & & \quad $\gamma=0.26 \pm 1.46$ \quad [$2.88$] & & & & \\ & & & \quad $\mu=0.63 \pm 1.26$ \quad [$-1.49$] & & & &  \tabularnewline
\hline \end{tabular}
\caption{Same as Table~\ref{TableConstraintsHigh}, but with parameter constraints obtained using the full combination of datasets \texttt{CC+SN+SH0ES+BAO+CMB}.} 
\label{TableConstraintsFull} \par\end{centering} \end{table*}

\begin{figure*} \includegraphics[scale=0.35]{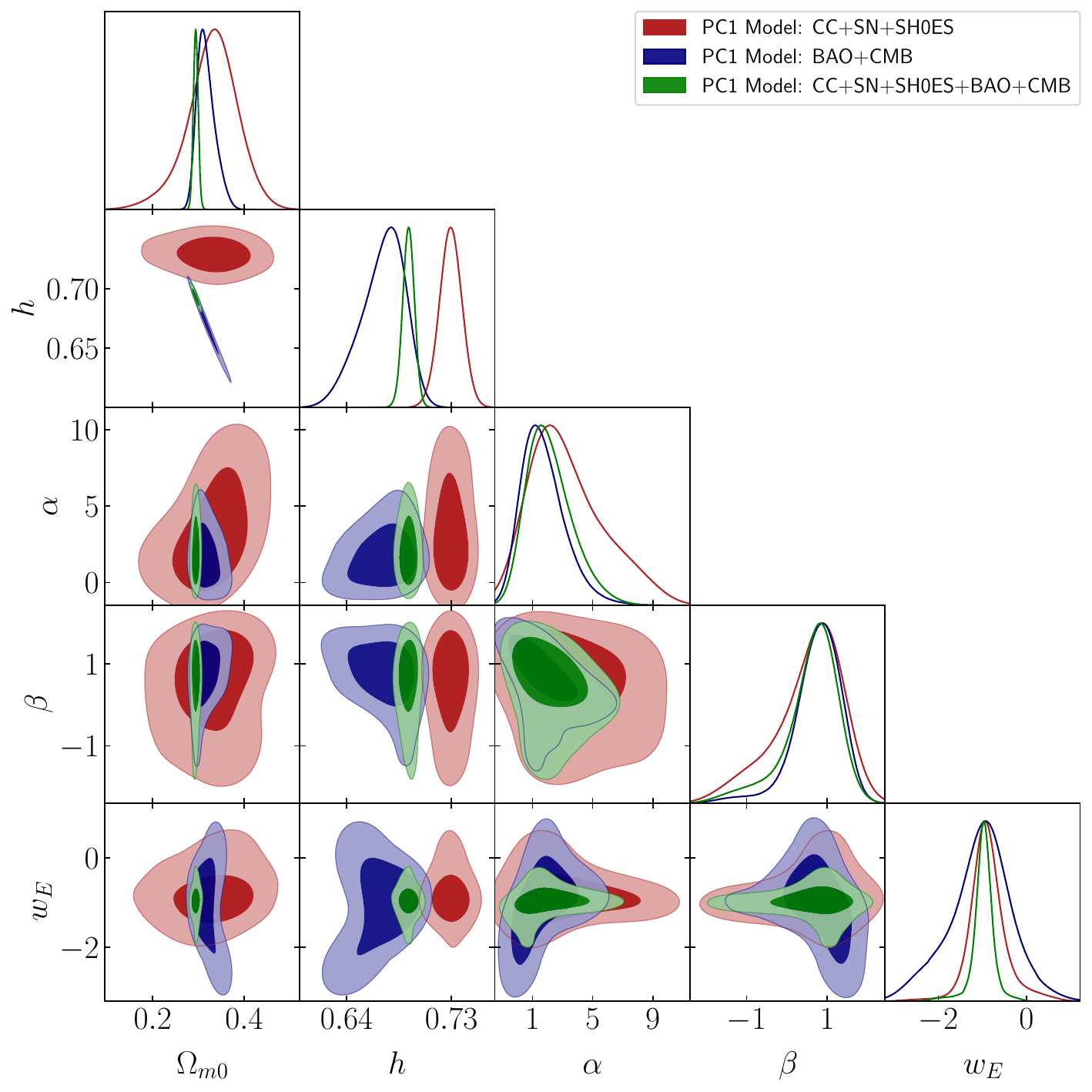} \includegraphics[scale=0.35]{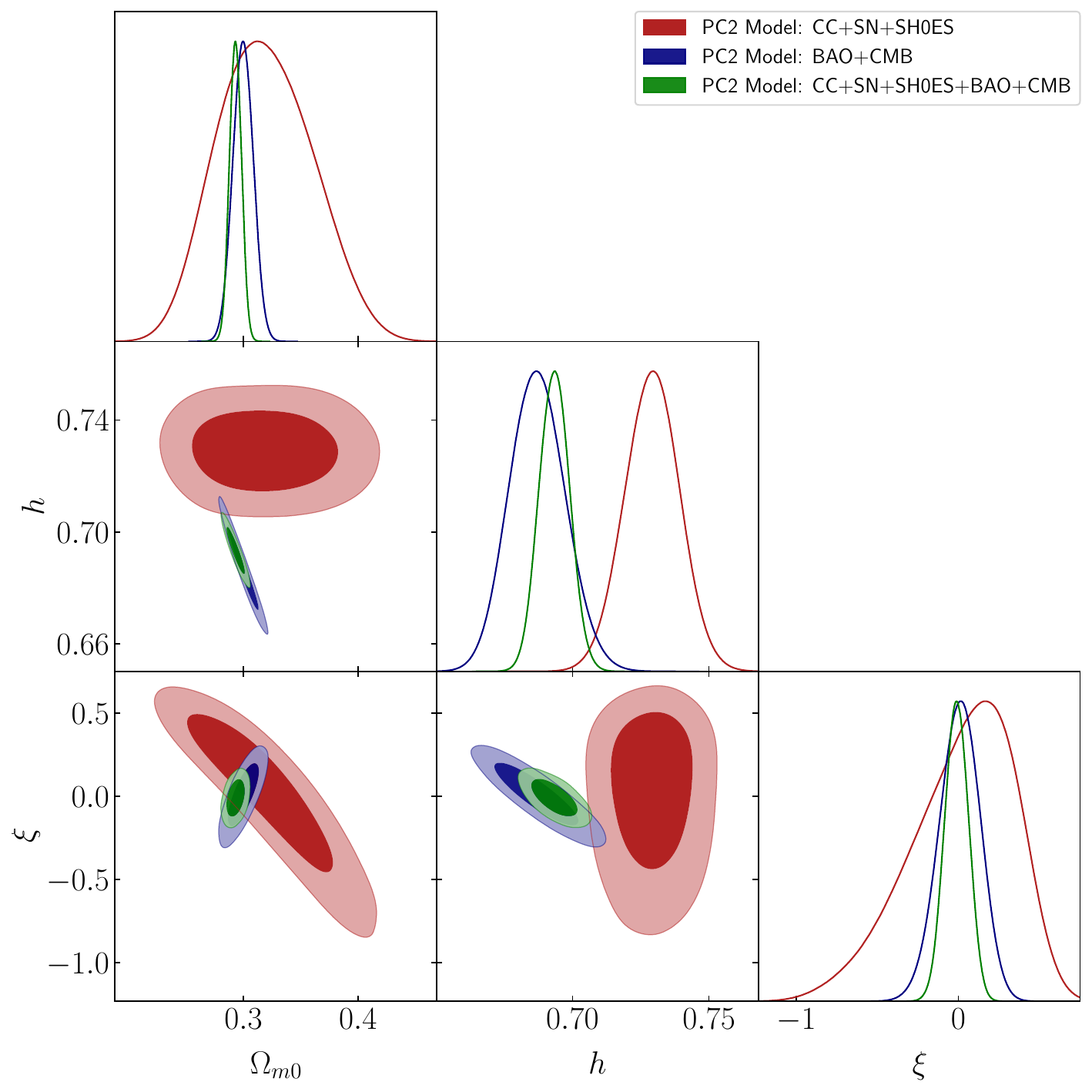} \\ \includegraphics[scale=0.35]{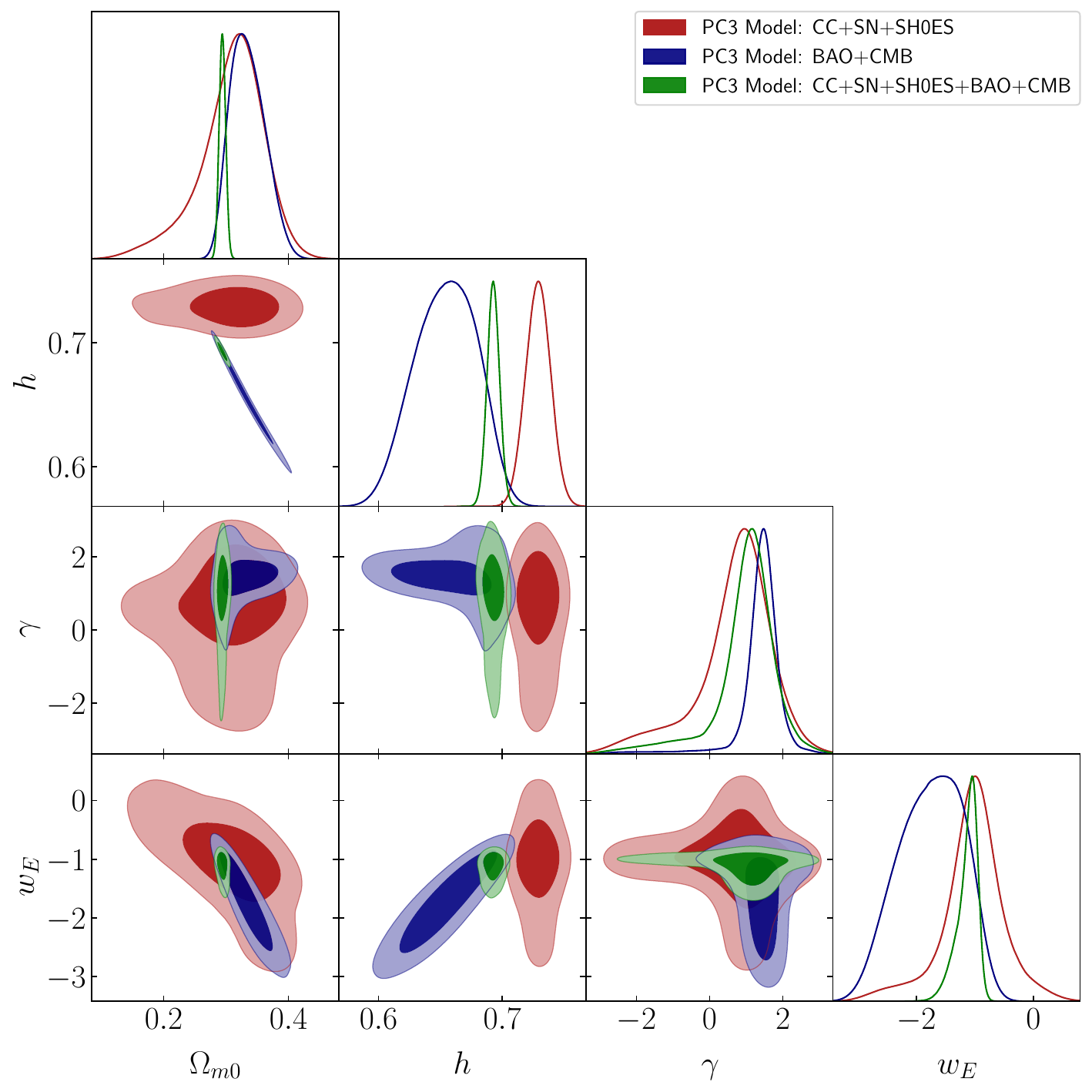} \includegraphics[scale=0.35]{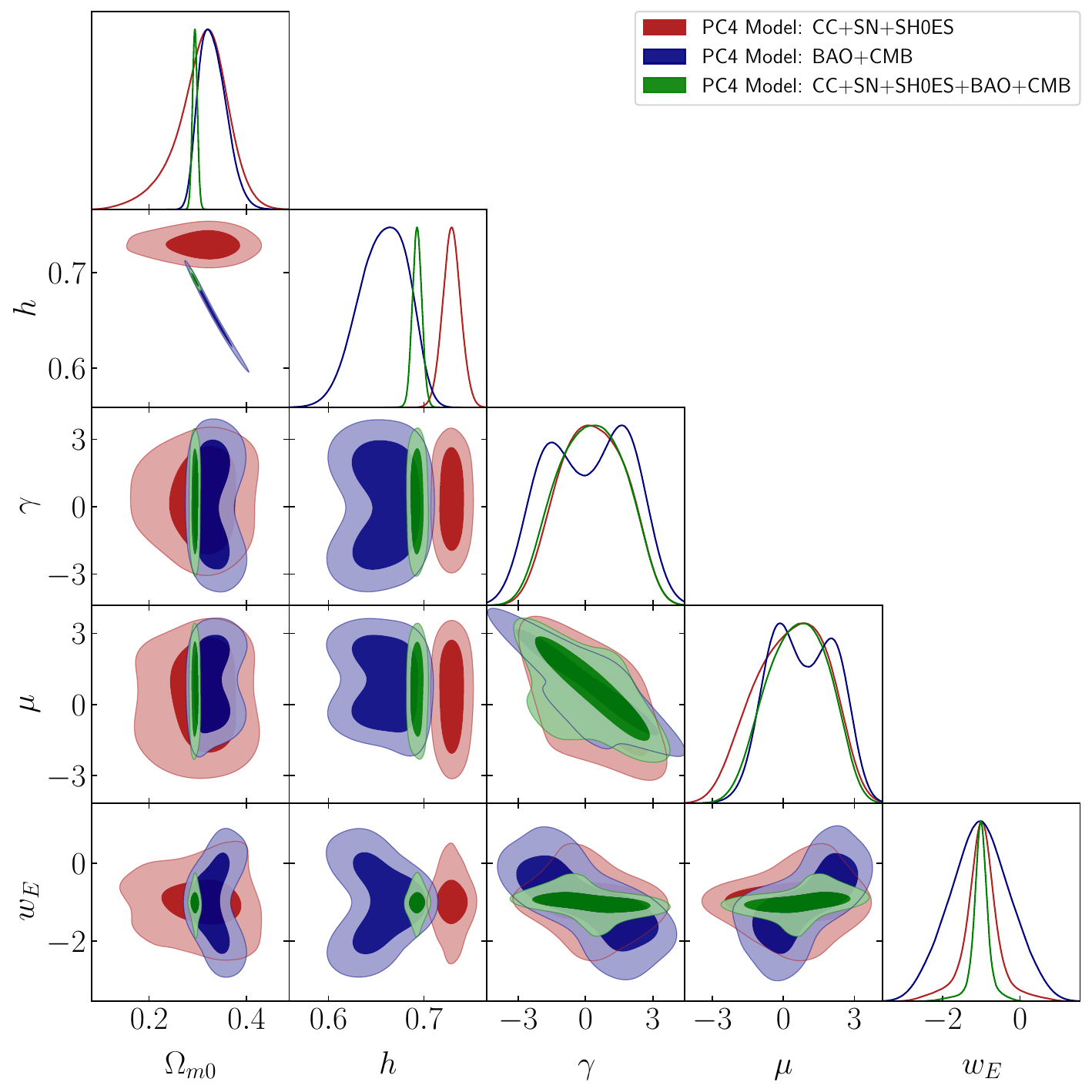} \caption{Triangular plots showing the 1D and 2D posterior distributions for the PC models using the \texttt{CC+SN+SH0ES}, \texttt{BAO+CMB}, and full \texttt{CC+SN+SH0ES+BAO+CMB} dataset combinations. The panels correspond to PC1 (top left), PC2 (top right), PC3 (bottom left), and PC4 (bottom right).} \label{fig:contourplots} \end{figure*}


\section{Cosmological results}\label{sec:cosmological-results}

Tables~\ref{TableConstraintsHigh} and~\ref{TableConstraintsLow} summarise the constraints on the cosmological parameters obtained from analyses based on the \texttt{CC+SN+SH0ES} and \texttt{BAO+CMB} dataset combinations, respectively.  
These complementary analyses are designed to explicitly investigate the Hubble tension by isolating late-time and early-time information within different cosmological models.

From the inferred values of $H_0$ and their corresponding $1\,\sigma$ uncertainties, we find that the tension between the two dataset combinations reaches the level of $4.3\,\sigma$ in the $\Lambda$CDM scenario and $5.2\,\sigma$ in the $w_0w_a$CDM model.  
Within the PC framework, the tension is partially alleviated, being reduced to $2.4\,\sigma$, $3.0\,\sigma$, $2.6\,\sigma$, and $2.4\,\sigma$ for the PC1, PC2, PC3, and PC4 models, respectively.  
This mitigation is achieved without introducing strong departures from the standard expansion history, highlighting the ability of particle-creation mechanisms to improve consistency between late-time observations and high-redshift constraints.

However, it is worth noting that the alleviation of the $H_0$ tension is mainly driven by the larger uncertainties on $H_0$ inferred from the \texttt{BAO+CMB} combination in the PC models, while the central values remain close to the corresponding $\Lambda$CDM value. Indeed, in the \texttt{BAO+CMB} analysis, $H_0$ is relatively weakly constrained within the $w_0w_a$CDM and PC models. This behaviour is expected, as these scenarios are constructed to deviate from $\Lambda$CDM only at late times, making them predominantly sensitive to low-redshift observables rather than to early-Universe physics.

To maximise the constraining power and improve the overall performance of the models, we present a joint analysis combining all available datasets, \texttt{CC+SN+SH0ES+BAO+CMB}. The corresponding parameter constraints are reported in Table~\ref{TableConstraintsFull}, together with 
the total $\chi^2_{\text{min}}$, the reduced $\chi^2_{\text{red}}$, the logarithm of the Bayesian evidence $\ln{P\left(D|M_i\right)}$, and the logarithm of the Bayes factor $\ln B$, with $B\equiv B_{M_i-\Lambda \textrm{CDM}}$, quantifying the statistical preference of each model $M_i$ relative to $\Lambda$CDM.

Based on the results reported in Table~\ref{TableConstraintsFull}, we find that the PC models achieve a lower minimum $\chi^2_{\min}$ than $\Lambda$CDM, with a reduction of $\Delta\chi^2_{\min}\simeq 8$, at the cost of up to three additional parameters. Despite this improvement, the reduced $\chi^2_{\text{red}}$ remains essentially unchanged ($\chi^2_{\text{red}}=0.894$ for $\Lambda$CDM versus $0.890$ for the PC models), indicating a comparable overall goodness of fit. The logarithm of the Bayes factor reported in Table~\ref{TableConstraintsFull} shows that PC models are mildly favoured over $\Lambda$CDM, with consistently positive values of $\ln B$. However, according to the empirical Jeffreys scale, the statistical evidence remains weak for PC2 and PC4, and inconclusive for PC1 and PC3. Notably, PC models are generally preferred over the $w_0w_a$CDM parametrization.

A closer inspection of Tables~\ref{TableConstraintsHigh} and~\ref{TableConstraintsLow} indicates that this improvement is primarily driven by the \texttt{BAO+CMB} dataset combination. In this case, all models except PC1 yield positive (though inconclusive) values of $\ln B$. Conversely, low-redshift datasets lead to nearly identical $\chi^2_{\min}$ values across all models, resulting in reduced Bayesian evidence for extensions of $\Lambda$CDM due to the larger parameter space. Overall, \texttt{BAO+CMB} data show a mild preference for PC scenarios, while low-redshift observations remain largely insensitive and penalize additional model complexity. Taken together, these results indicate that PC models are currently slightly favoured over $\Lambda$CDM. Their ability to achieve a comparable goodness of fit, while providing a physically motivated extension of late-time cosmological dynamics, supports particle creation as a viable and statistically competitive alternative to the standard model.

The marginalised posterior distributions for the flat $\Lambda$CDM and $w_0w_a$CDM models are shown in Fig.~\ref{fig:contour-LCDM-and-w0waCDM}, while the corresponding triangle plots for the PC models are displayed in Fig.~\ref{fig:contourplots}. All posterior analyses and corner plots were produced using the \texttt{GetDist} package~\cite{Lewis:2019xzd}.

\begin{figure*} \includegraphics[scale=0.58]{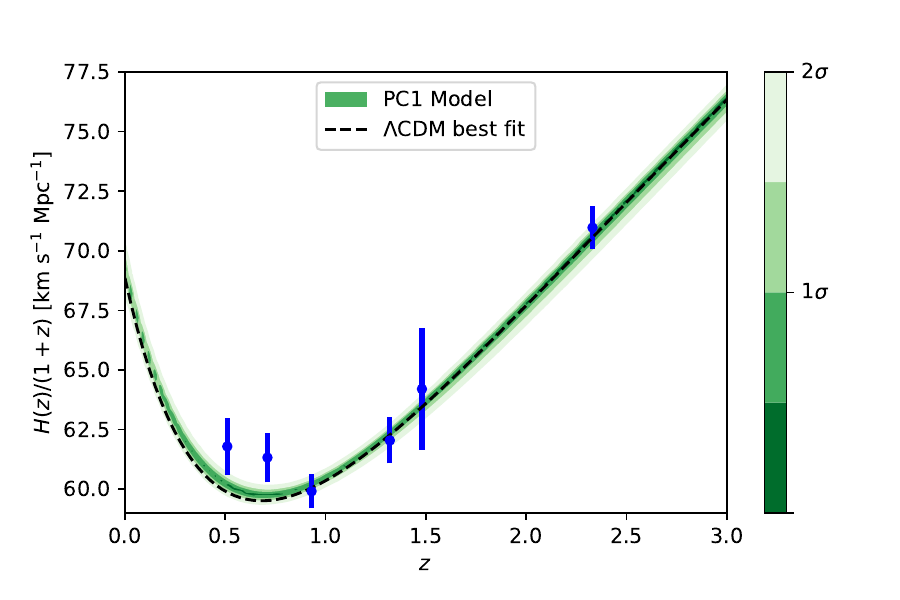} \includegraphics[scale=0.58]{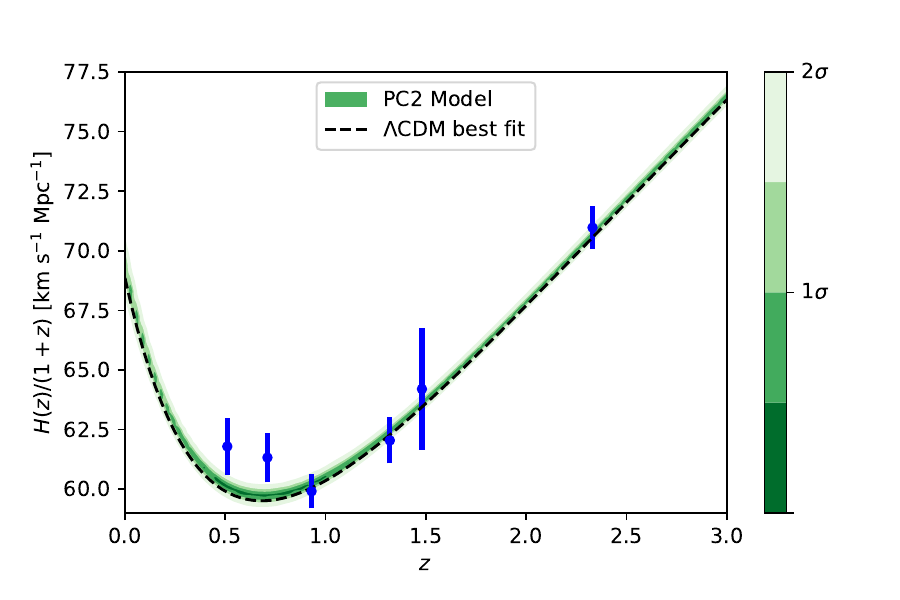} \\ \includegraphics[scale=0.58]{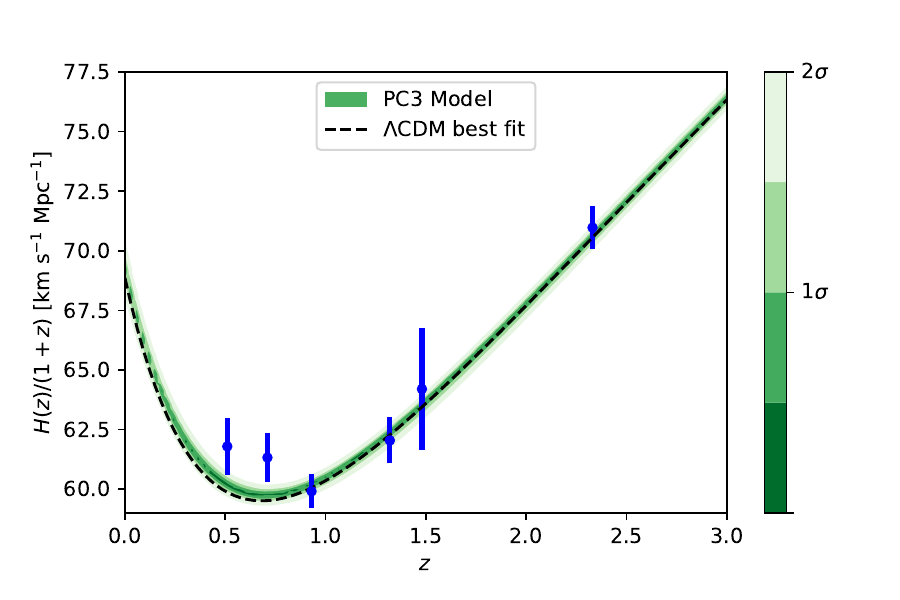} \includegraphics[scale=0.58]{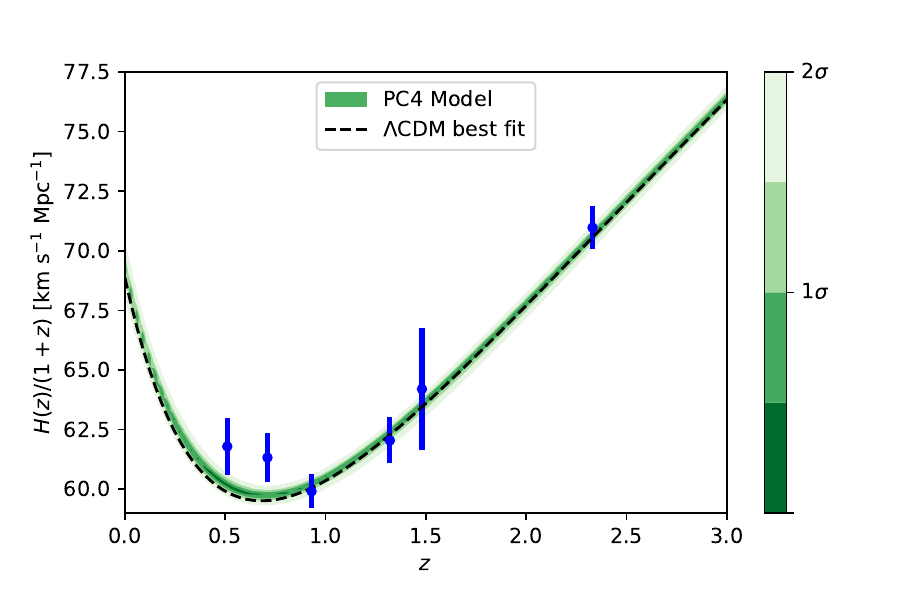} \caption{Plots of the function $H(z)/(1+z)$ for the PC models. The upper panels show the PC1 (left) and PC2 (right) models, while the lower panels show the PC3 (left) and PC4 (right) models. The functional posterior distributions are shown up to the $2\,\sigma$ level, based on the parameter constraints reported in Table~\ref{TableConstraintsFull} for the \texttt{CC+SN+SH0ES+BAO+CMB} dataset combination. For comparison, the dashed curve corresponds to the $\Lambda$CDM model, using the best-fit parameter values from Table~\ref{TableConstraintsFull}. The blue data points with error bars indicate the DESI DR2 BAO measurements, included as an illustrative subset of the data to demonstrate the agreement with observations.} 
\label{fig:Hubble_PC} \end{figure*}

Figure~\ref{fig:Hubble_PC} shows the evolution of the quantity $H(z)/(1+z)$ for the four PC models over the redshift range $0<z<3$, computed from the marginalised posterior distributions obtained using the full dataset combination. The shaded regions represent the $95\%$ confidence intervals, obtained with the \texttt{fgivenx} package~\cite{fgivenx}.  
For comparison, the $\Lambda$CDM prediction corresponding to the best-fit parameters in Table~\ref{TableConstraintsFull} is also shown.  
Despite the additional degrees of freedom introduced by particle creation, all PC models closely track the $\Lambda$CDM expansion history, even at late times, reflecting the strong constraints imposed by the combined datasets.

A closer inspection of the constraints reported in Table~\ref{TableConstraintsFull} reveals that, for PC1, PC3, and PC4, the parameter $w_E$ is fully consistent with the $\Lambda$CDM limit $w_E=-1$ at the $1\,\sigma$ level, while for PC2 the same conclusion applies to the parameter $\xi$, whose $\Lambda$CDM value $\xi=0$ is also recovered within $1\,\sigma$.  
This result further confirms that the PC models naturally converge towards $\Lambda$CDM in the region of parameter space preferred by current observations.

The requirement of an accelerating Universe is satisfied by all PC models. Using the best-fit parameters from Table~\ref{TableConstraintsFull}, one can explicitly verify that the theoretical acceleration conditions derived in Sect.~\ref{sec:constraints} are fulfilled, \textit{i.e.}, the conditions given in Eq.~\eqref{eq:condition-wE}, or equivalently Eqs.~\eqref{eq:condition-wE-model1}, \eqref{eq:condition-wE-model3}, and~\eqref{eq:condition-wE-model4} for the PC1, PC3, and PC4 models. For the PC2 model, instead, we evaluate the effective equation-of-state parameter of dark energy and verify that $w^{\textrm{eff}}_{\textrm{DE}}(z)<-1/3$.

\begin{figure*} \includegraphics[scale=0.58]{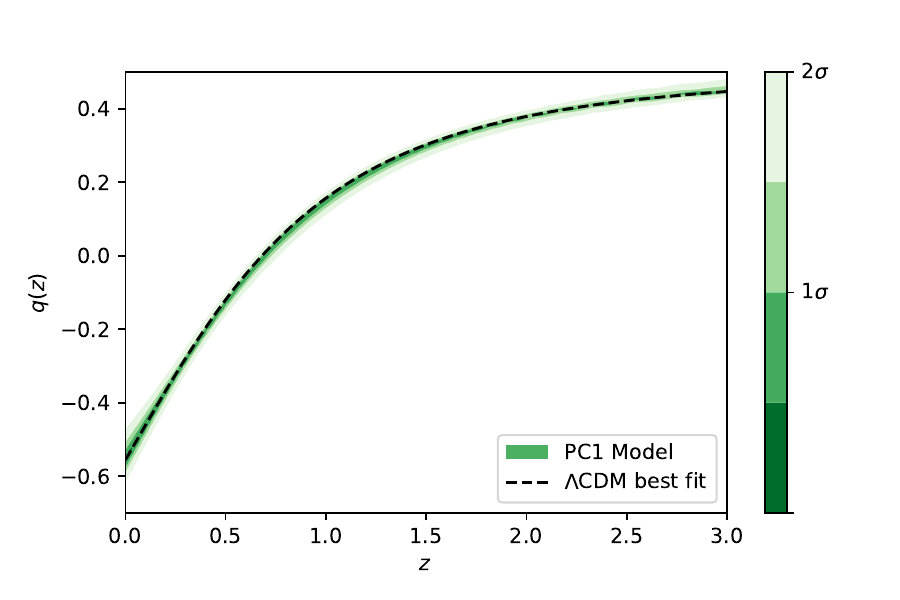} \includegraphics[scale=0.58]{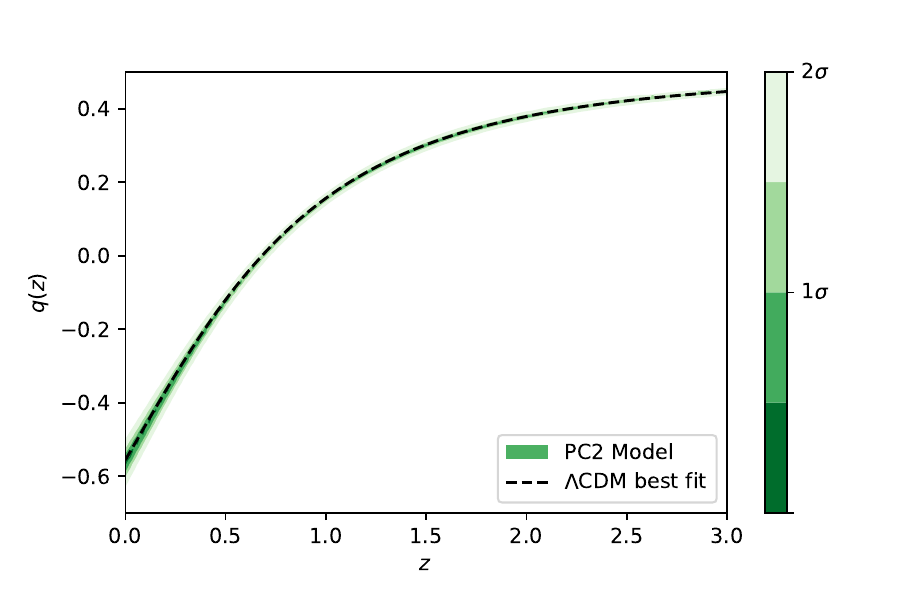} \\ \includegraphics[scale=0.58]{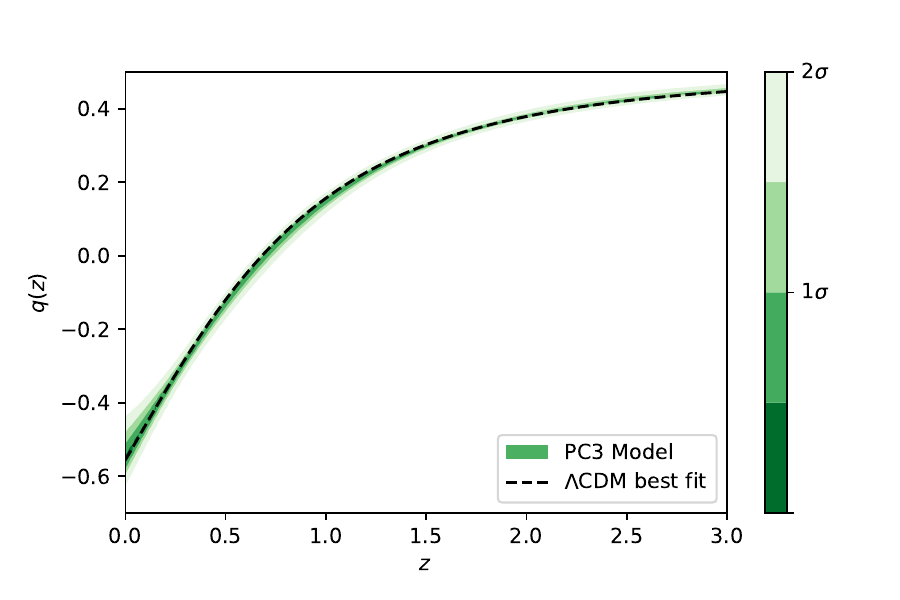} \includegraphics[scale=0.58]{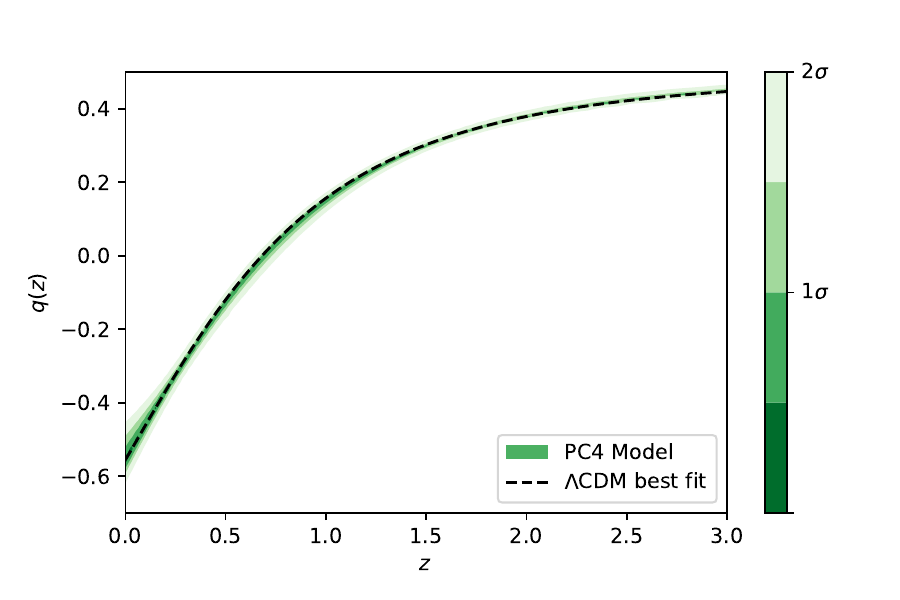} \caption{Plots of the deceleration parameter $q(z)$ for the PC models. The upper panels show the PC1 (left) and PC2 (right) models, while the lower panels show the PC3 (left) and PC4 (right) models. The functional posterior distributions are shown up to the $2\,\sigma$ level, based on the parameter constraints reported in Table~\ref{TableConstraintsFull} for the \texttt{CC+SN+SH0ES+BAO+CMB} dataset combination. For comparison, the dashed curve corresponds to the $\Lambda$CDM model, using the best-fit parameter values from Table~\ref{TableConstraintsFull}.}  
\label{fig:deceleration} \end{figure*}

Table~\ref{TableDerivedParameters} reports the present-day values of the deceleration parameter $q_0$ for all models using the combined \texttt{CC+SN+SH0ES+BAO+CMB} dataset, while Fig.~\ref{fig:deceleration} shows its redshift evolution, $q(z)$, obtained from the functional posterior distributions. In this context, we used Eqs.~\eqref{eq:q0-PC1}--\eqref{eq:q0-PC4} and~\eqref{eq:deceleration-parameter-with-Gamma}.  
All scenarios exhibit a transition from decelerated to accelerated expansion at $z<1$ and remain consistent with $\Lambda$CDM within $2\,\sigma$.  
Moreover, both $q_0$ and the present-day effective dark-energy equation of state, $w^{\textrm{eff}}_{\textrm{DE}}(0)$, agree with the $\Lambda$CDM expectations at the $1\,\sigma$ level for all PC models.

\begin{table*} \begin{centering} \begin{tabular}{|c|c|c|c|c|c|c|} \hline \multicolumn{7}{|c|}{Derived parameters for the CC+SN+SH0ES+BAO+CMB analysis}\tabularnewline 
\hline  & $\Lambda$CDM & $w_0w_a$CDM & PC1 & PC2 & PC3 & PC4 \tabularnewline 
\hline $q_0$ & $-0.56\pm0.01$ & $-0.48\pm 0.06$ & $-0.55\pm0.03$ & $-0.56\pm 0.03$ & $-0.54^{+0.03}_{-0.05}$ & $-0.54^{+0.03}_{-0.04}$ \tabularnewline 
\hline $w^{\textrm{eff}}_{\textrm{DE}}(0)$ & $-1$ & $-0.93\pm 0.05$ & $-0.99^{+0.02}_{-0.03}$ & $-1.00\pm 0.02$ & $-0.98^{+0.03}_{-0.04}$ & $-0.99^{+0.02}_{-0.04}$ \tabularnewline
\hline \end{tabular}
\caption{Derived parameters (mean values and 68 $\%$ confidence-level uncertainties) for the flat $\Lambda$CDM, $w_0w_a$CDM, and PC models, obtained from the statistical distributions of cosmological parameters inferred from the \texttt{CC+SN+SH0ES+BAO+CMB} dataset combination (see Table~\ref{TableConstraintsFull}). The first row reports the deceleration parameter evaluated at the present epoch, $z=0$ (see Eqs.~\eqref{eq:q0-PC1}, \eqref{eq:q0-PC2}, \eqref{eq:q0-PC3}, and~\eqref{eq:q0-PC4}). The second row reports the effective dark-energy equation-of-state parameter evaluated at $z=0$ (see Eqs.~\eqref{eq:w-eff-DE-model1}, \eqref{eq:w-eff-DE-model2}, \eqref{eq:w-eff-DE-model3}, and~\eqref{eq:w-eff-DE-model4}).}\label{TableDerivedParameters} \par\end{centering} \end{table*}

An important outcome of our analysis concerns the physical nature of the created component.  
From Table~\ref{TableConstraintsFull}, we find that in all four PC models the additional component satisfies $w_E<-1/3$, implying behaviour compatible with dark energy in the standard cosmological scenario.  
This result is particularly significant, as the equation-of-state parameter $w_E$ was treated as a completely free quantity, following an agnostic approach with no prior assumptions on the nature of the created particles.  
The data therefore indicate that, even within a particle-creation framework, a dark-energy-like component is required to account for the late-time dynamics of the Universe.

\begin{figure*} \includegraphics[scale=0.25]{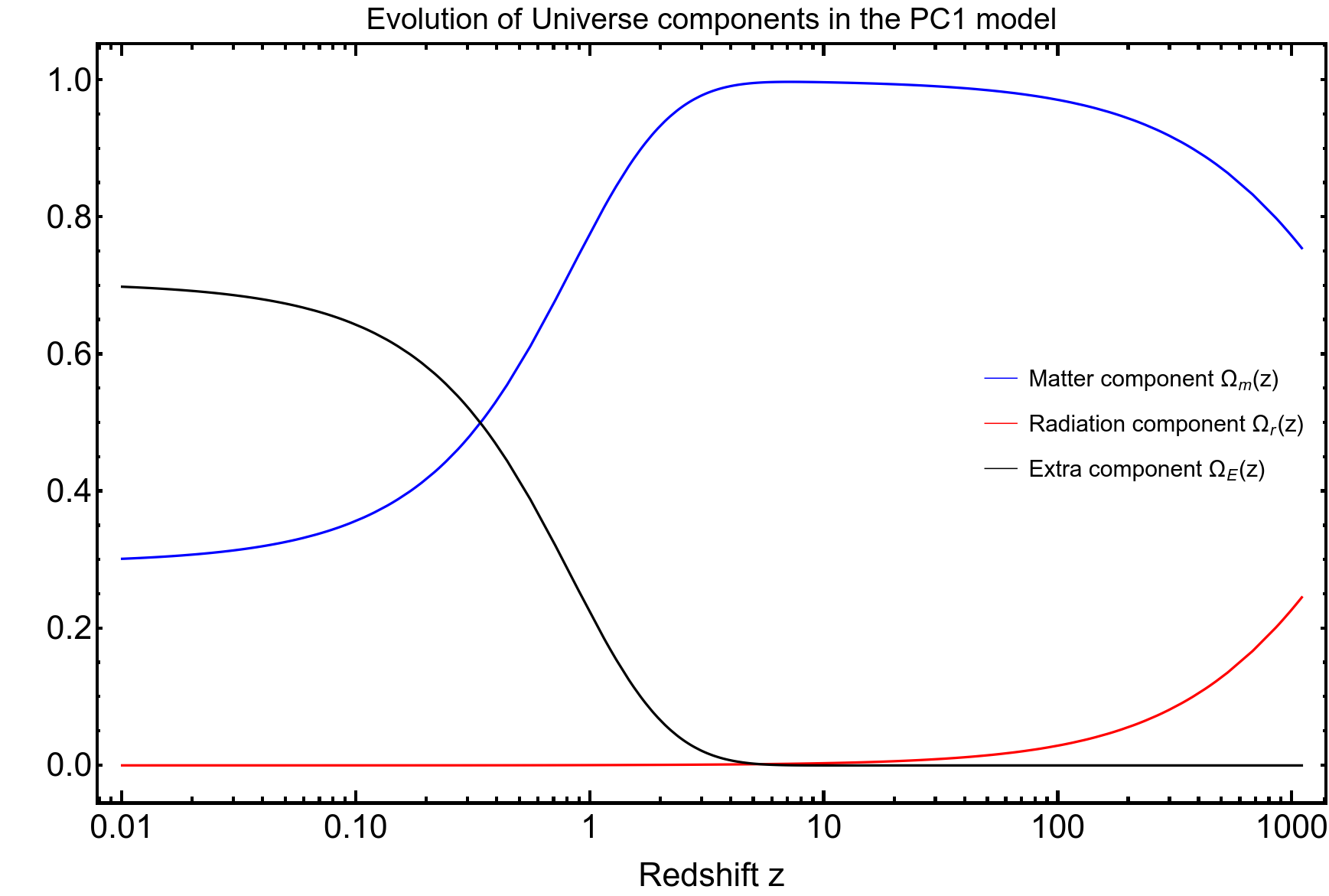} 
\caption{Redshift evolution of the cosmological components in the PC1 model, computed using the best-fit parameter values from Table~\ref{TableConstraintsFull} obtained in the joint \texttt{CC+SN+SH0ES+BAO+CMB} analysis. The fractional energy densities $\Omega_m(z)$, $\Omega_r(z)$, and $\Omega_E(z)$ are shown by the blue, red, and black curves, respectively.} \label{fig:components} \end{figure*}

In this context, Fig.~\ref{fig:components} illustrates the redshift evolution of the fractional energy densities $\Omega_m(z)$, $\Omega_r(z)$, and $\Omega_E(z)$ (defined as usual as $\Omega_i(z)=\rho_i(z)/\rho_{\rm crit}(z)$ for component $i$, with $\rho_{\rm crit}(z)=3H^2(z)/\chi$). The plot refers to the PC1 model, but the same qualitative behaviour applies to the other PC models.  
The contribution of the created component dominates the energy budget in the local Universe, while becoming rapidly negligible at high redshift compared to matter and radiation, which follow their standard scaling.  
This behaviour provides further evidence that the newly created particles effectively act as a dark-energy component, leaving early-Universe physics unaltered.

Conversely, scenarios corresponding to the production of pressureless matter particles ($w_E=0$) for $0< z< 3$ are strongly disfavoured by the data, being excluded at the $4.7\,\sigma$, $5.6\,\sigma$, and $5.4\,\sigma$ levels for the PC1, PC3, and PC4 models, respectively.  
This represents a substantial departure from earlier studies, which primarily focused on the creation of cold dark matter particles, and highlights the importance of confronting particle-creation scenarios with current high-precision cosmological data.  
At the same time, the constraints on $w_E$ do not allow us to distinguish conclusively between an effective quintessence-like regime and phantom behaviour, differently from~\cite{Navone:2025gxr}, where the dark energy particle production is associated with the bulk viscosity. 

\begin{figure*}
    \includegraphics[scale=0.58]{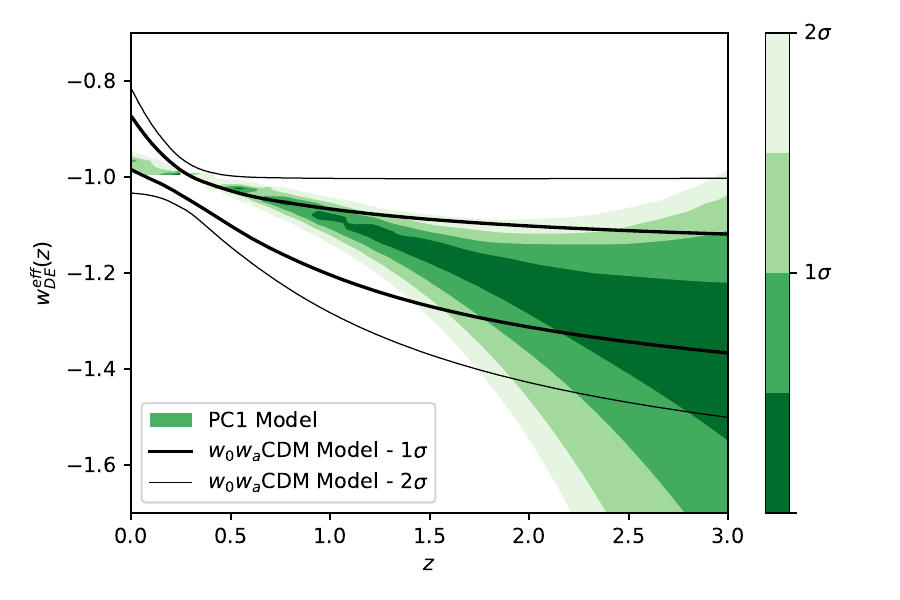} \includegraphics[scale=0.58]{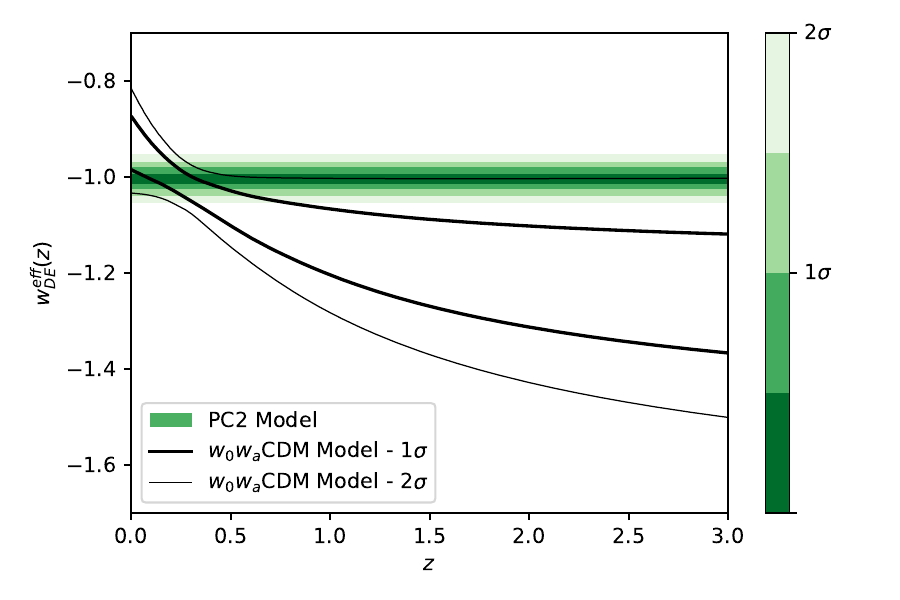} \\ \includegraphics[scale=0.58]{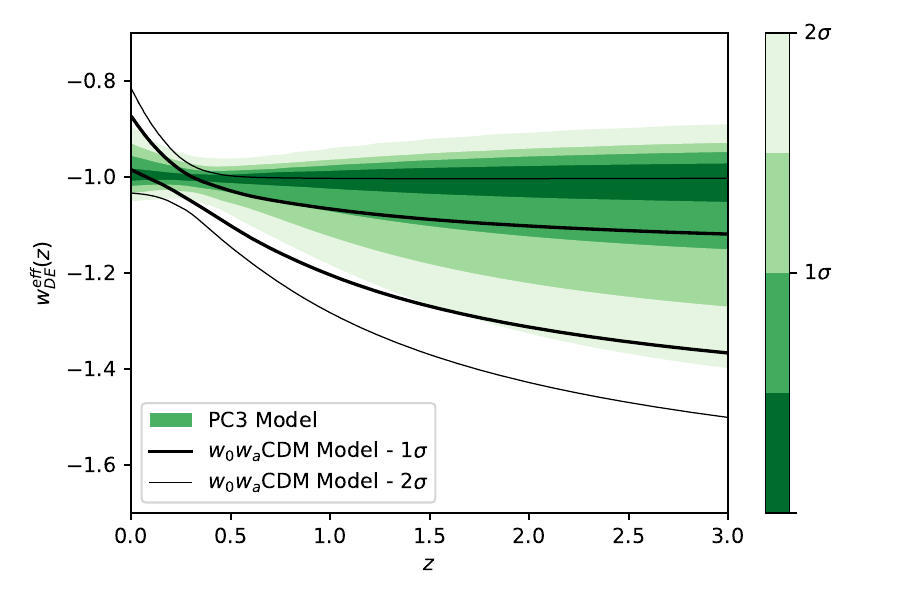} \includegraphics[scale=0.58]{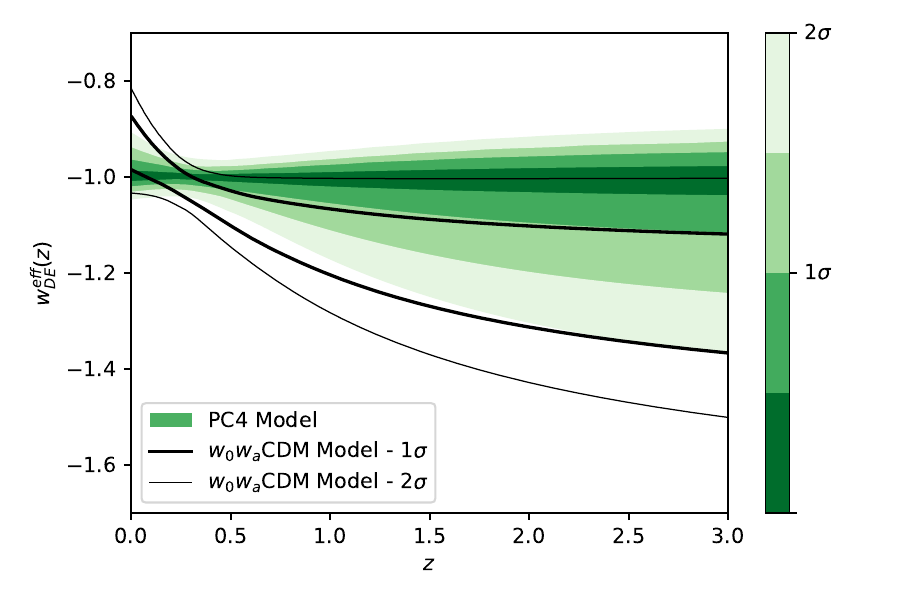} \caption{Effective dark-energy equation-of-state parameter $w^{\textrm{eff}}_{\textrm{DE}}(z)$ for the PC models. The upper panels show the PC1 (left) and PC2 (right) models, while the lower panels show the PC3 (left) and PC4 (right) models. The functional posterior distributions are shown up to the $2\,\sigma$ level, based on the parameter constraints reported in Table~\ref{TableConstraintsFull} for the \texttt{CC+SN+SH0ES+BAO+CMB} dataset combination. For comparison with a dynamical dark-energy scenario, the $1\,\sigma$ (black curves) and $2\,\sigma$ (grey curves) confidence contours associated with the $w_0w_a$CDM model are also shown in each panel.} \label{fig:effective-w}
\end{figure*}

The effective dark-energy equation of state, $w^{\textrm{eff}}_{\textrm{DE}}(z)$, is shown in Fig.~\ref{fig:effective-w} for all PC models and compared with the corresponding confidence regions obtained within the $w_0w_a$CDM framework. We use Eqs.~\eqref{eq:w-eff-DE-model1}--\eqref{eq:w-eff-DE-model4} and the functional posterior distributions derived from the \texttt{CC+SN+SH0ES+BAO+CMB} analysis. Table~\ref{TableDerivedParameters} also reports the present-day values ($z=0$) of the effective equation-of-state parameter for all PC models.

In all cases, the condition $w^{\textrm{eff}}_{\textrm{DE}}(z)<-1/3$ is satisfied, confirming the presence of accelerated expansion.  
The PC2 model, which is equivalent to a $w$CDM scenario with a constant equation of state, yields a value of $w^{\textrm{eff}}_{\textrm{DE}}$ fully consistent with a cosmological constant at the $1\,\sigma$ level.  
Similarly, the PC3 and PC4 models remain compatible with $\Lambda$CDM within $1\,\sigma$ across the full redshift range considered. PC3 and PC4 exhibit only a mild redshift dependence of $w^{\textrm{eff}}_{\textrm{DE}}(z)$ when compared with the dynamical dark-energy behaviour obtained in the CPL parametrization (black curves in Fig.~\ref{fig:effective-w}).

The PC1 model stands out as particularly interesting, as it displays a more pronounced dynamical evolution of $w^{\textrm{eff}}_{\textrm{DE}}(z)$ at low redshift, indicating a possible deviation from $\Lambda$CDM while remaining fully consistent with accelerated expansion. This behaviour is also reflected in the evolution of $H(z)/(1+z)$ and $q(z)$ shown in Figs.~\ref{fig:Hubble_PC} and~\ref{fig:deceleration}. Moreover, the PC1 model exhibits a transition from an effective quintessence regime to a phantom-like behaviour at low redshift.  
This feature makes PC1 a useful framework for exploring departures from $\Lambda$CDM and effective dynamical dark-energy scenarios, particularly in light of recent results from the DESI collaboration~\cite{DESI:2025zgx}.

\section{Conclusions} \label{sec:conclusions}

In this work, we have investigated a class of cosmological models based on particle creation (PC), aimed at providing an effective description of late-time cosmic acceleration and exploring possible implications for the $H_0$ tension. By adopting an agnostic approach to the nature of the created component and confronting the models with a comprehensive set of cosmological observations, we have assessed their viability in comparison with the standard $\Lambda$CDM model and dynamical dark-energy scenarios.

Using combinations of the \texttt{CC}, \texttt{SN}, \texttt{SH0ES}, \texttt{BAO}, and \texttt{CMB} datasets, we have shown that all the PC models considered are fully consistent with current observations and provide an excellent description of the late-time expansion history of the Universe. When individual dataset combinations are analysed separately, the PC models partially alleviate the $H_0$ tension with respect to $\Lambda$CDM and $w_0w_a$CDM, although a residual discrepancy remains (see the discussion in Sect.~\ref{sec:cosmological-results}). 
While \texttt{BAO+CMB} data show a mild preference for the particle-creation framework, low-redshift observations remain largely insensitive in terms of goodness of fit, effectively penalizing the additional model complexity. In the joint analysis, PC scenarios provide fits comparable to, and in terms of minimum $\chi^2_{\textrm{min}}$ slightly better than, those of $\Lambda$CDM. However, according to the empirical Jeffreys scale, the Bayesian evidence in favour of PC models over $\Lambda$CDM remains weak or inconclusive (see Table~\ref{TableConstraintsFull}). Overall, the combined dataset indicates that PC models are mildly favoured relative to $\Lambda$CDM.

A key outcome of our analysis is that the intrinsic equation-of-state parameter of the created component is robustly constrained by the data to satisfy $w_E < -1/3$ in all PC models, implying that the newly created particles effectively behave as a dark-energy-like component. This result is particularly significant, since the nature of the extra component was not specified \textit{a priori} and $w_E$ was treated as a free parameter. Our findings therefore indicate that, even within scenarios based on particle creation in the late Universe, a dark-energy-like component is still required by observations to explain the late-time dynamics of the Universe. Moreover, the hypothesis of particle creation in the form of pressureless matter ($w_E = 0$) at low redshift is strongly disfavoured by the data at high statistical significance, in contrast with several earlier studies that focused on cold dark matter production~\cite{Freaza:2002ic,Pan:2016jli,Cardenas:2020grl,Elizalde:2024rvg}.

From a dynamical perspective, the PC models describe an accelerating Universe, with present-day values of the deceleration parameter and effective dark-energy equation of state fully consistent with $\Lambda$CDM within $1\,\sigma$. The reconstructed evolutions of $H(z)/(1+z)$, $q(z)$, and $w^{\textrm{eff}}_{\textrm{DE}}(z)$ closely follow the $\Lambda$CDM predictions, with only mild deviations allowed at late times. In particular, all PC models exhibit a transition from deceleration to acceleration at $z<1$, in agreement with standard cosmology.

An important conceptual difference with respect to phenomenological dynamical dark-energy models, such as $w_0w_a$CDM, lies in the redshift range over which deviations from $\Lambda$CDM are admitted. By construction, we impose $g(0)=1$ and recover the $\Lambda$CDM expansion history at $z \geq 3 $, in order to preserve the sound horizon at the drag epoch and allow for a consistent use of CMB data as a high-redshift BAO probe. As a consequence, departures from $\Lambda$CDM are confined to the interval $0<z<3$, where the continuity equation is solved numerically and particle-creation effects become relevant. This leads to a weak but non-trivial effective dynamical dark-energy behaviour, distinct from the asymptotic evolution inherent to the $w_0w_a$CDM model. Indeed, within the CPL parametrization one has $w(z)\to w_0$ at $z=0$ and $w(z)\to w_0+w_a$ asymptotically as $z\to\infty$. By contrast, our results show that the PC models remain extremely close to $\Lambda$CDM at both low and high redshift, while allowing for mild, localized deviations at intermediate redshifts.

Among the PC models explored, PC1 emerges as particularly interesting, as it allows for a more pronounced evolution of the effective dark-energy equation of state $w^{\textrm{eff}}_{\textrm{DE}}(z)$ while remaining consistent with accelerated expansion and with all background observables. This feature makes it a useful framework for exploring controlled departures from $\Lambda$CDM and for providing a physically motivated alternative to purely phenomenological dynamical dark-energy models, particularly in light of recent and forthcoming high-precision measurements from large-scale structure surveys such as DESI~\cite{DESI:2025zgx}.

Finally, we note that while particle-creation models provide a consistent and physically motivated description of late-time cosmic acceleration, the parametrization of the particle-creation rate remains phenomenological. Further progress will require a deeper theoretical understanding of the underlying microphysical mechanisms, as well as an extension of the analysis to perturbations and structure formation. Moreover, allowing particle creation to operate over a broader redshift range than $0< z < 3$ and solving the continuity equation numerically also in the early Universe may help to further develop the physical description of these models. In this work, however, we have focused on low redshifts due to numerical limitations.
Future observational data, in particular from next-generation galaxy surveys and improved local measurements of $H_0$, will be crucial to further test this class of models and to assess their viability as alternatives to the standard cosmological paradigm.


\begin{acknowledgments}
The work of TS is supported by the Della Riccia foundation grant 2025. MDA is supported by the Leverhulme Trust.
EDV is supported by a Royal Society Dorothy Hodgkin Research Fellowship.
This article/publication is based upon work from COST Action CA21136 Addressing observational tensions in cosmology with systematics and fundamental physics (CosmoVerse) supported by COST (European Cooperation in Science and Technology). L.A.E.\ acknowledges financial support from the T\"{u}rkiye Bilimsel ve Teknolojik Ara\c{s}t{\i}rma Kurumu (T\"{U}B\.{I}TAK, Scientific and Technological Research Council of T\"{u}rkiye) through grant no.\ 124N627. TS and MDA acknowledge the IT Services at The University of Sheffield for providing High Performance Computing resources.
\end{acknowledgments}



\appendix

\section{Preliminary analysis for choosing priors} \label{sec:appendix}

\begin{figure*} \includegraphics[scale=0.25]{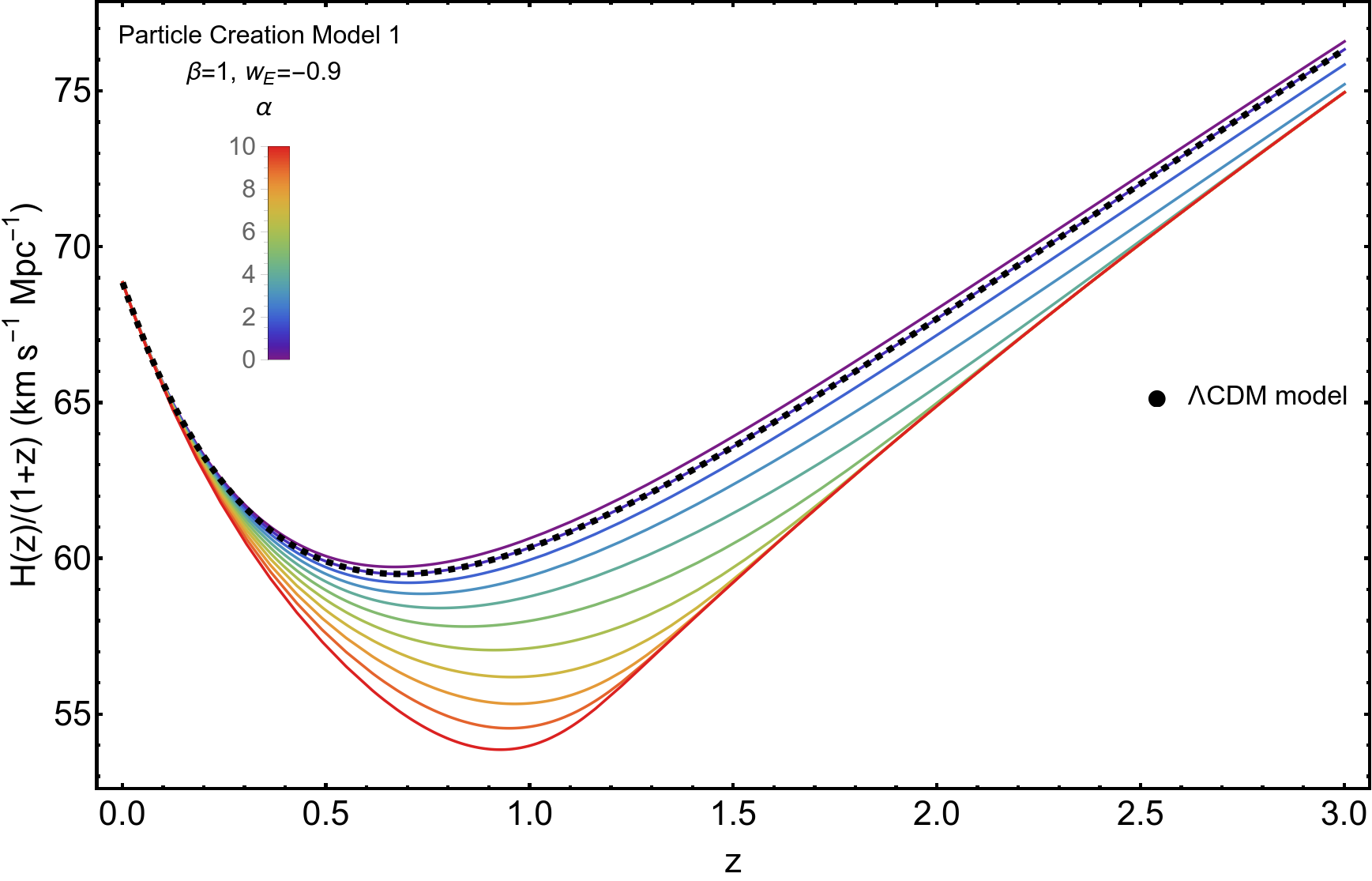} \includegraphics[scale=0.25]{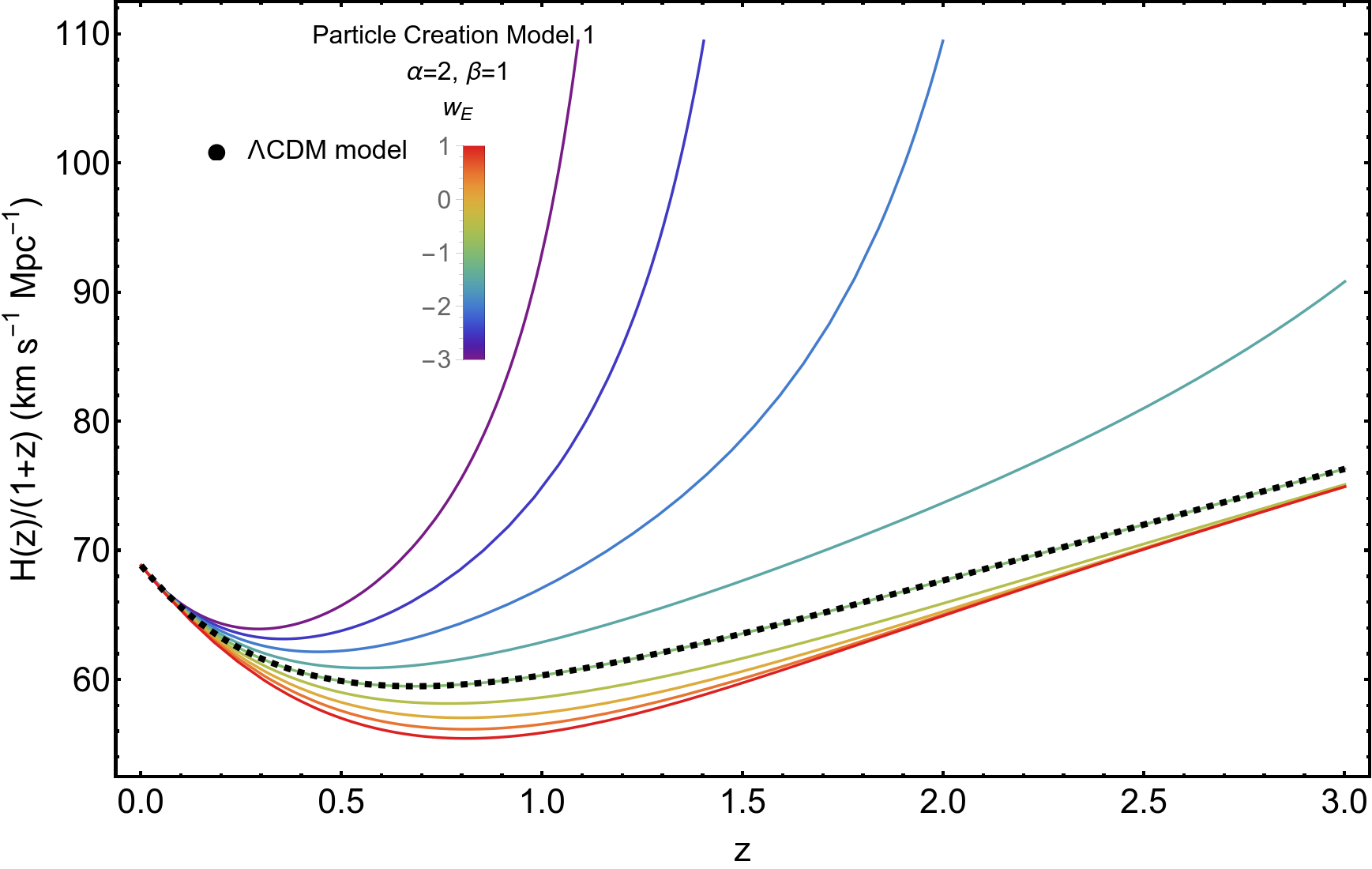}  \caption{Left panel: plot of the function $\frac{H(z)}{1+z}$ in the PC1 model for different values of $\alpha$. The parameter $w_E$ was fixed to the value $-0.9$. The trivial case $w_E=-1$ is not interesting, since it ends up with the $\Lambda$CDM model, as it can be seen also from Eq.~\eqref{eq:continuity-g-with-Gamma-in-z}. Right panel: plot of the function $\frac{H(z)}{1+z}$ in the PC1 model for different values of $w_E$. The extra parameters were set $\alpha=2$ and $\beta=1$. For both panels, $H_0$, $\Omega_{m0}$ are fixed to their best-fit values within the corresponding $\Lambda$CDM model, which are reported in Table~\ref{TableConstraintsFull} (\texttt{CC+SN+SH0ES+BAO+CMB} constraints). The dashed line indicates the flat $\Lambda$CDM model for a comparison. } \label{fig:model1-varying-alpha} \end{figure*}
\begin{figure*} \includegraphics[scale=0.25]{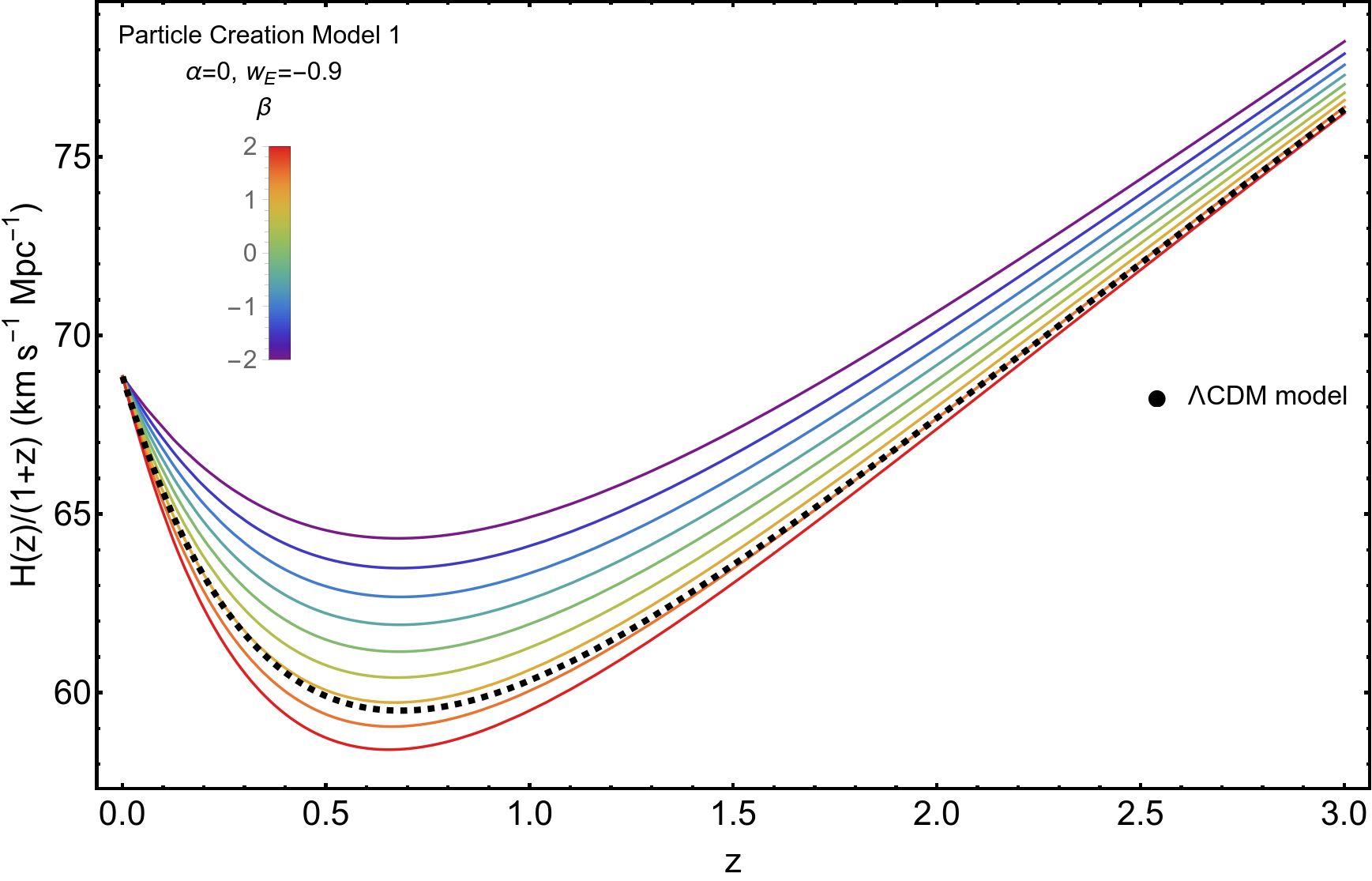} \includegraphics[scale=0.25]{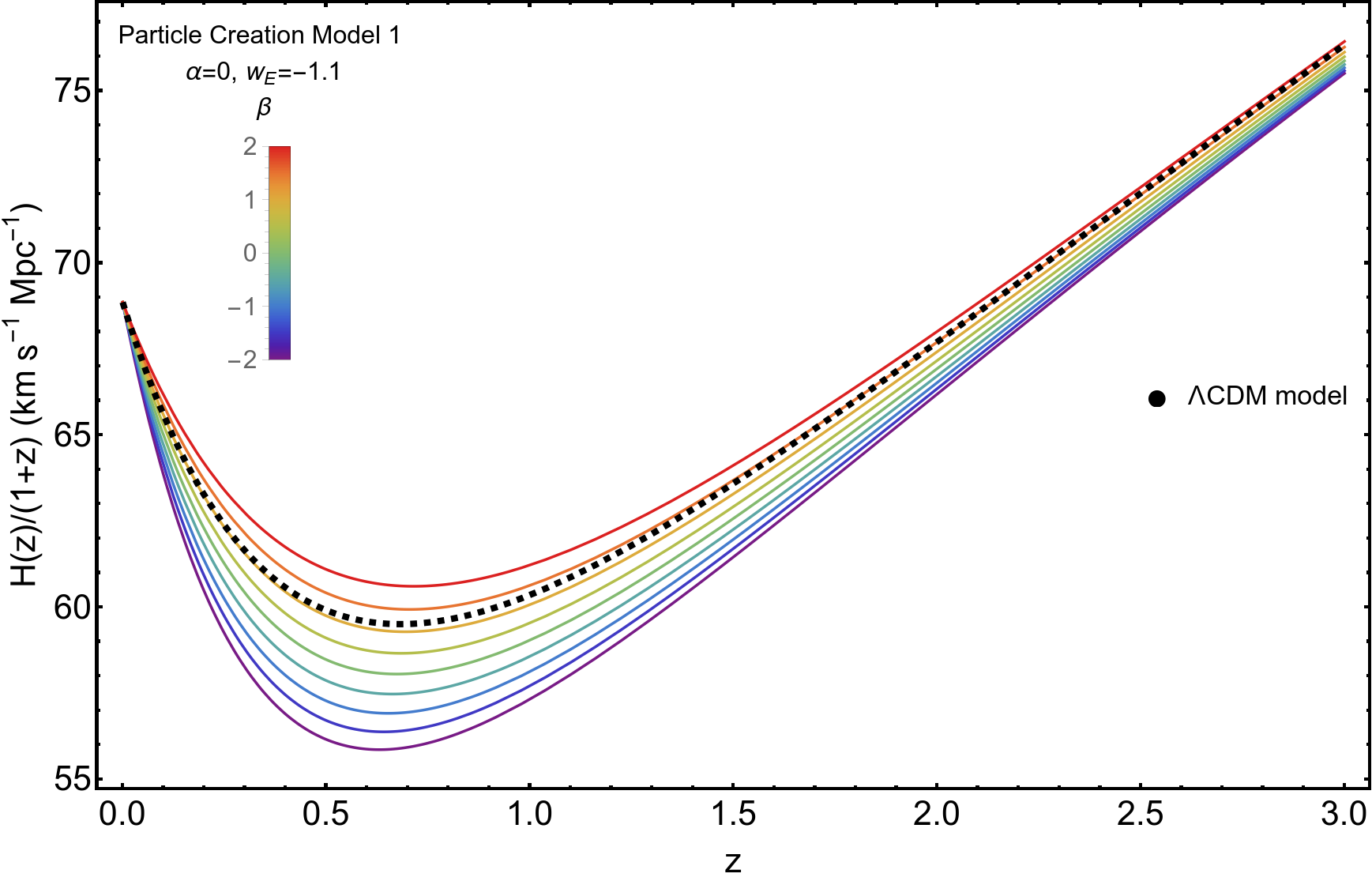} \\  \includegraphics[scale=0.25]{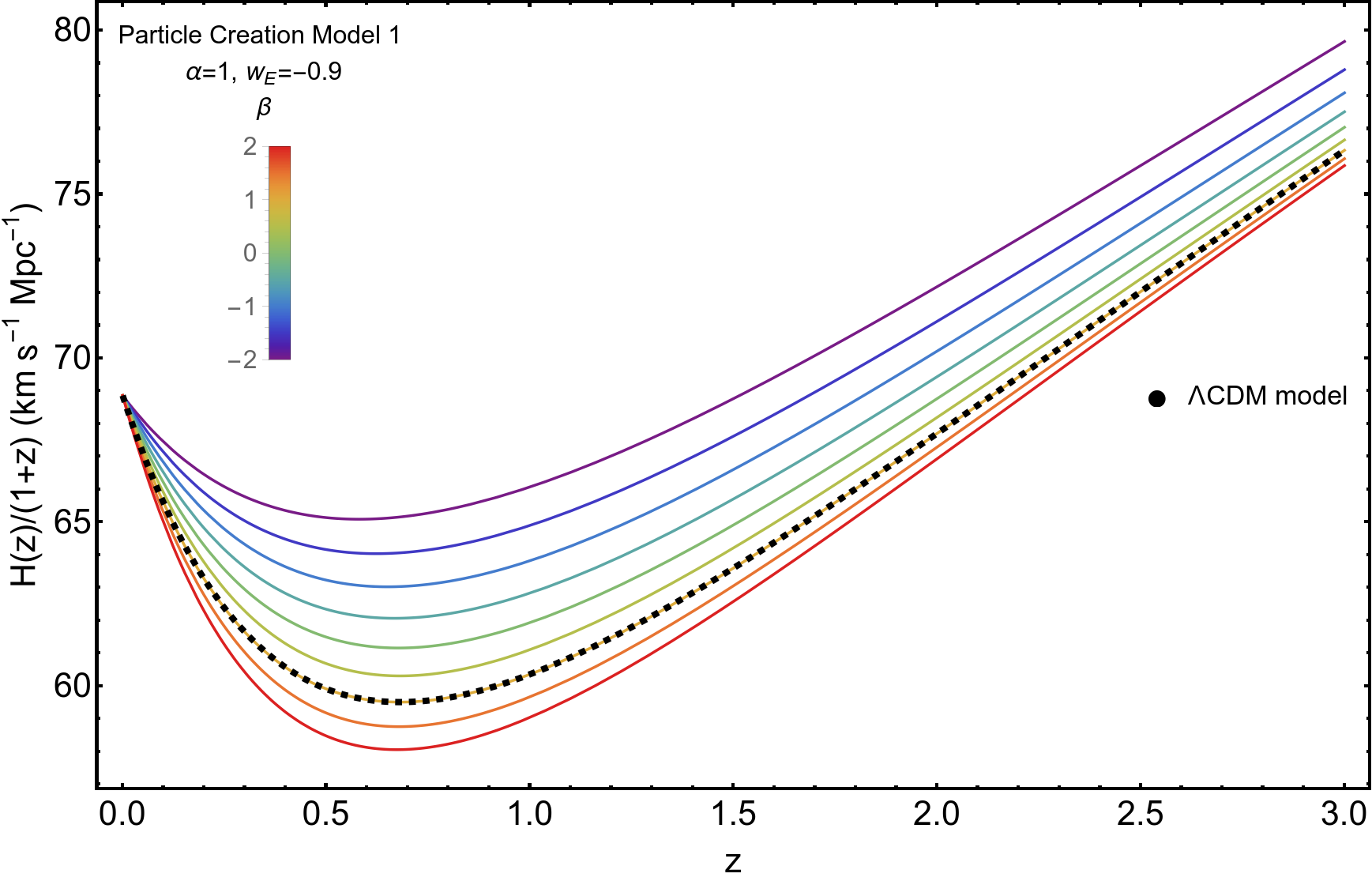} \includegraphics[scale=0.25]{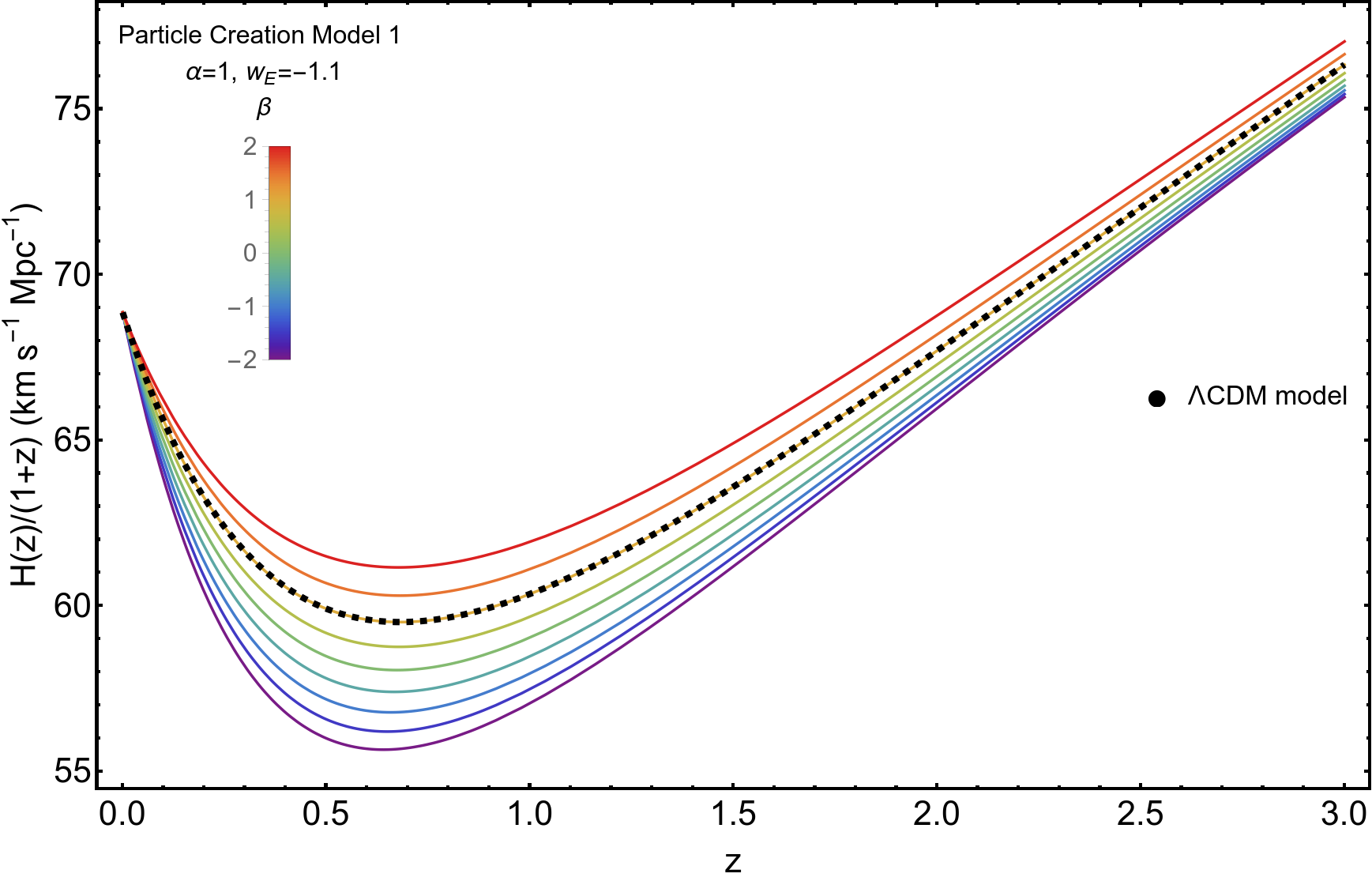} \\  \includegraphics[scale=0.25]{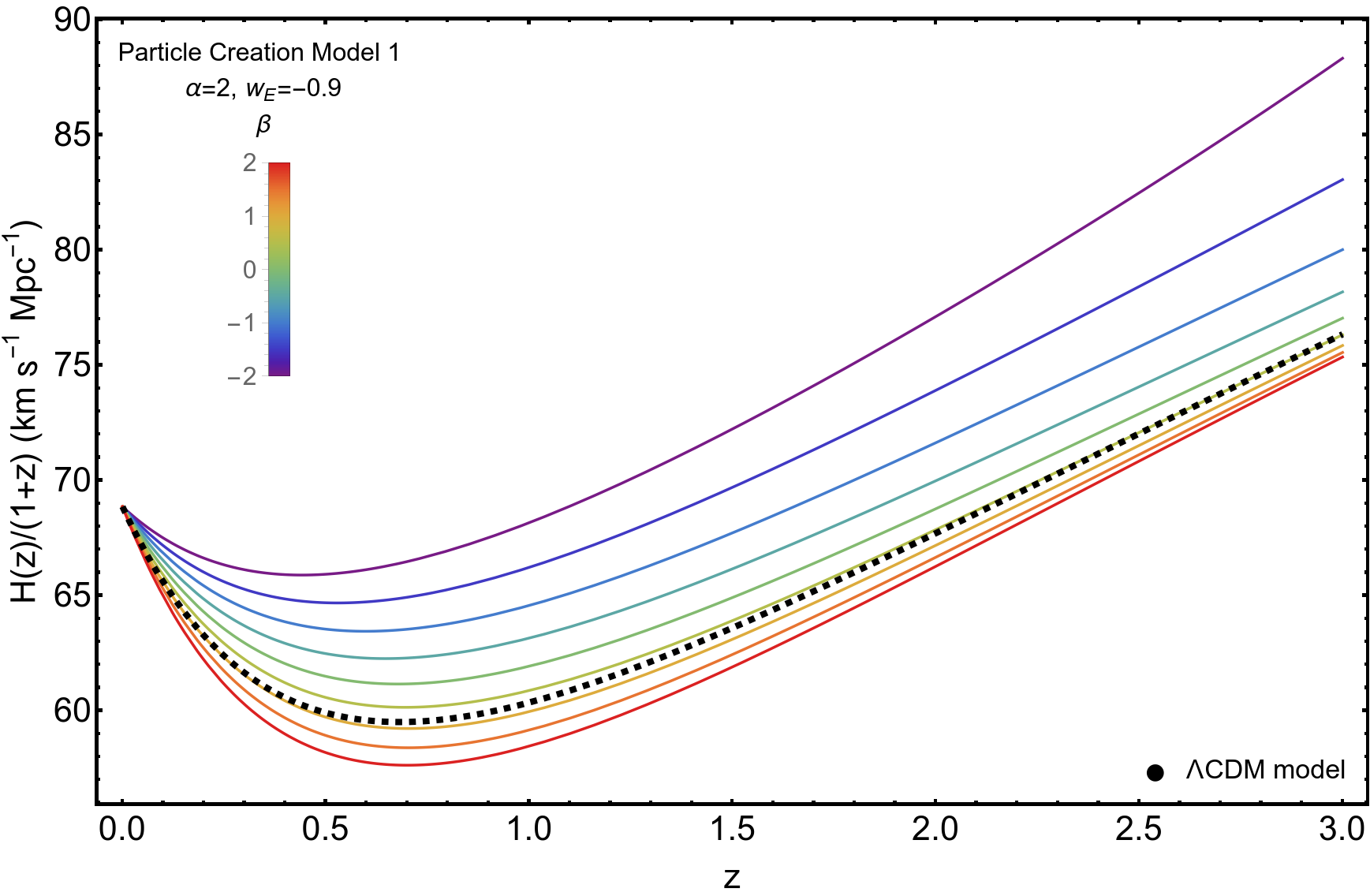} \includegraphics[scale=0.25]{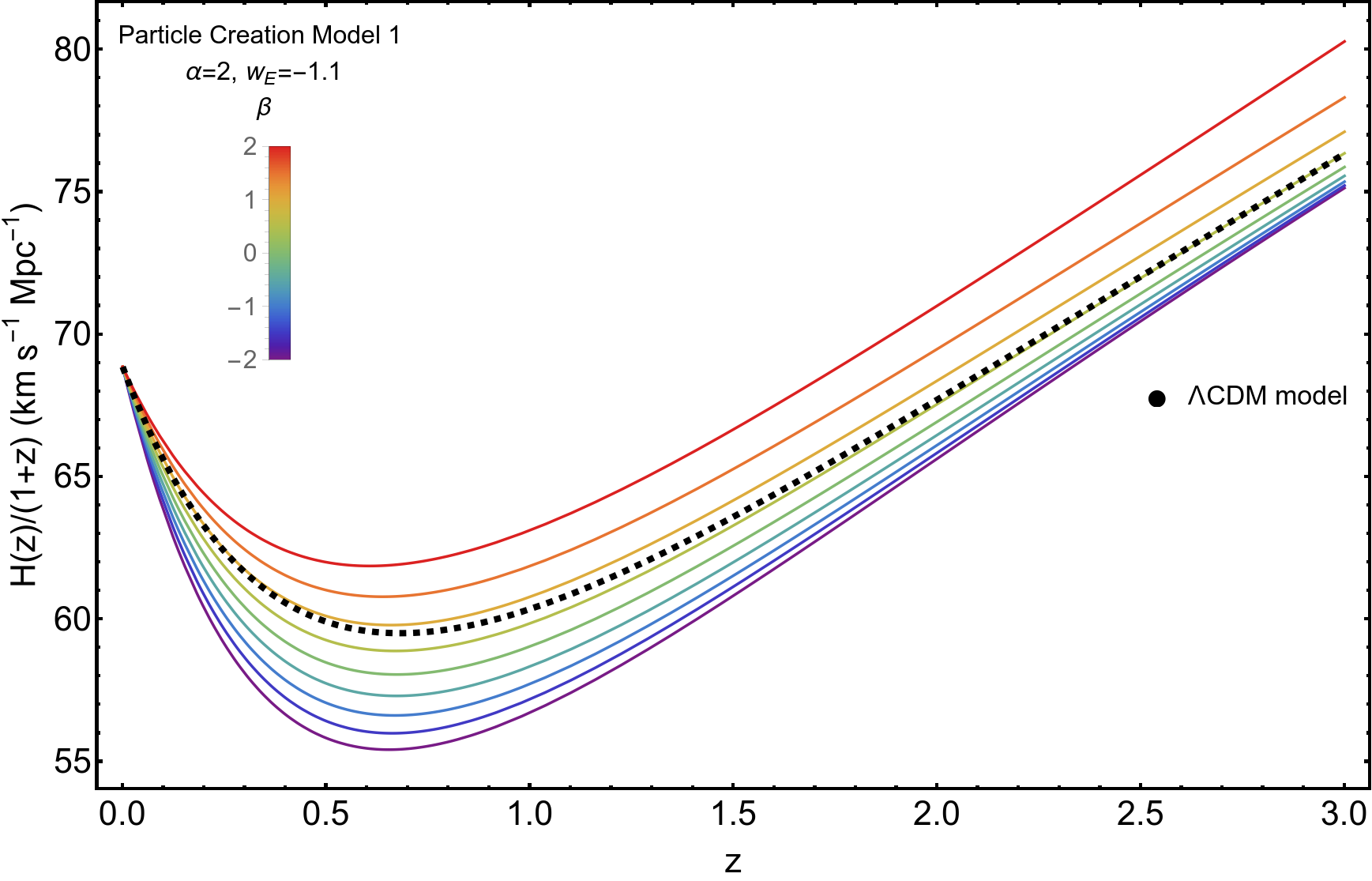} \caption{Plots of the function $\frac{H(z)}{1+z}$ in the PC1 model for different values of $\beta$. The parameter $w_E$ was fixed to the value $-0.9$ and $-1.1$ on the left and right panels, respectively. The trivial case $w_E=-1$ is not interesting, since it ends up with the $\Lambda$CDM model, as it can be seen also from Eq.~\eqref{eq:continuity-g-with-Gamma-in-z}. The upper, central, and lower panels refer to $\alpha=0,1,2$, respectively. The other cosmological parameters are fixed to their best-fit values within the corresponding $\Lambda$CDM model, which are reported in Table~\ref{TableConstraintsFull} (\texttt{CC+SN+SH0ES+BAO+CMB} constraints). The dashed line indicates the flat $\Lambda$CDM model for a comparison.} \label{fig:model1-varying-beta} \end{figure*}

The priors for all parameters used in the analysis are chosen to be flat (agnostic) and can be found in Table~\ref{TablePriors}. These were chosen following a preliminary exploration of the PC models, aimed at evaluating the impact of different parameter values on the Hubble function. For illustration, Figs.~\ref{fig:model1-varying-alpha} and~\ref{fig:model1-varying-beta} show this procedure for the PC1 model. A similar analysis was carried out for each PC model.

\begin{figure*} \includegraphics[scale=0.25]{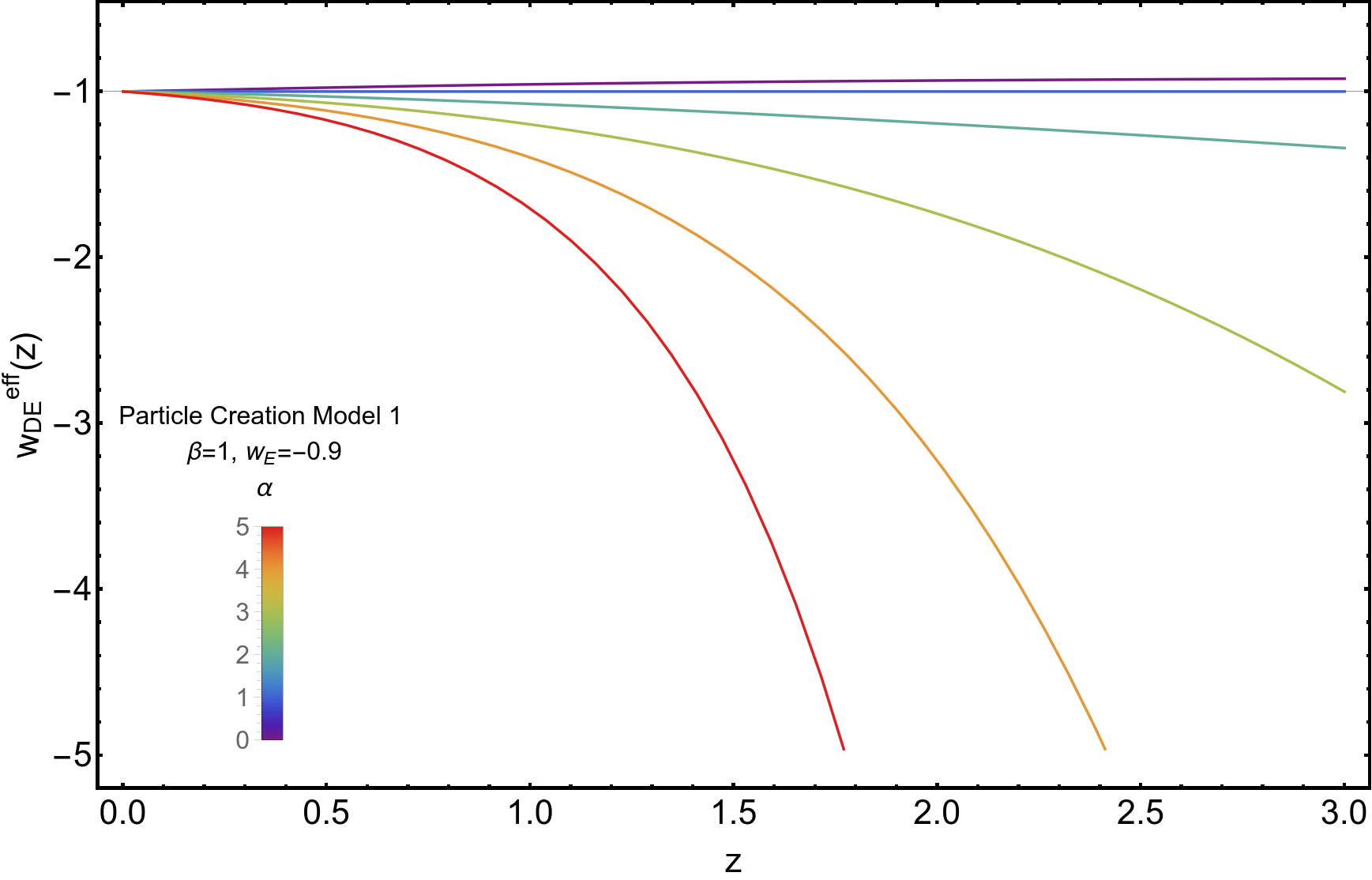} \caption{Plot of the effective equation of state parameter of dark energy $w^{\textrm{eff}}_{\textrm{DE}}(z)$ in the PC1 model for different values of $\alpha$. The extra parameters are set $w_E=-0.9$ and $\beta=1$, while $H_0$, $\Omega_{m0}$ are fixed to their best-fit values within the corresponding $\Lambda$CDM model, which are reported in Table~\ref{TableConstraintsFull} (\texttt{CC+SN+SH0ES+BAO+CMB} constraints). The grey curve indicates the cosmological constant scenario with $w^{\textrm{eff}}_{\textrm{DE}}=-1$.} \label{fig:model1-w-effective-varying-alpha-priors} \end{figure*}

Likewise, Fig.~\ref{fig:model1-w-effective-varying-alpha-priors} shows the dependence of the effective dark-energy equation of state on the model parameters within PC1, forming part of the prior-selection process. For reference, we also use the flat (zero-curvature) $\Lambda$CDM model, considering the best-fit values reported in Table~\ref{TableConstraintsFull} and obtained from \texttt{CC+SN+SH0ES+BAO+CMB} analysis.

\bibliographystyle{apsrev4-2}
\bibliography{biblio,biblio_2}

@misc{simplemc,
  author = "A. Slosar and J. A. Vazquez",
  year = 2020,
  howpublished = "\url{https://github.com/ja-vazquez/SimpleMC}"
}

@ARTICLE{2020MNRAS.493.3132S,
       author = {{Speagle}, Joshua S.},
        title = "{DYNESTY: a dynamic nested sampling package for estimating Bayesian posteriors and evidences}",
      journal = {"Mon. Not. Roy. Astron. Soc."},
         year = 2020,
        month = apr,
       volume = {493},
       number = {3},
        pages = {3132-3158},
          doi = {10.1093/mnras/staa278},
archivePrefix = {arXiv},
       eprint = {1904.02180},
 primaryClass = {astro-ph.IM},
       adsurl = {https://ui.adsabs.harvard.edu/abs/2020MNRAS.493.3132S}
}

@INPROCEEDINGS{2004AIPC..735..395S,
       author = {{Skilling}, John},
        title = "{Nested Sampling}",
    booktitle = {Bayesian Inference and Maximum Entropy Methods in Science and Engineering: 24th International Workshop on Bayesian Inference and Maximum Entropy Methods in Science and Engineering},
         year = 2004,
       editor = {{Fischer}, Rainer and {Preuss}, Roland and {Toussaint}, Udo Von},
       series = {American Institute of Physics Conference Series},
       volume = {735},
        month = nov,
    publisher = {AIP},
        pages = {395-405},
          doi = {10.1063/1.1835238},
       adsurl = {https://ui.adsabs.harvard.edu/abs/2004AIPC..735..395S}
}

@inproceedings{Schiavone:2022shz,
    author = "Schiavone, Tiziano and Montani, Giovanni and Dainotti, Maria Giovanna and De Simone, Biagio and Rinaldi, Enrico and Lambiase, Gaetano",
    title = "{Running Hubble constant from the SNe Ia Pantheon sample?}",
    booktitle = "{17th Italian-Korean Symposium on Relativistic Astrophysics}",
    eprint = "2205.07033",
    archivePrefix = "arXiv",
    primaryClass = "astro-ph.CO",
    reportNumber = "RIKEN-ITHEMS-Report-22",
    month = "5",
    year = "2022"
}

@article{Malekjani:2023ple,
    author = "Malekjani, Mohammad and Conville, Ruair{\'\i} Mc and Colg{\'a}in, Eoin {\'O}. and Pourojaghi, Saeed and Sheikh-Jabbari, M. M.",
    title = "{On redshift evolution and negative dark energy density in Pantheon + Supernovae}",
    eprint = "2301.12725",
    archivePrefix = "arXiv",
    primaryClass = "astro-ph.CO",
    doi = "10.1140/epjc/s10052-024-12667-z",
    journal = "Eur. Phys. J. C",
    volume = "84",
    number = "3",
    pages = "317",
    year = "2024"
}

@article{Krishnan:2020obg,
    author = "Krishnan, C. and Colg{\'a}in, Eoin {\'O}. and Ruchika and Sen, Anjan A. and Sheikh-Jabbari, M. M. and Yang, Tao",
    title = "{Is there an early Universe solution to Hubble tension?}",
    eprint = "2002.06044",
    archivePrefix = "arXiv",
    primaryClass = "astro-ph.CO",
    doi = "10.1103/PhysRevD.102.103525",
    journal = "Phys. Rev. D",
    volume = "102",
    number = "10",
    pages = "103525",
    year = "2020"
}

@article{Colgain:2022rxy,
    author = "Colg{\'a}in, Eoin {\'O}. and Sheikh-Jabbari, M. M. and Solomon, Rance and Dainotti, Maria G. and Stojkovic, Dejan",
    title = "{Putting flat {\ensuremath{\Lambda}}CDM in the (Redshift) bin}",
    eprint = "2206.11447",
    archivePrefix = "arXiv",
    primaryClass = "astro-ph.CO",
    doi = "10.1016/j.dark.2024.101464",
    journal = "Phys. Dark Univ.",
    volume = "44",
    pages = "101464",
    year = "2024"
}

@article{Colgain:2022tql,
    author = "Colg{\'a}in, Eoin {\'O}. and Sheikh-Jabbari, M. M. and Solomon, Rance",
    title = "{High redshift {\ensuremath{\Lambda}}CDM cosmology: To bin or not to bin?}",
    eprint = "2211.02129",
    archivePrefix = "arXiv",
    primaryClass = "astro-ph.CO",
    doi = "10.1016/j.dark.2023.101216",
    journal = "Phys. Dark Univ.",
    volume = "40",
    pages = "101216",
    year = "2023"
}

@article{Jia:2022ycc,
    author = "Jia, X. D. and Hu, J. P. and Wang, F. Y.",
    title = "{Evidence of a decreasing trend for the Hubble constant}",
    eprint = "2212.00238",
    archivePrefix = "arXiv",
    primaryClass = "astro-ph.CO",
    doi = "10.1051/0004-6361/202346356",
    journal = "Astron. Astrophys.",
    volume = "674",
    pages = "A45",
    year = "2023"
}

@article{Wang:2025xvi,
    author = "Wang, Yi-Ying and Li, Yin-Jie and Fan, Yi-Zhong",
    title = "{Evidence for dynamical dark energy with an evolving Hubble constant}",
    eprint = "2510.14390",
    archivePrefix = "arXiv",
    primaryClass = "astro-ph.CO",
    doi = "10.1051/0004-6361/202558451",
    journal = "Astron. Astrophys.",
    volume = "707",
    pages = "A189",
    year = "2026"
}

@article{H0LiCOW:2019pvv,
    author = "Wong, Kenneth C. and others",
    collaboration = "H0LiCOW",
    title = "{H0LiCOW {\textendash} XIII. A 2.4 per cent measurement of H0 from lensed quasars: 5.3{\ensuremath{\sigma}} tension between early- and late-Universe probes}",
    eprint = "1907.04869",
    archivePrefix = "arXiv",
    primaryClass = "astro-ph.CO",
    doi = "10.1093/mnras/stz3094",
    journal = "Mon. Not. Roy. Astron. Soc.",
    volume = "498",
    number = "1",
    pages = "1420--1439",
    year = "2020"
}

@article{Kazantzidis:2020tko,
    author = "Kazantzidis, L. and Perivolaropoulos, L.",
    title = "{Hints of a Local Matter Underdensity or Modified Gravity in the Low $z$ Pantheon data}",
    eprint = "2004.02155",
    archivePrefix = "arXiv",
    primaryClass = "astro-ph.CO",
    doi = "10.1103/PhysRevD.102.023520",
    journal = "Phys. Rev. D",
    volume = "102",
    number = "2",
    pages = "023520",
    year = "2020"
}

@article{Dainotti:2023yrk,
    author = "Dainotti, Maria and De Simone, Biagio and Montani;, Giovanni and Schiavone;, Tiziano and Lambiase., Gaetano",
    title = "{The Hubble constant tension: current status and future perspectives through new cosmological probes}",
    eprint = "2301.10572",
    archivePrefix = "arXiv",
    primaryClass = "astro-ph.CO",
    doi = "10.22323/1.436.0235",
    journal = "PoS",
    volume = "CORFU2022",
    pages = "235",
    year = "2023"
}

@BOOK{Weinberg:2008zzc,
       author = {{Weinberg}, Steven},
        title = "{Cosmology}",
         year = 2008,
    publisher = "Oxford University Press",
      address = "Oxford, UK"
}

@article{Elizalde:2024rvg,
    author = "Elizalde, Emilio and Khurshudyan, Martiros and Odintsov, Sergei D.",
    title = "{Can we learn from matter creation to solve the $H_{0}$ tension problem?}",
    eprint = "2407.20285",
    archivePrefix = "arXiv",
    primaryClass = "gr-qc",
    doi = "10.1140/epjc/s10052-024-13146-1",
    journal = "Eur. Phys. J. C",
    volume = "84",
    number = "8",
    pages = "782",
    year = "2024"
}

@article{Cardenas:2020grl,
    author = "C\'ardenas, V\'\i{}ctor H. and Cruz, Miguel and Lepe, Samuel and Nojiri, Shin'ichi and Odintsov, Sergei D.",
    title = "{Challenging matter creation models in the phantom divide}",
    eprint = "2004.02337",
    archivePrefix = "arXiv",
    primaryClass = "gr-qc",
    doi = "10.1103/PhysRevD.101.083530",
    journal = "Phys. Rev. D",
    volume = "101",
    number = "8",
    pages = "083530",
    year = "2020"
}

@article{Calvao:1991wg,
    author = "Calvao, M. O. and Lima, J. A. S. and Waga, I.",
    title = "{On the thermodynamics of matter creation in cosmology}",
    reportNumber = "IF-UFRJ-91-23",
    doi = "10.1016/0375-9601(92)90437-Q",
    journal = "Phys. Lett. A",
    volume = "162",
    pages = "223--226",
    year = "1992"
}

@article{CostaNetto:2025vew,
    author = "Costa Netto, Jos\'e Medeiros da and Silva, Heydson Henrique Brito da",
    title = "{A thermodynamic model for dark energy including particle creation or destruction processes}",
    eprint = "2503.00087",
    archivePrefix = "arXiv",
    primaryClass = "gr-qc",
    doi = "10.1016/j.cjph.2025.02.007",
    journal = "Chin. J. Phys.",
    volume = "94",
    pages = "684--689",
    year = "2025"
}

@article{Lima:2025yza,
    author = "Lima, P. W. R. and Lima, J. A. S. and Jesus, J. F.",
    title = "{New accelerating cosmology without dark energy: the particle creation approach and the reduced relativistic gas}",
    eprint = "2502.14139",
    archivePrefix = "arXiv",
    primaryClass = "astro-ph.CO",
    doi = "10.1140/epjc/s10052-025-14115-y",
    journal = "Eur. Phys. J. C",
    volume = "85",
    number = "4",
    pages = "449",
    year = "2025"
}

@article{Cardenas:2025sqf,
    author = "C{\'a}rdenas, V{\'\i}ctor H. and Lepe, Samuel",
    title = "{Matter creation, adiabaticity and phantom behavior}",
    eprint = "2501.14509",
    archivePrefix = "arXiv",
    primaryClass = "astro-ph.CO",
    doi = "10.1140/epjc/s10052-025-14154-5",
    journal = "Eur. Phys. J. C",
    volume = "85",
    number = "4",
    pages = "411",
    year = "2025"
}

@article{Marciu:2024gqv,
    author = "Marciu, Mihai",
    title = "{Matter-geometry interplay in new scalar tensor theories of gravity}",
    eprint = "2410.04584",
    archivePrefix = "arXiv",
    primaryClass = "gr-qc",
    doi = "10.1140/epjc/s10052-024-13558-z",
    journal = "Eur. Phys. J. C",
    volume = "84",
    number = "11",
    pages = "1191",
    year = "2024",
    note = "[Erratum: Eur.Phys.J.C 84, 1285 (2024)]"
}

@article{Cipriano:2024jng,
    author = "Cipriano, Ricardo A. C. and Ganiyeva, Nailya and Harko, Tiberiu and Lobo, Francisco S. N. and Pinto, Miguel A. S. and Rosa, Jo\~ao Lu\'\i{}s",
    title = "{Energy-Momentum Squared Gravity: A Brief Overview}",
    eprint = "2408.14106",
    archivePrefix = "arXiv",
    primaryClass = "gr-qc",
    doi = "10.3390/universe10090339",
    journal = "Universe",
    volume = "10",
    number = "9",
    pages = "339",
    year = "2024"
}

@article{Montani:2024xys,
    author = "Montani, Giovanni and Carlevaro, Nakia and De Angelis, Mariaveronica",
    title = "{Modified Gravity in the Presence of Matter Creation: Scenario for the Late Universe}",
    eprint = "2407.12409",
    archivePrefix = "arXiv",
    primaryClass = "gr-qc",
    doi = "10.3390/e26080662",
    journal = "Entropy",
    volume = "26",
    number = "8",
    pages = "662",
    year = "2024"
}

@article{Singh:2024nsh,
    author = {Singh, G. P. and Garg, Romanshu and Singh, Ashutosh},
    title = {A generalized $\Lambda$CDM model with parameterized Hubble parameter in particle creation, viscous and $f(R)$ model framework},
    journal = {International Journal of Geometric Methods in Modern Physics},
    volume = {0},
    number = {0},
    pages = {2550111},
    year = {0},
    doi = {10.1142/S0219887825501117},
    eprint = "2405.15626"
}

@article{Bouali:2023fid,
    author = "Bouali, Amine and Chaudhary, Himanshu and Harko, Tiberiu and Lobo, Francisco S. N. and Ouali, Taoufik and Pinto, Miguel A. S.",
    title = "{Observational constraints and cosmological implications of scalar\textendash{}tensor f(R, T) gravity}",
    eprint = "2309.15497",
    archivePrefix = "arXiv",
    primaryClass = "gr-qc",
    doi = "10.1093/mnras/stad2998",
    journal = "Mon. Not. Roy. Astron. Soc.",
    volume = "526",
    number = "3",
    pages = "4192--4208",
    year = "2023"
}

@article{Trevisani:2023wpw,
    author = "Trevisani, S. R. G. and Lima, J. A. S.",
    title = "{Gravitational matter creation, multi-fluid cosmology and kinetic theory}",
    eprint = "2303.01974",
    archivePrefix = "arXiv",
    primaryClass = "astro-ph.CO",
    doi = "10.1140/epjc/s10052-023-11301-8",
    journal = "Eur. Phys. J. C",
    volume = "83",
    number = "3",
    pages = "244",
    year = "2023"
}

@article{Cardenas:2023zmn,
    author = "C\'ardenas, V\'\i{}ctor H. and Cruz, Miguel and Lepe, Samuel",
    title = "{Generalized second law of thermodynamics for the matter creation scenario and emergence of phantom regime}",
    eprint = "2302.10155",
    archivePrefix = "arXiv",
    primaryClass = "gr-qc",
    doi = "10.1140/epjp/s13360-024-05447-x",
    journal = "Eur. Phys. J. Plus",
    volume = "139",
    number = "7",
    pages = "642",
    year = "2024"
}

@article{Akarsu:2023nyl,
    author = "Akarsu, Ozgur and Uzun, N. Merve",
    title = "{Cosmological models in scale-independent energy-momentum squared gravity}",
    eprint = "2301.11204",
    archivePrefix = "arXiv",
    primaryClass = "gr-qc",
    doi = "10.1016/j.dark.2023.101194",
    journal = "Phys. Dark Univ.",
    volume = "40",
    pages = "101194",
    year = "2023"
}

@article{Ganjizadeh:2022mxe,
    author = "Ganjizadeh, S. and Amani, Alireza and Ramzanpour, M. A.",
    title = "{Observational Hubble parameter data constraints on the interactive model of gravity with particle creation}",
    eprint = "2208.07710",
    archivePrefix = "arXiv",
    primaryClass = "gr-qc",
    reportNumber = "CPC-2022-0297.R2",
    doi = "10.1088/1674-1137/ac8c22",
    journal = "Chin. Phys. C",
    volume = "46",
    number = "12",
    pages = "125104",
    year = "2022"
}

@article{Cardenas:2020exv,
    author = "C\'ardenas, V\'\i{}ctor H. and Cruz, Miguel and Lepe, Samuel",
    title = "{Cosmic expansion with matter creation and bulk viscosity}",
    eprint = "2008.12403",
    archivePrefix = "arXiv",
    primaryClass = "gr-qc",
    doi = "10.1103/PhysRevD.102.123543",
    journal = "Phys. Rev. D",
    volume = "102",
    number = "12",
    pages = "123543",
    year = "2020"
}

@article{Bolotin:2020qbx,
    author = "Bolotin, Yu. L. and Cherkaskiy, V. A. and Konchatnyi, M. I. and Pan, Supriya and Yang, Weiqiang",
    title = "{Do current observations support transient acceleration of our universe?}",
    eprint = "2008.09602",
    archivePrefix = "arXiv",
    primaryClass = "gr-qc",
    doi = "10.1142/S0218271822500365",
    journal = "Int. J. Mod. Phys. D",
    volume = "31",
    number = "05",
    pages = "2250036",
    year = "2022"
}

@article{Freaza:2002ic,
    author = "Freaza, M. P. and de Souza, R. S. and Waga, I.",
    title = "{Cosmic acceleration and matter creation}",
    doi = "10.1103/PhysRevD.66.103502",
    journal = "Phys. Rev. D",
    volume = "66",
    pages = "103502",
    year = "2002"
}

@article{Schiavone:2024heb,
    author = "Schiavone, Tiziano and Montani, Giovanni",
    title = "{Evolution of an effective Hubble constant in f (R) modified gravity}",
    eprint = "2408.01410",
    archivePrefix = "arXiv",
    primaryClass = "gr-qc",
    doi = "10.1393/ncc/i2025-25105-3",
    journal = "Nuovo Cim. C",
    volume = "48",
    number = "3",
    pages = "105",
    year = "2025"
}

@article{DeSimone:2024lvy,
    author = "De Simone, B. and van Putten, M. H. P. M. and Dainotti, M. G. and Lambiase, G.",
    title = "{A doublet of cosmological models to challenge the H0 tension in the Pantheon Supernovae Ia catalog}",
    eprint = "2411.05744",
    archivePrefix = "arXiv",
    primaryClass = "astro-ph.CO",
    doi = "10.1016/j.jheap.2024.12.003",
    journal = "JHEAp",
    volume = "45",
    pages = "290--298",
    year = "2025"
}

@article{Dainotti:2023ebr,
    author = "Dainotti, Maria Giovanna and Bargiacchi, Giada and Bogdan, Malgorzata and Capozziello, Salvatore and Nagataki, Shigehiro",
    title = "{Reduced uncertainties up to 43{\%} on the Hubble constant and the matter density with the SNe Ia with a new statistical analysis}",
    eprint = "2303.06974",
    journal = "arXiv",
    archivePrefix = "arXiv",
    primaryClass = "astro-ph.CO",
    month = "3",
    year = "2023"
}

@article{Dainotti:2024gca,
    author = "Dainotti, M. G. and Bargiacchi, G. and Bogdan, M. and Capozziello, S. and Nagataki, S.",
    title = "{On the statistical assumption on the distance moduli of Supernovae Ia and its impact on the determination of cosmological parameters}",
    doi = "10.1016/j.jheap.2024.01.001",
    journal = "JHEAp",
    volume = "41",
    pages = "30--41",
    year = "2024"
}

@article{Montani:2024ntj,
    author = "Montani, Giovanni and Carlevaro, Nakia and Dainotti, Maria G.",
    title = "{Running Hubble constant: Evolutionary Dark Energy}",
    eprint = "2411.07060",
    archivePrefix = "arXiv",
    primaryClass = "gr-qc",
    doi = "10.1016/j.dark.2025.101847",
    journal = "Phys. Dark Univ.",
    volume = "48",
    pages = "101847",
    year = "2025"
}

@article{Montani:2024pou,
    author = "Montani, Giovanni and Carlevaro, Nakia and Escamilla, Luis A. and Di Valentino, Eleonora",
    title = "{Kinetic model for dark energy{\textemdash}dark matter interaction: Scenario for the hubble tension}",
    eprint = "2404.15977",
    archivePrefix = "arXiv",
    primaryClass = "gr-qc",
    doi = "10.1016/j.dark.2025.101848",
    journal = "Phys. Dark Univ.",
    volume = "48",
    pages = "101848",
    year = "2025"
}

@article{Montani:2025jkk,
    author = "Montani, Giovanni and De Angelis, Mariaveronica and Dainotti, Maria Giovanna",
    title = "{Decay of dark energy into dark matter in a metric f(R) gravity: Effective running Hubble constant}",
    eprint = "2506.13288",
    archivePrefix = "arXiv",
    primaryClass = "astro-ph.CO",
    doi = "10.1016/j.dark.2025.101969",
    journal = "Phys. Dark Univ.",
    volume = "49",
    pages = "101969",
    year = "2025"
}

@article{Dainotti:2025qxz,
    author = "Dainotti, M. G. and others",
    title = "{A New Master Supernovae Ia sample and the investigation of the Hubble tension}",
    eprint = "2501.11772",
    archivePrefix = "arXiv",
    primaryClass = "astro-ph.CO",
    reportNumber = "KEK-TH-2711, KEK-Cosmo-0378",
    doi = "10.1016/j.jheap.2025.100405",
    journal = "JHEAp",
    volume = "48",
    pages = "100405",
    year = "2025"
}

@article{Fazzari:2025mww,
    author = "Fazzari, E. and Dainotti, M. G. and Montani, G. and Melchiorri, A.",
    title = "{The effective running Hubble constant in SNe Ia as a marker for the dark energy nature}",
    eprint = "2506.04162",
    archivePrefix = "arXiv",
    primaryClass = "astro-ph.CO",
    doi = "10.1016/j.jheap.2025.100459",
    journal = "JHEAp",
    volume = "49",
    pages = "100459",
    year = "2026"
}

@article{Montani:2025rcy,
    author = "Montani, Giovanni and Fazzari, Elisa and Carlevaro, Nakia and Dainotti, Maria Giovanna",
    title = "{Two Dynamical Scenarios for Binned Master Sample Interpretation}",
    eprint = "2507.14048",
    archivePrefix = "arXiv",
    primaryClass = "astro-ph.CO",
    doi = "10.3390/e27090895",
    journal = "Entropy",
    volume = "27",
    number = "9",
    pages = "895",
    year = "2025"
}

@article{fgivenx,
    doi = {10.21105/joss.00849},
    url = {http://dx.doi.org/10.21105/joss.00849},
    year  = {2018},
    month = {Aug},
    publisher = {The Open Journal},
    volume = {3},
    number = {28},
    author = {Will Handley},
    title = {fgivenx: Functional Posterior Plotter},
    journal = {The Journal of Open Source Software}
}

@article{Lima:2007kk,
    author = "Lima, J. A. S. and Calvao, M. O. and Waga, I.",
    title = "{Cosmology, Thermodynamics and Matter Creation}",
    eprint = "0708.3397",
    archivePrefix = "arXiv",
    primaryClass = "astro-ph",
    month = "8",
    journal = "e-print",
    year = "2007"
}

@article{Navone:2025gxr,
    author = "Navone, Iolanda and Dainotti, Maria Giovanna and Fazzari, Elisa and Montani, Giovanni and Maki, Naoto and Kohri, Kazunori",
    title = "{Creation of Viscous Dark Energy by the Hubble Flow: Comparison with SNe Ia Master Sample Binned Data}",
    eprint = "2511.16130",
    archivePrefix = "arXiv",
    primaryClass = "astro-ph.CO",
    month = "11",
    year = "2025",
    journal = "e-print"
}

@article{Valletta:2025bgu,
    author = "Valletta, A. and Montani, G. and Dainotti, M. G. and Fazzari, E.",
    title = "{On the metric f(R) gravity viability in accounting for the binned supernovae data}",
    eprint = "2512.19568",
    archivePrefix = "arXiv",
    primaryClass = "gr-qc",
    doi = "10.1016/j.jheap.2026.100612",
    journal = "JHEAp",
    volume = "53",
    pages = "100612",
    year = "2026"
}

@article{Montani:2025nmz,
    author = "Montani, Giovanni and Escamilla, Luis A. and Carlevaro, Nakia and Di Valentino, Eleonora",
    title = "{Decay of $f(R)$ quintessence into dark matter: Mitigating the Hubble tension?}",
    eprint = "2512.20193",
    archivePrefix = "arXiv",
    primaryClass = "astro-ph.CO",
    doi = "10.1103/mn69-1dn6",
    journal = "Phys. Rev. D",
    volume = "113",
    number = "2",
    pages = "023507",
    year = "2026"
}

@article{DOnofrio:2025cuk,
    author = "D'Onofrio, Simone and Odintsov, Sergei and Schiavone, Tiziano",
    title = "{Exponential $f(R)$ cosmology with massive neutrinos as a dynamical dark energy framework}",
    eprint = "2511.06924",
    archivePrefix = "arXiv",
    primaryClass = "gr-qc",
    month = "11",
    year = "2025",
    journal = "e-print"
}

@article{Efstratiou:2025iqi,
    author = "Efstratiou, Dimitrios and Paraskevas, Evangelos Achilleas and Perivolaropoulos, Leandros",
    title = "{Addressing the DESI DR2 Phantom-Crossing Anomaly and Enhanced $H_0$ Tension with Reconstructed Scalar-Tensor Gravity}",
    eprint = "2511.04610",
    archivePrefix = "arXiv",
    primaryClass = "astro-ph.CO",
    month = "11",
    year = "2025",
    journal = "e-print"
}

@article{Odintsov:2024woi,
    author = "Odintsov, Sergei D. and S{\'a}ez-Chill{\'o}n G{\'o}mez, Diego and Sharov, German S.",
    title = "{Modified gravity/dynamical dark energy vs $\Lambda $CDM: is the game over?}",
    eprint = "2412.09409",
    archivePrefix = "arXiv",
    primaryClass = "gr-qc",
    doi = "10.1140/epjc/s10052-025-14013-3",
    journal = "Eur. Phys. J. C",
    volume = "85",
    number = "3",
    pages = "298",
    year = "2025"
}

@article{Jia:2024wix,
    author = "Jia, X. D. and Hu, J. P. and Yi, S. X. and Wang, F. Y.",
    title = "{Uncorrelated Estimations of H$_{0}$ Redshift Evolution from DESI Baryon Acoustic Oscillation Observations}",
    eprint = "2406.02019",
    archivePrefix = "arXiv",
    primaryClass = "astro-ph.CO",
    doi = "10.3847/2041-8213/ada94d",
    journal = "Astrophys. J. Lett.",
    volume = "979",
    number = "2",
    pages = "L34",
    year = "2025"
}

@article{Jia:2025poj,
    author = "Jia, X. D. and Hu, J. P. and Gao, D. H. and Yi, S. X. and Wang, F. Y.",
    title = "{The Hubble Tension Resolved by the DESI Baryon Acoustic Oscillations Measurements}",
    eprint = "2509.17454",
    archivePrefix = "arXiv",
    primaryClass = "astro-ph.CO",
    doi = "10.3847/2041-8213/ae1965",
    journal = "Astrophys. J. Lett.",
    volume = "994",
    number = "1",
    pages = "L22",
    year = "2025"
}

@article{Kalita:2025jqz,
    author = "Kalita, Surajit and Uniyal, Akhil and Bulik, Tomasz and Mizuno, Yosuke",
    title = "{Revealing Limitation in the Standard Cosmological Model: A Redshift-dependent Hubble Constant from Fast Radio Bursts}",
    eprint = "2506.14947",
    archivePrefix = "arXiv",
    primaryClass = "astro-ph.CO",
    doi = "10.3847/1538-4357/ae261b",
    journal = "Astrophys. J.",
    volume = "996",
    number = "1",
    pages = "50",
    year = "2026"
}

@article{Parker:1968mv,
    author = "Parker, L.",
    title = "{Particle creation in expanding universes}",
    doi = "10.1103/PhysRevLett.21.562",
    journal = "Phys. Rev. Lett.",
    volume = "21",
    pages = "562--564",
    year = "1968"
}

@article{Parker:1969au,
    author = "Parker, Leonard",
    title = "{Quantized fields and particle creation in expanding universes. 1.}",
    doi = "10.1103/PhysRev.183.1057",
    journal = "Phys. Rev.",
    volume = "183",
    pages = "1057--1068",
    year = "1969"
}

@article{Dai:2026pvx,
    author = "Dai, Xinyi and Yang, Yupeng and Wang, Yicheng and Qu, Yankun and Yi, Shuangxi and Wang, Fayin",
    title = "{Redshift evolution of the Hubble constant: Constraints and new insights from an interacting dark energy model}",
    eprint = "2602.22840",
    archivePrefix = "arXiv",
    primaryClass = "astro-ph.CO",
    doi = "10.1103/zg3l-yt32",
    journal = "Phys. Rev. D",
    volume = "113",
    number = "6",
    pages = "063514",
    year = "2026"
}

@article{Zeldovich:1971mw,
    author = "Zel'dovich, Ya. B. and Starobinsky, Alexei A.",
    title = "{Particle Production and Vacuum Polarization in an Anisotropic Gravitational Field}",
    journal = "Sov. Phys. JETP",
    volume = "34",
    number = "6",
    pages = "1159--1166",
    year = "1972"
}

@article{Turner:1997npq,
    author = "Turner, Michael S. and White, Martin J.",
    title = "{CDM models with a smooth component}",
    eprint = "astro-ph/9701138",
    archivePrefix = "arXiv",
    reportNumber = "FERMILAB-PUB-97-002-A",
    doi = "10.1103/PhysRevD.56.R4439",
    journal = "Phys. Rev. D",
    volume = "56",
    number = "8",
    pages = "R4439",
    year = "1997"
}

@article{Li:2025owk,
    author = "Li, Tian-Nuo and Du, Guo-Hong and Li, Yun-He and Wu, Peng-Ju and Jin, Shang-Jie and Zhang, Jing-Fei and Zhang, Xin",
    title = "{Probing the sign-changeable interaction between dark energy and dark matter with DESI baryon acoustic oscillations and DES supernovae data}",
    eprint = "2501.07361",
    archivePrefix = "arXiv",
    primaryClass = "astro-ph.CO",
    doi = "10.1007/s11433-025-2771-5",
    journal = "Sci. China Phys. Mech. Astron.",
    volume = "69",
    number = "1",
    pages = "210413",
    year = "2026"
}

@article{Li:2026xaz,
    author = "Li, Tian-Nuo and Giar{\`e}, William and Du, Guo-Hong and Li, Yun-He and Di Valentino, Eleonora and Zhang, Jing-Fei and Zhang, Xin",
    title = "{Strong Evidence for Dark Sector Interactions}",
    eprint = "2601.07361",
    archivePrefix = "arXiv",
    primaryClass = "astro-ph.CO",
    journal = "e-print",
    month = "1",
    year = "2026"
}

@article{Zhang:2025dwu,
    author = "Zhang, Yi-Min and Li, Tian-Nuo and Du, Guo-Hong and Zhou, Sheng-Han and Gao, Li-Yang and Zhang, Jing-Fei and Zhang, Xin",
    title = "{Alleviating the $H_0$ tension through new interacting dark energy model in light of DESI DR2}",
    eprint = "2510.12627",
    archivePrefix = "arXiv",
    primaryClass = "astro-ph.CO",
    journal = "e-print",
    month = "10",
    year = "2025"
}

@article{Erdem:2024vsr,
    author = "Erdem, Recai",
    title = "{Gravitational Particle Production and the Hubble Tension}",
    eprint = "2402.16791",
    archivePrefix = "arXiv",
    primaryClass = "gr-qc",
    doi = "10.3390/universe10090338",
    journal = "Universe",
    volume = "10",
    number = "9",
    pages = "338",
    year = "2024"
}

@article{Erdem:2025xtr,
    author = "Erdem, Recai",
    title = "{Gravitational particle production, the cosmological tensions and fast radio bursts}",
    eprint = "2508.19770",
    archivePrefix = "arXiv",
    primaryClass = "gr-qc",
    journal = "e-print",
    month = "8",
    year = "2025"
}

@article{Trotta:2008qt,
    author = "Trotta, Roberto",
    title = "{Bayes in the sky: Bayesian inference and model selection in cosmology}",
    eprint = "0803.4089",
    archivePrefix = "arXiv",
    primaryClass = "astro-ph",
    doi = "10.1080/00107510802066753",
    journal = "Contemp. Phys.",
    volume = "49",
    pages = "71--104",
    year = "2008"
}

@article{Yang:2025uyv,
    author = "Yang, Weiqiang and Zhang, Sibo and Mena, Olga and Pan, Supriya and Di Valentino, Eleonora",
    title = "{Dark Energy Is Not That Into You: Variable Couplings after DESI DR2 BAO}",
    eprint = "2508.19109",
    archivePrefix = "arXiv",
    primaryClass = "astro-ph.CO",
    month = "8",
    journal = "e-print",
    year = "2025"
}

@article{Abdalla:2022yfr,
 archiveprefix = {arXiv},
 author = {Abdalla, Elcio and others},
 doi = {10.1016/j.jheap.2022.04.002},
 eprint = {2203.06142},
 journal = {JHEAp},
 pages = {49--211},
 primaryclass = {astro-ph.CO},
 reportnumber = {FERMILAB-CONF-22-192-SCD},
 title = {{Cosmology intertwined: A review of the particle physics, astrophysics, and cosmology associated with the cosmological tensions and anomalies}},
 volume = {34},
 year = {2022}
}

@article{ACT:2020gnv,
 archiveprefix = {arXiv},
 author = {Aiola, Simone and others},
 collaboration = {ACT},
 doi = {10.1088/1475-7516/2020/12/047},
 eprint = {2007.07288},
 journal = {JCAP},
 pages = {047},
 primaryclass = {astro-ph.CO},
 title = {{The Atacama Cosmology Telescope: DR4 Maps and Cosmological Parameters}},
 volume = {12},
 year = {2020}
}

@article{Bargiacchi:2023jse,
 archiveprefix = {arXiv},
 author = {Bargiacchi, G. and Dainotti, M. G. and Nagataki, S. and Capozziello, S.},
 doi = {10.1093/mnras/stad763},
 eprint = {2303.07076},
 journal = {Mon. Not. Roy. Astron. Soc.},
 number = {3},
 pages = {3909--3924},
 primaryclass = {astro-ph.CO},
 title = {{Gamma-ray bursts, quasars, baryonic acoustic oscillations, and supernovae Ia: new statistical insights and cosmological constraints}},
 volume = {521},
 year = {2023}
}

@article{Benisty:2024lmj,
 archiveprefix = {arXiv},
 author = {Benisty, David and Pan, Supriya and Staicova, Denitsa and Di Valentino, Eleonora and Nunes, Rafael C.},
 doi = {10.1051/0004-6361/202449883},
 eprint = {2403.00056},
 journal = {Astron. Astrophys.},
 pages = {A156},
 primaryclass = {astro-ph.CO},
 title = {{Late-time constraints on interacting dark energy: Analysis independent of H0, rd, and MB}},
 volume = {688},
 year = {2024}
}

@article{Bernui:2023byc,
 archiveprefix = {arXiv},
 author = {Bernui, Armando and Di Valentino, Eleonora and Giar\`e, William and Kumar, Suresh and Nunes, Rafael C.},
 doi = {10.1103/PhysRevD.107.103531},
 eprint = {2301.06097},
 journal = {Phys. Rev. D},
 number = {10},
 pages = {103531},
 primaryclass = {astro-ph.CO},
 title = {{Exploring the H0 tension and the evidence for dark sector interactions from 2D BAO measurements}},
 volume = {107},
 year = {2023}
}

@article{Beutler:2011hx,
 archiveprefix = {arXiv},
 author = {Beutler, Florian and Blake, Chris and Colless, Matthew and Jones, D. Heath and Staveley-Smith, Lister and Campbell, Lachlan and Parker, Quentin and Saunders, Will and Watson, Fred},
 doi = {10.1111/j.1365-2966.2011.19250.x},
 eprint = {1106.3366},
 journal = {Mon. Not. Roy. Astron. Soc.},
 pages = {3017--3032},
 primaryclass = {astro-ph.CO},
 title = {{The 6dF Galaxy Survey: Baryon Acoustic Oscillations and the Local Hubble Constant}},
 volume = {416},
 year = {2011}
}

@article{BOSS:2016wmc,
 archiveprefix = {arXiv},
 author = {Alam, Shadab and others},
 collaboration = {BOSS},
 doi = {10.1093/mnras/stx721},
 eprint = {1607.03155},
 journal = {Mon. Not. Roy. Astron. Soc.},
 number = {3},
 pages = {2617--2652},
 primaryclass = {astro-ph.CO},
 title = {{The clustering of galaxies in the completed SDSS-III Baryon Oscillation Spectroscopic Survey: cosmological analysis of the DR12 galaxy sample}},
 volume = {470},
 year = {2017}
}

@article{Brout:2022vxf,
 archiveprefix = {arXiv},
 author = {Brout, Dillon and others},
 doi = {10.3847/1538-4357/ac8e04},
 eprint = {2202.04077},
 journal = {Astrophys. J.},
 number = {2},
 pages = {110},
 primaryclass = {astro-ph.CO},
 title = {{The Pantheon+ Analysis: Cosmological Constraints}},
 volume = {938},
 year = {2022}
}

@article{Castello:2023zjr,
 archiveprefix = {arXiv},
 author = {Castello, Sveva and Mancarella, Michele and Grimm, Nastassia and Sobral-Blanco, Daniel and Tutusaus, Isaac and Bonvin, Camille},
 doi = {10.1088/1475-7516/2024/05/003},
 eprint = {2311.14425},
 journal = {JCAP},
 pages = {003},
 primaryclass = {astro-ph.CO},
 title = {{Gravitational redshift constraints on the effective theory of interacting dark energy}},
 volume = {05},
 year = {2024}
}

@article{Chevallier:2000qy,
 archiveprefix = {arXiv},
 author = {Chevallier, Michel and Polarski, David},
 doi = {10.1142/S0218271801000822},
 eprint = {gr-qc/0009008},
 journal = {Int. J. Mod. Phys. D},
 pages = {213--224},
 title = {{Accelerating universes with scaling dark matter}},
 volume = {10},
 year = {2001}
}

@article{Cipriano:2023yhv,
 archiveprefix = {arXiv},
 author = {Cipriano, Ricardo A. C. and Harko, Tiberiu and Lobo, Francisco S. N. and Pinto, Miguel A. S. and Rosa, Jo\~ao Lu\'\i{}s},
 doi = {10.1016/j.dark.2024.101463},
 eprint = {2310.15018},
 journal = {Phys. Dark Univ.},
 pages = {101463},
 primaryclass = {gr-qc},
 title = {{Gravitationally induced matter creation in scalar\textendash{}tensor f(R,T\ensuremath{\mu}\ensuremath{\nu}T\ensuremath{\mu}\ensuremath{\nu}) gravity}},
 volume = {44},
 year = {2024}
}

@article{Dainotti:2023bwq,
 archiveprefix = {arXiv},
 author = {Dainotti, Maria Giovanna and Bargiacchi, Giada and Bogdan, Ma\l{}gorzata and Lenart, Aleksander \L{}ukasz and Iwasaki, Kazunari and Capozziello, Salvatore and Zhang, Bing and Fraija, Nissim},
 doi = {10.3847/1538-4357/acd63f},
 eprint = {2305.10030},
 journal = {Astrophys. J.},
 number = {1},
 pages = {63},
 primaryclass = {astro-ph.CO},
 title = {{Reducing the Uncertainty on the Hubble Constant up to 35\% with an Improved Statistical Analysis: Different Best-fit Likelihoods for Type Ia Supernovae, Baryon Acoustic Oscillations, Quasars, and Gamma-Ray Bursts}},
 volume = {951},
 year = {2023}
}

@article{Dainotti:2024aha,
 archiveprefix = {arXiv},
 author = {Dainotti, Maria Giovanna and Lenart, Aleksander Lukasz and Yengejeh, Mina Godsi and Chakraborty, Satyajit and Fraija, Nissim and Di Valentino, Eleonora and Montani, Giovanni},
 doi = {10.1016/j.dark.2024.101428},
 eprint = {2401.12847},
 journal = {Phys. Dark Univ.},
 pages = {101428},
 primaryclass = {astro-ph.HE},
 title = {{A new binning method to choose a standard set of Quasars}},
 volume = {44},
 year = {2024}
}

@article{DES:2021wwk,
 archiveprefix = {arXiv},
 author = {Abbott, T. M. C. and others},
 collaboration = {DES},
 doi = {10.1103/PhysRevD.105.023520},
 eprint = {2105.13549},
 journal = {Phys. Rev. D},
 number = {2},
 pages = {023520},
 primaryclass = {astro-ph.CO},
 reportnumber = {FERMILAB-PUB-21-221-AE, DES-2020-0617},
 title = {{Dark Energy Survey Year 3 results: Cosmological constraints from galaxy clustering and weak lensing}},
 volume = {105},
 year = {2022}
}

@article{DiValentino:2017iww,
 archiveprefix = {arXiv},
 author = {Di Valentino, Eleonora and Melchiorri, Alessandro and Mena, Olga},
 doi = {10.1103/PhysRevD.96.043503},
 eprint = {1704.08342},
 journal = {Phys. Rev. D},
 number = {4},
 pages = {043503},
 primaryclass = {astro-ph.CO},
 title = {{Can interacting dark energy solve the $H_0$ tension?}},
 volume = {96},
 year = {2017}
}

@article{DiValentino:2019ffd,
 archiveprefix = {arXiv},
 author = {Di Valentino, Eleonora and Melchiorri, Alessandro and Mena, Olga and Vagnozzi, Sunny},
 doi = {10.1016/j.dark.2020.100666},
 eprint = {1908.04281},
 journal = {Phys. Dark Univ.},
 pages = {100666},
 primaryclass = {astro-ph.CO},
 title = {{Interacting dark energy in the early 2020s: A promising solution to the $H_0$ and cosmic shear tensions}},
 volume = {30},
 year = {2020}
}

@article{DiValentino:2020leo,
 archiveprefix = {arXiv},
 author = {Di Valentino, Eleonora and Mena, Olga},
 doi = {10.1093/mnrasl/slaa175},
 eprint = {2009.12620},
 journal = {Mon. Not. Roy. Astron. Soc.},
 number = {1},
 pages = {L22--L26},
 primaryclass = {astro-ph.CO},
 title = {{A fake Interacting Dark Energy detection?}},
 volume = {500},
 year = {2020}
}

@article{DiValentino:2020vnx,
 archiveprefix = {arXiv},
 author = {Di Valentino, Eleonora},
 doi = {10.1093/mnras/stab187},
 eprint = {2011.00246},
 journal = {Mon. Not. Roy. Astron. Soc.},
 number = {2},
 pages = {2065--2073},
 primaryclass = {astro-ph.CO},
 reportnumber = {IPPP/20/72},
 title = {{A combined analysis of the $H_0$ late time direct measurements and the impact on the Dark Energy sector}},
 volume = {502},
 year = {2021}
}

@article{DiValentino:2020zio,
 archiveprefix = {arXiv},
 author = {Di Valentino, Eleonora and others},
 doi = {10.1016/j.astropartphys.2021.102605},
 eprint = {2008.11284},
 journal = {Astropart. Phys.},
 pages = {102605},
 primaryclass = {astro-ph.CO},
 reportnumber = {FERMILAB-PUB-21-590-PPD},
 title = {{Snowmass2021 - Letter of interest cosmology intertwined II: The hubble constant tension}},
 volume = {131},
 year = {2021}
}

@article{DiValentino:2021izs,
 archiveprefix = {arXiv},
 author = {Di Valentino, Eleonora and Mena, Olga and Pan, Supriya and Visinelli, Luca and Yang, Weiqiang and Melchiorri, Alessandro and Mota, David F. and Riess, Adam G. and Silk, Joseph},
 doi = {10.1088/1361-6382/ac086d},
 eprint = {2103.01183},
 journal = {Class. Quant. Grav.},
 number = {15},
 pages = {153001},
 primaryclass = {astro-ph.CO},
 reportnumber = {IPPP/20/108},
 title = {{In the realm of the Hubble tension\textemdash{}a review of solutions}},
 volume = {38},
 year = {2021}
}

@article{DiValentino:2022fjm,
 author = {Di Valentino, Eleonora},
 doi = {10.3390/universe8080399},
 journal = {Universe},
 number = {8},
 pages = {399},
 title = {{Challenges of the Standard Cosmological Model}},
 volume = {8},
 year = {2022}
}

@book{DiValentino:2024yew,
 doi = {10.1007/978-981-99-0177-7},
 editor = {Di Valentino, Eleonora and Brout, Dillon},
 isbn = {978-981-99-0176-0, 978-981-99-0179-1, 978-981-99-0177-7},
 publisher = {Springer},
 series = {Springer Series in Astrophysics and Cosmology},
 title = {{The Hubble Constant Tension}},
 year = {2024}
}

@article{eBOSS:2020yzd,
 archiveprefix = {arXiv},
 author = {Alam, Shadab and others},
 collaboration = {eBOSS},
 doi = {10.1103/PhysRevD.103.083533},
 eprint = {2007.08991},
 journal = {Phys. Rev. D},
 number = {8},
 pages = {083533},
 primaryclass = {astro-ph.CO},
 title = {{Completed SDSS-IV extended Baryon Oscillation Spectroscopic Survey: Cosmological implications from two decades of spectroscopic surveys at the Apache Point Observatory}},
 volume = {103},
 year = {2021}
}

@article{Escamilla:2023shf,
 archiveprefix = {arXiv},
 author = {Escamilla, Luis A. and Akarsu, Ozgur and Di Valentino, Eleonora and Vazquez, J. Alberto},
 doi = {10.1088/1475-7516/2023/11/051},
 eprint = {2305.16290},
 journal = {JCAP},
 pages = {051},
 primaryclass = {astro-ph.CO},
 title = {{Model-independent reconstruction of the interacting dark energy kernel: Binned and Gaussian process}},
 volume = {11},
 year = {2023}
}

@article{Escamilla:2024xmz,
 archiveprefix = {arXiv},
 author = {Escamilla, Luis A. and Fiorucci, Donatella and Montani, Giovanni and Di Valentino, Eleonora},
 doi = {10.1016/j.dark.2024.101652},
 eprint = {2408.04354},
 journal = {Phys. Dark Univ.},
 pages = {101652},
 primaryclass = {astro-ph.CO},
 title = {{Exploring the Hubble tension with a late time Modified Gravity scenario}},
 volume = {46},
 year = {2024}
}

@article{Forconi:2023hsj,
 archiveprefix = {arXiv},
 author = {Forconi, Matteo and Giar\`e, William and Mena, Olga and Ruchika and Di Valentino, Eleonora and Melchiorri, Alessandro and Nunes, Rafael C.},
 doi = {10.1088/1475-7516/2024/05/097},
 eprint = {2312.11074},
 journal = {JCAP},
 pages = {097},
 primaryclass = {astro-ph.CO},
 title = {{A double take on early and interacting dark energy from JWST}},
 volume = {05},
 year = {2024}
}

@article{Zlatev:1998tr,
    author = "Zlatev, Ivaylo and Wang, Li-Min and Steinhardt, Paul J.",
    title = "{Quintessence, cosmic coincidence, and the cosmological constant}",
    eprint = "astro-ph/9807002",
    archivePrefix = "arXiv",
    doi = "10.1103/PhysRevLett.82.896",
    journal = "Phys. Rev. Lett.",
    volume = "82",
    pages = "896--899",
    year = "1999"
}

@article{DESI:2025qqy,
    author = "Andrade, U. and others",
    collaboration = "DESI",
    title = "{Validation of the DESI DR2 measurements of baryon acoustic oscillations from galaxies and quasars}",
    eprint = "2503.14742",
    archivePrefix = "arXiv",
    primaryClass = "astro-ph.CO",
    reportNumber = "FERMILAB-PUB-25-0162-PPD",
    doi = "10.1103/kdys-w8vl",
    journal = "Phys. Rev. D",
    volume = "112",
    number = "8",
    pages = "083512",
    year = "2025"
}

@article{CosmoVerseNetwork:2025alb,
    author = "Di Valentino, Eleonora and others",
    collaboration = "CosmoVerse Network",
    title = "{The CosmoVerse White Paper: Addressing observational tensions in cosmology with systematics and fundamental physics}",
    eprint = "2504.01669",
    archivePrefix = "arXiv",
    primaryClass = "astro-ph.CO",
    doi = "10.1016/j.dark.2025.101965",
    journal = "Phys. Dark Univ.",
    volume = "49",
    pages = "101965",
    year = "2025"
}

@article{Gao:2021xnk,
 archiveprefix = {arXiv},
 author = {Gao, Li-Yang and Zhao, Ze-Wei and Xue, She-Sheng and Zhang, Xin},
 doi = {10.1088/1475-7516/2021/07/005},
 eprint = {2101.10714},
 journal = {JCAP},
 pages = {005},
 primaryclass = {astro-ph.CO},
 title = {{Relieving the H 0 tension with a new interacting dark energy model}},
 volume = {07},
 year = {2021}
}

@article{Gao:2022ahg,
 archiveprefix = {arXiv},
 author = {Gao, Li-Yang and Xue, She-Sheng and Zhang, Xin},
 doi = {10.1088/1674-1137/ad2b52},
 eprint = {2212.13146},
 journal = {Chin. Phys. C},
 number = {5},
 pages = {051001},
 primaryclass = {astro-ph.CO},
 title = {{Dark energy and matter interacting scenario to relieve H $_{0}$ and S $_{8}$ tensions*}},
 volume = {48},
 year = {2024}
}

@misc{Giare:2023xoc,
 archiveprefix = {arXiv},
 author = {Giar\`e, William},
 doi = {10.1007/978-981-99-0177-7_36},
 eprint = {2305.16919},
 month = {5},
 primaryclass = {astro-ph.CO},
 title = {{CMB Anomalies and the Hubble Tension}},
 year = {2023}
}

@article{Giare:2024ytc,
 archiveprefix = {arXiv},
 author = {Giar\`e, William and Zhai, Yuejia and Pan, Supriya and Di Valentino, Eleonora and Nunes, Rafael C. and van de Bruck, Carsten},
 doi = {10.1103/PhysRevD.110.063527},
 eprint = {2404.02110},
 journal = {Phys. Rev. D},
 number = {6},
 pages = {063527},
 primaryclass = {astro-ph.CO},
 title = {{Tightening the reins on nonminimal dark sector physics: Interacting dark energy with dynamical and nondynamical equation of state}},
 volume = {110},
 year = {2024}
}

@article{Gohar:2020bod,
 archiveprefix = {arXiv},
 author = {Gohar, Hussain and Salzano, Vincenzo},
 doi = {10.1140/epjc/s10052-021-09086-9},
 eprint = {2008.09635},
 journal = {Eur. Phys. J. C},
 number = {4},
 pages = {338},
 primaryclass = {gr-qc},
 title = {{Cosmological Constraints on Entropic Cosmology with Matter Creation}},
 volume = {81},
 year = {2021}
}

@article{Halder:2024uao,
 archiveprefix = {arXiv},
 author = {Halder, Sudip and de Haro, Jaume and Saha, Tapan and Pan, Supriya},
 doi = {10.1103/PhysRevD.109.083522},
 eprint = {2403.01397},
 journal = {Phys. Rev. D},
 number = {8},
 pages = {083522},
 primaryclass = {gr-qc},
 title = {{Phase space analysis of sign-shifting interacting dark energy models}},
 volume = {109},
 year = {2024}
}

@article{Hoerning:2023hks,
    author = "Hoerning, Gabriel A. and Landim, Ricardo G. and Ponte, Luiza O. and Rolim, Raphael P. and Abdalla, Filipe B. and Abdalla, Elcio",
    title = "{Constraints on interacting dark energy revisited: Implications for the Hubble tension}",
    eprint = "2308.05807",
    archivePrefix = "arXiv",
    primaryClass = "astro-ph.CO",
    doi = "10.1103/6zrh-8fmv",
    journal = "Phys. Rev. D",
    volume = "112",
    number = "2",
    pages = "023523",
    year = "2025"
}

@article{Hu:2023jqc,
 archiveprefix = {arXiv},
 author = {Hu, Jian-Ping and Wang, Fa-Yin},
 doi = {10.3390/universe9020094},
 eprint = {2302.05709},
 journal = {Universe},
 number = {2},
 pages = {94},
 primaryclass = {astro-ph.CO},
 title = {{Hubble Tension: The Evidence of New Physics}},
 volume = {9},
 year = {2023}
}

@article{vanderWesthuizen:2025rip,
    author = "van der Westhuizen, Marcel and Abebe, Amare and Di Valentino, Eleonora",
    title = "{III. Interacting Dark Energy: Summary of models, Pathologies, and Constraints}",
    eprint = "2509.04496",
    archivePrefix = "arXiv",
    primaryClass = "gr-qc",
    doi = "10.1016/j.dark.2025.102121",
    journal = "Phys. Dark Univ.",
    volume = "50",
    pages = "102121",
    year = "2025"
}

@article{AtacamaCosmologyTelescope:2025blo,
    author = "Louis, Thibaut and others",
    collaboration = "Atacama Cosmology Telescope",
    title = "{The Atacama Cosmology Telescope: DR6 power spectra, likelihoods and {\ensuremath{\Lambda}}CDM parameters}",
    eprint = "2503.14452",
    archivePrefix = "arXiv",
    primaryClass = "astro-ph.CO",
    reportNumber = "FERMILAB-PUB-25-0071-PPD",
    doi = "10.1088/1475-7516/2025/11/062",
    journal = "JCAP",
    volume = "11",
    pages = "062",
    year = "2025"
}

@misc{SPT-3G:2025bzu,
    author = "Camphuis, E. and others",
    collaboration = "SPT-3G",
    title = "{SPT-3G D1: CMB temperature and polarization power spectra and cosmology from 2019 and 2020 observations of the SPT-3G Main field}",
    eprint = "2506.20707",
    archivePrefix = "arXiv",
    primaryClass = "astro-ph.CO",
    reportNumber = "FERMILAB-PUB-25-0144-PPD",
    month = "6",
    year = "2025"
}

@misc{H0DN:2025lyy,
    author = "Casertano, Stefano and others",
    collaboration = "H0DN",
    title = "{The Local Distance Network: a community consensus report on the measurement of the Hubble constant at 1{\%} precision}",
    eprint = "2510.23823",
    archivePrefix = "arXiv",
    primaryClass = "astro-ph.CO",
    month = "10",
    year = "2025"
}

@misc{DESI:2025wyn,
    author = "Gu, Gan and others",
    collaboration = "DESI",
    title = "{Dynamical Dark Energy in light of the DESI DR2 Baryonic Acoustic Oscillations Measurements}",
    eprint = "2504.06118",
    archivePrefix = "arXiv",
    primaryClass = "astro-ph.CO",
    reportNumber = "FERMILAB-PUB-25-0235-PPD",
    doi = "10.1038/s41550-025-02669-6",
    month = "4",
    year = "2025"
}

@article{DES:2024jxu,
    author = "Abbott, T. M. C. and others",
    collaboration = "DES",
    title = "{The Dark Energy Survey: Cosmology Results with \ensuremath{\sim}1500 New High-redshift Type Ia Supernovae Using the Full 5 yr Data Set}",
    eprint = "2401.02929",
    archivePrefix = "arXiv",
    primaryClass = "astro-ph.CO",
    reportNumber = "FERMILAB-PUB-23-0821-PPD, DES-2023-805",
    doi = "10.3847/2041-8213/ad6f9f",
    journal = "Astrophys. J. Lett.",
    volume = "973",
    number = "1",
    pages = "L14",
    year = "2024"
}

@article{Wright:2025xka,
    author = "Wright, Angus H. and others",
    title = "{KiDS-Legacy: Cosmological constraints from cosmic shear with the complete Kilo-Degree Survey}",
    eprint = "2503.19441",
    archivePrefix = "arXiv",
    primaryClass = "astro-ph.CO",
    doi = "10.1051/0004-6361/202554908",
    journal = "Astron. Astrophys.",
    volume = "703",
    pages = "A158",
    year = "2025"
}

@article{Silva:2025hxw,
    author = "Silva, Emanuelly and Sabogal, Miguel A. and Scherer, Mateus and Nunes, Rafael C. and Di Valentino, Eleonora and Kumar, Suresh",
    title = "{New constraints on interacting dark energy from DESI DR2 BAO observations}",
    eprint = "2503.23225",
    archivePrefix = "arXiv",
    primaryClass = "astro-ph.CO",
    doi = "10.1103/qqc6-76z4",
    journal = "Phys. Rev. D",
    volume = "111",
    number = "12",
    pages = "123511",
    year = "2025"
}

@article{Kamionkowski:2022pkx,
 archiveprefix = {arXiv},
 author = {Kamionkowski, Marc and Riess, Adam G.},
 doi = {10.1146/annurev-nucl-111422-024107},
 eprint = {2211.04492},
 journal = {Ann. Rev. Nucl. Part. Sci.},
 pages = {153--180},
 primaryclass = {astro-ph.CO},
 title = {{The Hubble Tension and Early Dark Energy}},
 volume = {73},
 year = {2023}
}

@article{Kumar:2016zpg,
 archiveprefix = {arXiv},
 author = {Kumar, Suresh and Nunes, Rafael C.},
 doi = {10.1103/PhysRevD.94.123511},
 eprint = {1608.02454},
 journal = {Phys. Rev. D},
 number = {12},
 pages = {123511},
 primaryclass = {astro-ph.CO},
 title = {{Probing the interaction between dark matter and dark energy in the presence of massive neutrinos}},
 volume = {94},
 year = {2016}
}

@article{Kumar:2017dnp,
 archiveprefix = {arXiv},
 author = {Kumar, Suresh and Nunes, Rafael C.},
 doi = {10.1103/PhysRevD.96.103511},
 eprint = {1702.02143},
 journal = {Phys. Rev. D},
 number = {10},
 pages = {103511},
 primaryclass = {astro-ph.CO},
 title = {{Echo of interactions in the dark sector}},
 volume = {96},
 year = {2017}
}

@article{Kumar:2021eev,
 archiveprefix = {arXiv},
 author = {Kumar, Suresh},
 doi = {10.1016/j.dark.2021.100862},
 eprint = {2102.12902},
 journal = {Phys. Dark Univ.},
 pages = {100862},
 primaryclass = {astro-ph.CO},
 title = {{Remedy of some cosmological tensions via effective phantom-like behavior of interacting vacuum energy}},
 volume = {33},
 year = {2021}
}

@article{Lewis:2019xzd,
    author = "Lewis, Antony",
    title = "{GetDist: a Python package for analysing Monte Carlo samples}",
    eprint = "1910.13970",
    archivePrefix = "arXiv",
    primaryClass = "astro-ph.IM",
    doi = "10.1088/1475-7516/2025/08/025",
    journal = "JCAP",
    volume = "08",
    pages = "025",
    year = "2025"
}

@article{Li:2024qso,
    author = "Li, Tian-Nuo and Wu, Peng-Ju and Du, Guo-Hong and Jin, Shang-Jie and Li, Hai-Li and Zhang, Jing-Fei and Zhang, Xin",
    title = "{Constraints on Interacting Dark Energy Models from the DESI Baryon Acoustic Oscillation and DES Supernovae Data}",
    eprint = "2407.14934",
    archivePrefix = "arXiv",
    primaryClass = "astro-ph.CO",
    doi = "10.3847/1538-4357/ad87f0",
    journal = "Astrophys. J.",
    volume = "976",
    number = "1",
    pages = "1",
    year = "2024"
}

@article{Linder:2002et,
 archiveprefix = {arXiv},
 author = {Linder, Eric V.},
 doi = {10.1103/PhysRevLett.90.091301},
 eprint = {astro-ph/0208512},
 journal = {Phys. Rev. Lett.},
 pages = {091301},
 title = {{Exploring the expansion history of the universe}},
 volume = {90},
 year = {2003}
}

@article{Liu:2024vlt,
 archiveprefix = {arXiv},
 author = {Liu, Yang and Yu, Hongwei and Wu, Puxun},
 doi = {10.1103/PhysRevD.110.L021304},
 eprint = {2406.02956},
 journal = {Phys. Rev. D},
 number = {2},
 pages = {L021304},
 primaryclass = {astro-ph.CO},
 title = {{Alleviating the Hubble-constant tension and the growth tension via a transition of absolute magnitude favored by the Pantheon+ sample}},
 volume = {110},
 year = {2024}
}

@article{Lucca:2020zjb,
 archiveprefix = {arXiv},
 author = {Lucca, Matteo and Hooper, Deanna C.},
 doi = {10.1103/PhysRevD.102.123502},
 eprint = {2002.06127},
 journal = {Phys. Rev. D},
 number = {12},
 pages = {123502},
 primaryclass = {astro-ph.CO},
 reportnumber = {ULB-TH/20-01},
 title = {{Shedding light on dark matter-dark energy interactions}},
 volume = {102},
 year = {2020}
}

@article{Mishra:2023ueo,
 archiveprefix = {arXiv},
 author = {Mishra, Keshav Ram and Pacif, Shibesh Kumar Jas and Kumar, Rajesh and Bamba, Kazuharu},
 doi = {10.1016/j.dark.2023.101211},
 eprint = {2301.08743},
 journal = {Phys. Dark Univ.},
 pages = {101211},
 primaryclass = {gr-qc},
 title = {{Cosmological implications of an interacting model of dark matter \& dark energy}},
 volume = {40},
 year = {2023}
}

@article{Montani:2023xpd,
 archiveprefix = {arXiv},
 author = {Montani, Giovanni and De Angelis, Mariaveronica and Bombacigno, Flavio and Carlevaro, Nakia},
 doi = {10.1093/mnrasl/slad159},
 eprint = {2306.11101},
 journal = {Mon. Not. Roy. Astron. Soc.},
 number = {1},
 pages = {L156--L161},
 primaryclass = {gr-qc},
 title = {{Metric f(R) gravity with dynamical dark energy as a scenario for the Hubble tension}},
 volume = {527},
 year = {2023}
}

@article{Montani:2023ywn,
 archiveprefix = {arXiv},
 author = {Montani, Giovanni and Carlevaro, Nakia and Dainotti, Maria Giovanna},
 doi = {10.1016/j.dark.2024.101486},
 eprint = {2311.04822},
 journal = {Phys. Dark Univ.},
 pages = {101486},
 primaryclass = {gr-qc},
 title = {{Slow-rolling scalar dynamics as solution for the Hubble tension}},
 volume = {44},
 year = {2024}
}

@article{Moresco:2016mzx,
 archiveprefix = {arXiv},
 author = {Moresco, Michele and Pozzetti, Lucia and Cimatti, Andrea and Jimenez, Raul and Maraston, Claudia and Verde, Licia and Thomas, Daniel and Citro, Annalisa and Tojeiro, Rita and Wilkinson, David},
 doi = {10.1088/1475-7516/2016/05/014},
 eprint = {1601.01701},
 journal = {JCAP},
 pages = {014},
 primaryclass = {astro-ph.CO},
 title = {{A 6\% measurement of the Hubble parameter at $z\sim0.45$: direct evidence of the epoch of cosmic re-acceleration}},
 volume = {05},
 year = {2016}
}

@article{Moresco:2020fbm,
 archiveprefix = {arXiv},
 author = {Moresco, Michele and Jimenez, Raul and Verde, Licia and Cimatti, Andrea and Pozzetti, Lucia},
 doi = {10.3847/1538-4357/ab9eb0},
 eprint = {2003.07362},
 journal = {Astrophys. J.},
 number = {1},
 pages = {82},
 primaryclass = {astro-ph.GA},
 title = {{Setting the Stage for Cosmic Chronometers. II. Impact of Stellar Population Synthesis Models Systematics and Full Covariance Matrix}},
 volume = {898},
 year = {2020}
}

@misc{Moresco:2023zys,
 archiveprefix = {arXiv},
 author = {Moresco, Michele},
 eprint = {2307.09501},
 month = {7},
 primaryclass = {astro-ph.CO},
 title = {{Addressing the Hubble tension with cosmic chronometers}},
 year = {2023}
}

@article{Murgia:2016ccp,
 archiveprefix = {arXiv},
 author = {Murgia, Riccardo and Gariazzo, Stefano and Fornengo, Nicolao},
 doi = {10.1088/1475-7516/2016/04/014},
 eprint = {1602.01765},
 journal = {JCAP},
 pages = {014},
 primaryclass = {astro-ph.CO},
 title = {{Constraints on the Coupling between Dark Energy and Dark Matter from CMB data}},
 volume = {04},
 year = {2016}
}

@article{Nojiri:2022ski,
 archiveprefix = {arXiv},
 author = {Nojiri, S. and Odintsov, S. D. and Oikonomou, V. K.},
 doi = {10.1016/j.nuclphysb.2022.115850},
 eprint = {2205.11681},
 journal = {Nucl. Phys. B},
 pages = {115850},
 primaryclass = {gr-qc},
 title = {{Integral F(R) gravity and saddle point condition as a remedy for the H0-tension}},
 volume = {980},
 year = {2022}
}

@article{Nunes:2016dlj,
 archiveprefix = {arXiv},
 author = {Nunes, Rafael C. and Pan, Supriya and Saridakis, Emmanuel N.},
 doi = {10.1103/PhysRevD.94.023508},
 eprint = {1605.01712},
 journal = {Phys. Rev. D},
 number = {2},
 pages = {023508},
 primaryclass = {astro-ph.CO},
 reportnumber = {PHYS.REV.-D94-(2016), 023508},
 title = {{New constraints on interacting dark energy from cosmic chronometers}},
 volume = {94},
 year = {2016}
}

@article{Nunes:2021zzi,
 archiveprefix = {arXiv},
 author = {Nunes, Rafael C. and Di Valentino, Eleonora},
 doi = {10.1103/PhysRevD.104.063529},
 eprint = {2107.09151},
 journal = {Phys. Rev. D},
 number = {6},
 pages = {063529},
 primaryclass = {astro-ph.CO},
 title = {{Dark sector interaction and the supernova absolute magnitude tension}},
 volume = {104},
 year = {2021}
}

@article{Odintsov:2020qzd,
 archiveprefix = {arXiv},
 author = {Odintsov, Sergei D. and S\'aez-Chill\'on G\'omez, Diego and Sharov, German S.},
 doi = {10.1016/j.nuclphysb.2021.115377},
 eprint = {2011.03957},
 journal = {Nucl. Phys. B},
 pages = {115377},
 primaryclass = {gr-qc},
 title = {{Analyzing the $H_0$ tension in $F(R)$ gravity models}},
 volume = {966},
 year = {2021}
}

@misc{Pan:2023mie,
 archiveprefix = {arXiv},
 author = {Pan, Supriya and Yang, Weiqiang},
 doi = {10.1007/978-981-99-0177-7_29},
 eprint = {2310.07260},
 month = {10},
 primaryclass = {astro-ph.CO},
 title = {{On the interacting dark energy scenarios - the case for Hubble constant tension}},
 year = {2023}
}

@article{Perivolaropoulos:2021jda,
 archiveprefix = {arXiv},
 author = {Perivolaropoulos, Leandros and Skara, Foteini},
 doi = {10.1016/j.newar.2022.101659},
 eprint = {2105.05208},
 journal = {New Astron. Rev.},
 pages = {101659},
 primaryclass = {astro-ph.CO},
 title = {{Challenges for \ensuremath{\Lambda}CDM: An update}},
 volume = {95},
 year = {2022}
}

@article{Perivolaropoulos:2024yxv,
    author = "Perivolaropoulos, Leandros",
    title = "{Hubble tension or distance ladder crisis?}",
    eprint = "2408.11031",
    archivePrefix = "arXiv",
    primaryClass = "astro-ph.CO",
    doi = "10.1103/PhysRevD.110.123518",
    journal = "Phys. Rev. D",
    volume = "110",
    number = "12",
    pages = "123518",
    year = "2024"
}

@article{Pinto:2022tlu,
 archiveprefix = {arXiv},
 author = {Pinto, Miguel A. S. and Harko, Tiberiu and Lobo, Francisco S. N.},
 doi = {10.1103/PhysRevD.106.044043},
 eprint = {2205.12545},
 journal = {Phys. Rev. D},
 number = {4},
 pages = {044043},
 primaryclass = {gr-qc},
 title = {{Gravitationally induced particle production in scalar-tensor f(R,T) gravity}},
 volume = {106},
 year = {2022}
}

@article{Planck:2016tof,
 archiveprefix = {arXiv},
 author = {Aghanim, N. and others},
 collaboration = {Planck},
 doi = {10.1051/0004-6361/201629504},
 eprint = {1608.02487},
 journal = {Astron. Astrophys.},
 pages = {A95},
 primaryclass = {astro-ph.CO},
 title = {{Planck intermediate results. LI. Features in the cosmic microwave background temperature power spectrum and shifts in cosmological parameters}},
 volume = {607},
 year = {2017}
}

@article{Planck:2018nkj,
 archiveprefix = {arXiv},
 author = {Aghanim, N. and others},
 collaboration = {Planck},
 doi = {10.1051/0004-6361/201833880},
 eprint = {1807.06205},
 journal = {Astron. Astrophys.},
 pages = {A1},
 primaryclass = {astro-ph.CO},
 title = {{Planck 2018 results. I. Overview and the cosmological legacy of Planck}},
 volume = {641},
 year = {2020}
}

@article{Planck:2018vyg,
 archiveprefix = {arXiv},
 author = {Aghanim, N. and others},
 collaboration = {Planck},
 doi = {10.1051/0004-6361/201833910},
 eprint = {1807.06209},
 journal = {Astron. Astrophys.},
 note = {[Erratum: Astron.Astrophys. 652, C4 (2021)]},
 pages = {A6},
 primaryclass = {astro-ph.CO},
 title = {{Planck 2018 results. VI. Cosmological parameters}},
 volume = {641},
 year = {2020}
}

@article{Pooya:2024wsq,
 archiveprefix = {arXiv},
 author = {Pooya, N. Nazari},
 doi = {10.1103/PhysRevD.110.043510},
 eprint = {2407.03766},
 journal = {Phys. Rev. D},
 number = {4},
 pages = {043510},
 primaryclass = {astro-ph.CO},
 title = {{Growth of matter perturbations in the interacting dark energy-dark matter scenarios}},
 volume = {110},
 year = {2024}
}

@article{Pourtsidou:2016ico,
 archiveprefix = {arXiv},
 author = {Pourtsidou, Alkistis and Tram, Thomas},
 doi = {10.1103/PhysRevD.94.043518},
 eprint = {1604.04222},
 journal = {Phys. Rev. D},
 number = {4},
 pages = {043518},
 primaryclass = {astro-ph.CO},
 title = {{Reconciling CMB and structure growth measurements with dark energy interactions}},
 volume = {94},
 year = {2016}
}

@article{Prigogine:1989zz,
 author = {Prigogine, I. and Geheniau, J. and Gunzig, E. and Nardone, P.},
 doi = {10.1007/BF00758981},
 journal = {Gen. Rel. Grav.},
 pages = {767--776},
 title = {{Thermodynamics and cosmology}},
 volume = {21},
 year = {1989}
}

@article{Riess:2021jrx,
 archiveprefix = {arXiv},
 author = {Riess, Adam G. and others},
 doi = {10.3847/2041-8213/ac5c5b},
 eprint = {2112.04510},
 journal = {Astrophys. J. Lett.},
 number = {1},
 pages = {L7},
 primaryclass = {astro-ph.CO},
 title = {{A Comprehensive Measurement of the Local Value of the Hubble Constant with 1 km s$^{-1}$ Mpc$^{-1}$ Uncertainty from the Hubble Space Telescope and the SH0ES Team}},
 volume = {934},
 year = {2022}
}

@article{Ross:2014qpa,
 archiveprefix = {arXiv},
 author = {Ross, Ashley J. and Samushia, Lado and Howlett, Cullan and Percival, Will J. and Burden, Angela and Manera, Marc},
 doi = {10.1093/mnras/stv154},
 eprint = {1409.3242},
 journal = {Mon. Not. Roy. Astron. Soc.},
 number = {1},
 pages = {835--847},
 primaryclass = {astro-ph.CO},
 title = {{The clustering of the SDSS DR7 main Galaxy sample \textendash{} I. A 4 per cent distance measure at $z = 0.15$}},
 volume = {449},
 year = {2015}
}

@article{Schiavone:2022wvq,
 archiveprefix = {arXiv},
 author = {Schiavone, Tiziano and Montani, Giovanni and Bombacigno, Flavio},
 doi = {10.1093/mnrasl/slad041},
 eprint = {2211.16737},
 journal = {Mon. Not. Roy. Astron. Soc.},
 number = {1},
 pages = {L72--L77},
 primaryclass = {gr-qc},
 title = {{f(R) gravity in the Jordan frame as a paradigm for the Hubble tension}},
 volume = {522},
 year = {2023}
}

@article{Schoneberg:2021qvd,
 archiveprefix = {arXiv},
 author = {Sch\"oneberg, Nils and Franco Abell\'an, Guillermo and P\'erez S\'anchez, Andrea and Witte, Samuel J. and Poulin, Vivian and Lesgourgues, Julien},
 doi = {10.1016/j.physrep.2022.07.001},
 eprint = {2107.10291},
 journal = {Phys. Rept.},
 pages = {1--55},
 primaryclass = {astro-ph.CO},
 title = {{The H0 Olympics: A fair ranking of proposed models}},
 volume = {984},
 year = {2022}
}

@article{Schoneberg:2024ifp,
 archiveprefix = {arXiv},
 author = {Sch\"oneberg, Nils},
 doi = {10.1088/1475-7516/2024/06/006},
 eprint = {2401.15054},
 journal = {JCAP},
 pages = {006},
 primaryclass = {astro-ph.CO},
 title = {{The 2024 BBN baryon abundance update}},
 volume = {06},
 year = {2024}
}

@article{Scolnic:2021amr,
 archiveprefix = {arXiv},
 author = {Scolnic, Dan and others},
 doi = {10.3847/1538-4357/ac8b7a},
 eprint = {2112.03863},
 journal = {Astrophys. J.},
 number = {2},
 pages = {113},
 primaryclass = {astro-ph.CO},
 title = {{The Pantheon+ Analysis: The Full Data Set and Light-curve Release}},
 volume = {938},
 year = {2022}
}

@article{Shah:2021onj,
 archiveprefix = {arXiv},
 author = {Shah, Paul and Lemos, Pablo and Lahav, Ofer},
 doi = {10.1007/s00159-021-00137-4},
 eprint = {2109.01161},
 journal = {Astron. Astrophys. Rev.},
 number = {1},
 pages = {9},
 primaryclass = {astro-ph.CO},
 title = {{A buyer\textquoteright{}s guide to the Hubble constant}},
 volume = {29},
 year = {2021}
}

@article{Silva:2024ift,
    author = "Silva, Emanuelly and Z{\'u}{\~n}iga-Bola{\~n}o, Ubaldo and Nunes, Rafael C. and Di Valentino, Eleonora",
    title = "{Non-linear matter power spectrum modeling in interacting dark energy cosmologies}",
    eprint = "2403.19590",
    archivePrefix = "arXiv",
    primaryClass = "astro-ph.CO",
    doi = "10.1140/epjc/s10052-024-13487-x",
    journal = "Eur. Phys. J. C",
    volume = "84",
    number = "10",
    pages = "1104",
    year = "2024"
}

@article{Vagnozzi:2019ezj,
 archiveprefix = {arXiv},
 author = {Vagnozzi, Sunny},
 doi = {10.1103/PhysRevD.102.023518},
 eprint = {1907.07569},
 journal = {Phys. Rev. D},
 number = {2},
 pages = {023518},
 primaryclass = {astro-ph.CO},
 title = {{New physics in light of the $H_0$ tension: An alternative view}},
 volume = {102},
 year = {2020}
}

@article{vanderWesthuizen:2023hcl,
 archiveprefix = {arXiv},
 author = {van der Westhuizen, Marcel A. and Abebe, Amare},
 doi = {10.1088/1475-7516/2024/01/048},
 eprint = {2302.11949},
 journal = {JCAP},
 pages = {048},
 primaryclass = {gr-qc},
 title = {{Interacting dark energy: clarifying the cosmological implications and viability conditions}},
 volume = {01},
 year = {2024}
}

@article{Verde:2019ivm,
 archiveprefix = {arXiv},
 author = {Verde, L. and Treu, T. and Riess, A. G.},
 doi = {10.1038/s41550-019-0902-0},
 eprint = {1907.10625},
 journal = {Nature Astron.},
 pages = {891},
 primaryclass = {astro-ph.CO},
 title = {{Tensions between the Early and the Late Universe}},
 volume = {3},
 year = {2019}
}

@article{Verde:2023lmm,
    author = {Verde, Licia and Sch{\"o}neberg, Nils and Gil-Mar{\'\i}n, H{\'e}ctor},
    title = "{A Tale of Many H0}",
    eprint = "2311.13305",
    archivePrefix = "arXiv",
    primaryClass = "astro-ph.CO",
    doi = "10.1146/annurev-astro-052622-033813",
    journal = "Ann. Rev. Astron. Astrophys.",
    volume = "62",
    number = "1",
    pages = "287--331",
    year = "2024"
}

@article{vonMarttens:2019ixw,
 archiveprefix = {arXiv},
 author = {von Marttens, Rodrigo and Lombriser, Lucas and Kunz, Martin and Marra, Valerio and Casarini, Luciano and Alcaniz, Jailson},
 doi = {10.1016/j.dark.2020.100490},
 eprint = {1911.02618},
 journal = {Phys. Dark Univ.},
 pages = {100490},
 primaryclass = {astro-ph.CO},
 title = {{Dark degeneracy I: Dynamical or interacting dark energy?}},
 volume = {28},
 year = {2020}
}

@article{Weinberg:1988cp,
 author = {Weinberg, Steven},
 doi = {10.1103/RevModPhys.61.1},
 editor = {Hsu, Jong-Ping and Fine, D.},
 journal = {Rev. Mod. Phys.},
 pages = {1--23},
 reportnumber = {UTTG-12-88},
 title = {{The Cosmological Constant Problem}},
 volume = {61},
 year = {1989}
}

@article{WMAP:2012fli,
 archiveprefix = {arXiv},
 author = {Bennett, C. L. and others},
 collaboration = {WMAP},
 doi = {10.1088/0067-0049/208/2/20},
 eprint = {1212.5225},
 journal = {Astrophys. J. Suppl.},
 pages = {20},
 primaryclass = {astro-ph.CO},
 title = {{Nine-Year Wilkinson Microwave Anisotropy Probe (WMAP) Observations: Final Maps and Results}},
 volume = {208},
 year = {2013}
}

@article{Yang:2018uae,
 archiveprefix = {arXiv},
 author = {Yang, Weiqiang and Mukherjee, Ankan and Di Valentino, Eleonora and Pan, Supriya},
 doi = {10.1103/PhysRevD.98.123527},
 eprint = {1809.06883},
 journal = {Phys. Rev. D},
 number = {12},
 pages = {123527},
 primaryclass = {astro-ph.CO},
 title = {{Interacting dark energy with time varying equation of state and the $H_0$ tension}},
 volume = {98},
 year = {2018}
}

@article{Yang:2020uga,
 archiveprefix = {arXiv},
 author = {Yang, Weiqiang and Di Valentino, Eleonora and Mena, Olga and Pan, Supriya and Nunes, Rafael C.},
 doi = {10.1103/PhysRevD.101.083509},
 eprint = {2001.10852},
 journal = {Phys. Rev. D},
 number = {8},
 pages = {083509},
 primaryclass = {astro-ph.CO},
 title = {{All-inclusive interacting dark sector cosmologies}},
 volume = {101},
 year = {2020}
}

@article{Yao:2023jau,
 author = {Yao, Yan-Hong and Meng, Xin-He},
 doi = {10.1016/j.dark.2022.101165},
 journal = {Phys. Dark Univ.},
 pages = {101165},
 title = {{Can interacting dark energy with dynamical coupling resolve the Hubble tension}},
 volume = {39},
 year = {2023}
}

@article{Zhai:2023yny,
 archiveprefix = {arXiv},
 author = {Zhai, Yuejia and Giar\`e, William and van de Bruck, Carsten and Di Valentino, Eleonora and Mena, Olga and Nunes, Rafael C.},
 doi = {10.1088/1475-7516/2023/07/032},
 eprint = {2303.08201},
 journal = {JCAP},
 pages = {032},
 primaryclass = {astro-ph.CO},
 title = {{A consistent view of interacting dark energy from multiple CMB probes}},
 volume = {07},
 year = {2023}
}

@article{Pan:2016jli,
    author = "Pan, Supriya and de Haro, Jaume and Paliathanasis, Andronikos and Slagter, Reinoud Jan",
    title = "{Evolution and Dynamics of a Matter creation model}",
    eprint = "1601.03955",
    archivePrefix = "arXiv",
    primaryClass = "gr-qc",
    doi = "10.1093/mnras/stw1034",
    journal = "Mon. Not. Roy. Astron. Soc.",
    volume = "460",
    number = "2",
    pages = "1445--1456",
    year = "2016"
}

@article{DESI:2025zgx,
    author = "Abdul Karim, M. and others",
    collaboration = "DESI",
    title = "{DESI DR2 results. II. Measurements of baryon acoustic oscillations and cosmological constraints}",
    eprint = "2503.14738",
    archivePrefix = "arXiv",
    primaryClass = "astro-ph.CO",
    reportNumber = "FERMILAB-PUB-25-0169-PPD",
    doi = "10.1103/tr6y-kpc6",
    journal = "Phys. Rev. D",
    volume = "112",
    number = "8",
    pages = "083515",
    year = "2025"
}

@article{DESI:2025fii,
    author = "Lodha, K. and others",
    collaboration = "DESI",
    title = "{Extended dark energy analysis using DESI DR2 BAO measurements}",
    eprint = "2503.14743",
    archivePrefix = "arXiv",
    primaryClass = "astro-ph.CO",
    reportNumber = "FERMILAB-PUB-25-0164-PPD",
    doi = "10.1103/w4c6-1r5j",
    journal = "Phys. Rev. D",
    volume = "112",
    number = "8",
    pages = "083511",
    year = "2025"
}

\end{document}